\documentclass[twocolumn]{svjour3}

\usepackage[latin1]{inputenc}
\usepackage{times}

\usepackage{natbib}
\usepackage{algorithm}
\usepackage{algpseudocode}
\usepackage{psfrag,amsmath, amssymb, amscd, amsfonts, latexsym, epsf, graphicx,amscd}
\usepackage{hyperref}

\usepackage[switch]{lineno}

\def\bbbr{{\mathbb R}} 
\def\bbbz{{\mathbb Z}}

\newcommand{\argmax}{\operatorname{argmax}}

\journalname{arXiv preprint}

\begin{document}

\title{\bf Time-causal and time-recursive wavelets%
   \thanks{The support from the Swedish Research Council 
     (contract 2022-02969) is gratefully acknowledged.}}

\titlerunning{Time-causal and time-recursive wavelets}

\author{Tony Lindeberg}

\institute{Tony Lindeberg,
                Computational Brain Science Lab,
                Division of Computational Science and Technology,
                KTH Royal Institute of Technology,
                SE-100 44 Stockholm, Sweden. 
                \email{tony@kth.se}
              ORCID: 0000-0002-9081-2170}
\date{}

\maketitle

\begin{abstract}
  \noindent
    This paper presents a framework for time-causal wavelet analysis.
  It targets real-time processing of temporal signals, where data from
  the future are not available.
  
  The study builds upon temporal scale-space theory, originating from
  a complete classification of temporal
  smoothing kernels that guarantee non-creation of new structures
  from finer to coarser temporal scale levels.
  We construct temporal wavelets from the temporal derivatives of a
  special time-causal smoothing kernel, referred to as the time-causal
  limit kernel, as arising from the classification of
  variation-diminishing smoothing transformations with the
  complementary requirement of temporal scale covariance, to guarantee
  self-similar handling of structures in the input signal at different
  temporal scales. This enables decomposition of
  the signal into different components at different scales, while
  adhering to temporal causality.

  The paper establishes theoretical foundations
  for these time-causal wavelet representations, and maps 
  structural relationships to the non-causal Ricker / Mexican hat wavelets. We also describe how
  efficient discrete approximations of the presented theory can be
  performed in terms of first-order recursive filters coupled in
  cascade, which enables numerically well-conditioned
  real-time processing with low resource usage.
  We characterize and quantify how the
  continuous scaling properties transfer to the discrete
  implementation, demonstrating how the proposed time-causal wavelet
  representation can reflect the duration of locally dominant temporal
  structures in the input signal.

  We propose this notion of time-causal wavelet
  analysis as a generic multi-purpose tool for signal processing tasks,
  where streams of signals are to be processed in real
  time, specifically for signals that may contain local variations over
  a rich span of temporal scales, or more generally for analysing
  physical or biophysical temporal phenomena, where a fully
  time-causal analysis is called for to be physically realistic, by not
  in any way accessing data from the future,
  as otherwise often done for prerecorded data.
  
\end{abstract}

\section{Introduction}
\label{sec-intro}

Temporal signals can inherently contain different types of structures
at different temporal scales. Making such multi-scale relationships in
the data explicit, and exploring these relationships,
can often substantially improve the performance of signal
processing pipelines. For the purpose of handling such scale
variations in not necessarily only temporal signals, the notion of wavelet
analysis has therefore been developed by the signal processing community,
to decompose signals into different components at different
scales
(Ricker \citeyear{Ric44-GeoPhys},
Morlet {\em et al.\/}
\citeyear{MorAreFouGla82I-GeomPhys,MorAreFouGla82II-GeomPhys},
Grossmann and Morlet \citeyear{GroMor84-SIAM},
Hosken \citeyear{Hos88-FirstBreak},
Mallat \citeyear{Mal89-PAMI,Mal99-book},
Heil and Walnut \citeyear{HeiWal99-SIAM},
Strang \citeyear{Str89},
Daubechies \citeyear{Dau92-book},
Meyer \citeyear{Mey92-book},
Chui \citeyear{Chu92-book},
Ruskain  {\em et al.\/}\ \citeyear{RusETAL92-book},
Donoho and Johnston \citeyear{DonJon93-Mardia},
Teolis \citeyear{Teo95-book},
Walker \citeyear{Wal99-book},
Debnath and Shah \citeyear{DebSha02-book},
Misiti {\em et al.\/}\ \citeyear{MisMisOppPog07-book}).
Due to their scale separating properties and their
computational efficiency, wavelet representations are used as a
powerful generic building block for representing the information contents in image data
and time-dependent signals for a very wide range of signal processing
methods and applications.

While there have been a few wavelet formulations that respect temporal causality
(Szu {\em et al.\/}\ \citeyear{SzuTelLoh92-OptEng},
V{\'a}zquez {\em et al.\/}\ \citeyear{VazMazMilGal05-JOSA}),
traditional wavelet theory is, however, largely
non-causal, making use of filters that are often symmetric or
antisymmetric with respect to the temporal origin.
Such an approach is, however, not feasible in real-time situations, when the future
cannot be accessed.
While one could, in principle, apply non-causal wavelet theory to
real-time signals by introducing explicit temporal buffers with
sufficiently long temporal delays,
beyond which the wavelets would be truncated to zero,
for time-critical applications, where one wants to minimize the
temporal delays, it is of interest to develop truly time-causal wavelets,
that do not make use of any information from the future.

The subject of this paper is to show how a principled temporal wavelet theory can
be developed based on the notion of time-causal temporal scale-space
representation, originally developed for video analysis in
Lindeberg (\citeyear{Lin16-JMIV}), and then dedicated for a pure
time-causal temporal domain in Lindeberg (\citeyear{Lin23-BICY}).
This temporal scale-space theory, which determines canonical filter
shapes for the
temporal wavelets, is based on convolution with a special
temporal smoothing kernel, denoted the time-causal limit kernel, and which
corresponds to the convolution of an infinite set of truncated
exponential kernels in cascade, with especially chosen time constants
to obtain temporal scale covariance.

Temporal scale covariance in this
respect corresponds to self-similarity over temporal scales, so that the
temporal scale-space representation computed for a rescaled signal
corresponds to a rescaling of the temporal scale-space representation
of the original signal, complemented with a shift in the dimension of
the temporal scale parameter. 
By combining this type of time-causal
temporal scale-space representation with temporal derivates, we can
construct mother wavelets for time-causal wavelet representations from
temporal derivatives of the time-causal limit kernel.
In this way, the resulting time-causal wavelets will have the ability
to handle temporal structures at different temporal scales in a provably
self-similar manner, and can in this way constitute the foundation for defining
time-causal and scale-covariant temporal basis functions,
to be used as computational primitives in more composed signal processing
or learning-based methods.

Furthermore,
by forming the differences between temporal scale-space
representations at adjacent temporal scales, we can construct
time-causal bandpass wavelet representations, which are here formally shown
to up to a scaling factor correspond to the time-causal wavelet
representation based on first-order temporal derivatives of the
time-causal limit kernel. From such a temporal bandpass
representation, reconstruction of the original signal is
straightforward, by mere addition over scales of the time-causal bandpass
representations, which opens up possibilities for real-time
manipulation of the signal as decomposed into a time-causal wavelet
representation. This situation is structurally very closely related to
the corresponding property of the Ricker / Mexican hat wavelet, based on
second-order derivatives of the non-causal Gaussian kernel, where
reconstruction, as will be shown here, can be directly performed by integrating the
corresponding second-order non-causal wavelet representations over the
scale parameter, or the differ\-ence-of-Gaussians wavelet, where
reconstruction is performed by instead summing up the
difference-of-Gaussians wavelet representations over the temporal scales.

By the time-causal limit kernel constituting the composition of a set
of truncated exponential kernels in cascade, the temporal smoothing
operation equivalently corresponds to applying a set of first-order
integrators in cascade. In terms of discrete implementation, such
an operation can be well approximated by a set of first-order recursive filters
coupled in cascade. Thereby, the resulting implementation is fully
time-recursive, meaning that no other temporal memory of the past is
needed, beyond the information contained in the temporal scale
channels themselves. Specifically, this implies straightforward and
very efficient
implementation schemes on regular signal processing architectures.

By combining such mother wavelets with different types of
normalizations with respect to the scale parameter, either in terms of
$L_p$-norms or the notion of scale-normalized derivatives,
scale selective signal processing algorithms can then be designed,
that lead to extrema over scales
in the magnitude of the resulting time-causal wavelet
representations at temporal scale levels proportional to
characteristic durations of different types of temporal structures in
the input data
(Lindeberg \citeyear{Lin17-JMIV,Lin18-SIIMS,Lin18-JMIV}),
with close relationships to the previously studied notion of modulus maxima
in non-causal wavelet representations (Mallat and Hwang \citeyear{MalHwa92-IT}).

In this way, we will demonstrate that significant components in more
traditional non-causal wavelet theory can be carried over to a
time-causal temporal domain, which opens up for a variety of
application areas, where either (i)~temporal signals are to be processed
in real time, such as for purposes of real-time monitoring, time
series analysis and prediction, including learning-based systems
as detailed further in Section~\ref{sec-summ-concl}, or as 
components in control-loop systems, that are to react to a continuous inflow
of information, as well as (ii)~analysis of biological or physical signals that
needs to be performed in a completely time-causal manner,
to be biologically or physically realistic, such as in mathematically
based models of biological or physical processes.

In this paper, we develop the underlying fundaments of the proposed
class of time-causal wavelets, with special emphasis on their theoretical
properties, combined with characterizations of scaling properties and
proofs-of-concepts for different types of temporal signals.

Beyond a very few early approaches, this is the first systematic
treatment of time-causal wavelet theory, with structurally close
relations to non-causal wavelets based on derivatives of the Gaussian
kernel. Specifically, we propose that the time-causal wavelet theory,
to be presented here, constitutes a canonical way to define wavelet
representations over a time-causal temporal domain, in a corresponding
way as non-causal wavelets based on derivatives of the Gaussian kernel,
such as the Ricker / Mexican hat wavelets,
can be regarded as canonical wavelet representations over a non-causal
temporal domain, according to the theory for variation-diminishing
convolution transformations,
to be described in Section~\ref{sec-char-1d-smooth-kernels}.

\subsection{Structure of the presentation}

The presentation is organized as follows:
Section~\ref{sec-methods} begins by showing how continuous
time-causal wavelets can be constructed based on temporal derivatives
of a special time-causal smoothing kernel, referred to as the
time-causal limit kernel, with the theoretical results summarized into
continuous time-causal wavelet representations in
Section~\ref{sec-time-caus-wavelets}.

Section~\ref{sec-disc-approx} then shows how this continuous kernel
can be canonically discretized in terms of a set of first-order
recursive filters coupled in cascade, complemented with time-causal
difference operators, to compute discrete approximations of temporal
derivatives of the time-causal limit kernel, thereby leading to a family
of discrete time-causal wavelet representations, as summarized in
Section~\ref{sec-disc-wavelet-repr}.

Section~\ref{sec-char-scaling-props} additionally provides a set of
characterizations of scaling properties of the resulting discrete
wavelet representations, quantifying how well the ideal 
scaling properties of the continuous time-causal wavelets carry over
to a discrete implementation, specifically as depending on the
distribution parameter $c$ of the time-causal limit kernel, which
specifies how densely the temporal scale levels are quantized, and
thereby leading to trade-off issues regarding the accuracy of 
computed features {\em vs.\/}\ the inherent temporal delays,
that by necessity will occur due to the time-causal processing of the
input signals.

Section~\ref{sec-exps} complements with examples of
results of 
computing the proposed time-causal wavelet representations
for different types of temporal signals, including experimental
verification of the exact reconstruction of the input signal from the
proposed subclass of pure time-causal bandpass representations.
Finally, Section~\ref{sec-summ-concl} concludes with a summary and
discussion, including relations to related multi-scale signal
processing and learning approaches, and proposals regarding further
application areas of the proposed time-causal wavelets.

\section{Methods}
\label{sec-methods}

\subsection{Continuous wavelet representations based on temporal
  derivatives of the time-causal limit kernel}
\label{sec-cont-theory}

In this section, we will after (i)~a first review of main concepts regarding
regular, not necessarily time-causal, wavelet theory,
describe (ii)~the time-causal limit kernel and how this kernel
is constrained with respect its shape based on
(iii)~a very general theoretical result regarding
provably variation-diminishing convolution transformations and the
desirable property of basing the time-causal wavelet analysis on a
(iv)~provably scale-covariant convolution kernel.
This material will then constitute the theoretical foundation
for defining time-causal wavelet representations in Section~\ref{sec-results}.

\subsubsection{Continuous wavelet representation}

\paragraph{Mother wavelet}

The notion of a continuous wavelet representation starts from a mother wavelet
$\chi(t)$, which is assumed to satisfy the admissibility criterion
(Mallat \citeyear{Mal99-book} Equation~(4.36))
\begin{equation}
  \label{eq-adm-crit-mother-wavelet}
  C_{\chi}
  = \int_{\omega = 0}^{\infty}
     \frac{|\hat{\chi}(\omega)|^2}{\omega} \, d\omega < \infty
\end{equation}
where
\begin{equation}
   \hat{\chi}(\omega) = \int_{t \in \bbbr} \chi(t) \, e^{-it \omega} \, dt
\end{equation}
denotes the Fourier transform of the mother wavelet $\chi(t)$.
Specifically, the admissibility criterion
(\ref{eq-adm-crit-mother-wavelet}) implies that
the integral of the mother wavelet should be zero%
\footnote{This can be seen from the fact that the admissibility
  criterion (\ref{eq-adm-crit-mother-wavelet}) implies that by
  necessity $\hat{\chi}(0) = 0$, which is also the integral of the mother
  wavelet.}
\begin{equation}
   \int_{t \in \bbbr} \chi(t) \, dt = 0.
\end{equation}
Furthermore, if the mother wavelet decays sufficiently fast to zero
when $|t| \rightarrow \infty$
\begin{equation}
  \label{eq-fast-decrease-mother-wavelet}
    \int_{t \in \bbbr} (1 + |t|) \, |\chi(t)| \, dt < \infty,
\end{equation}
then it holds that $C_{\chi} < \infty$
(Mallat \citeyear{Mal99-book} page~75).

\paragraph{Wavelet transform}

Given that the mother wavelet satisfies the admissibility criterion
and is normalized to unit $L_2$-norm
\begin{equation}
  \int_{t \in \bbbr} |\chi(t)|^2 \, dt = 1,
\end{equation}
the wavelet representation ${\mathcal W}_{\chi} f$ of any function $f \in
L_2(\bbbr)$ can then be
defined as (Heil and Walnut \citeyear{HeiWal99-SIAM} page~629)
\begin{equation}
  \label{eq-def-wavelet-transf}
  ({\mathcal W}_{\chi} f)(u, v) =
    \int_{t \in \bbbr} f(t) \, e^{-u/2} \, \overline{\chi(e^{-u} \, t - v)} \, dt,
\end{equation}
from which the original function can be recovered as
(Heil and Walnut \citeyear{HeiWal99-SIAM} page~629)
\begin{equation}
  \label{eq-def-inv-wavelet-transf}
  f(t) =
  \int_{u \in \bbbr} \int_{v \in \bbbr}
    ({\mathcal W}_{\chi} f)(u, v) \, e^{-u/2} \, \chi(e^{-u} \, t - v) \, du \, dv.
\end{equation}
Basically, in this representation, the variable $u$ constitutes a
logarithmic parameterization of the scaling factor $a = e^u$ in a scaling
transformation of the mother wavelet according to
\begin{equation}
  \label{eq-rescaled-mother-wavelet}
  \chi_a(t) = \frac{1}{a^b} \, \chi(\tfrac{t}{a})
\end{equation}
for some $a \in \bbbr_+$ and some $b \in \bbbr$.

The normalization of the mother wavelet $\chi(t)$ in the forward wavelet transform in
(\ref{eq-def-wavelet-transf}) and the inverse wavelet transform
(\ref{eq-def-inv-wavelet-transf}) corresponds to setting $b = 1/2$
in (\ref{eq-rescaled-mother-wavelet}).
As we will see later, other forms of normalizations of filters across
scales are commonly used in the related area of scale-space theory,
notably $b = 1$ to obtain maximum scale invariance, when computing
wavelet representations in terms of temporal derivatives of the
time-causal limit kernel over multiple
temporal scales, as we will consider in Section~\ref{sec-sc-norm-wavelet-repr}.

\begin{figure*}[hbtp]
  \begin{center}
    $\mathbf{c = 2}$
    
    \medskip
    
    \begin{tabular}{ccc}
      $n = 0$, $\sigma = 1$
      & $n = 1$, $\sigma = 1$
      & $n = 2$, $\sigma = 1$ \\
      \includegraphics[width=0.3\textwidth]{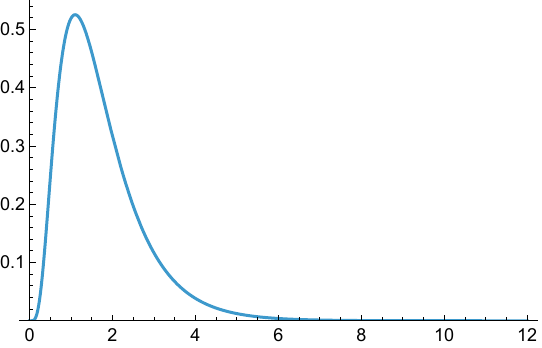}
      & \includegraphics[width=0.3\textwidth]{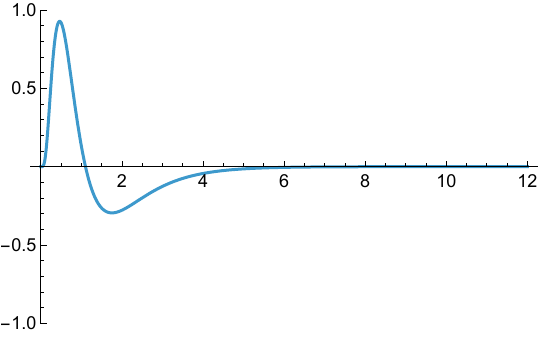}
      &
        \includegraphics[width=0.3\textwidth]{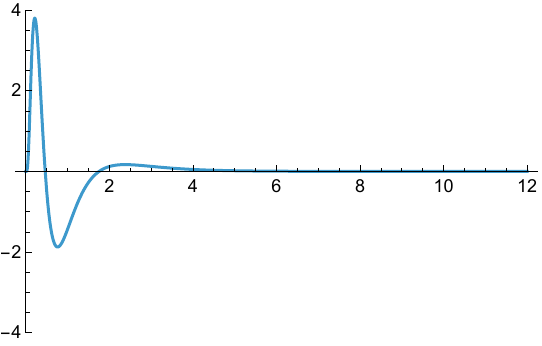}
      \\
      $n = 0$, $\sigma = 2$
      & $n = 1$, $\sigma = 2$
      & $n = 2$, $\sigma = 2$ \\
      \includegraphics[width=0.3\textwidth]{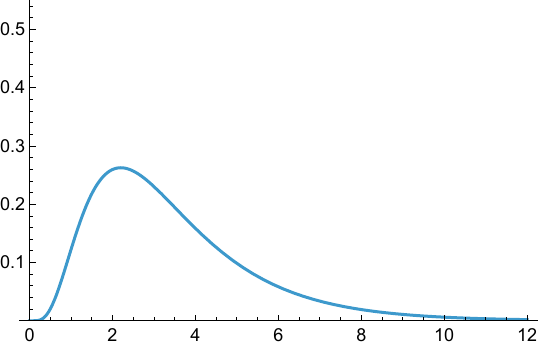}
      & \includegraphics[width=0.3\textwidth]{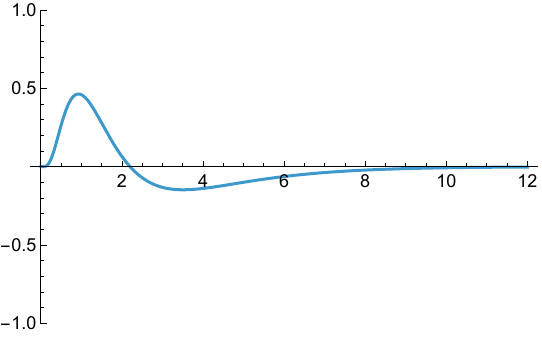}
      &
        \includegraphics[width=0.3\textwidth]{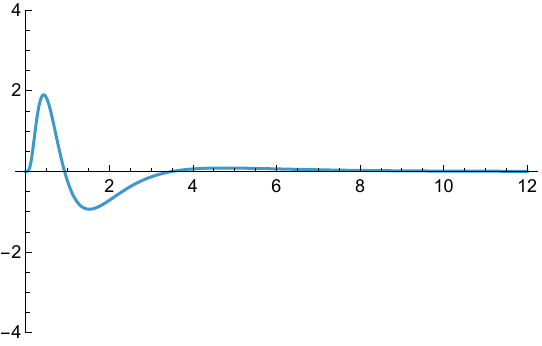}
      \\
    \end{tabular}      

    \bigskip
    \bigskip    

    $\mathbf{c = \sqrt{2}}$

    \medskip
  
    \begin{tabular}{ccc}
      $n = 0$, $\sigma = 1$
      & $n = 1$, $\sigma = 1$
      & $n = 2$, $\sigma = 1$ \\
      \includegraphics[width=0.3\textwidth]{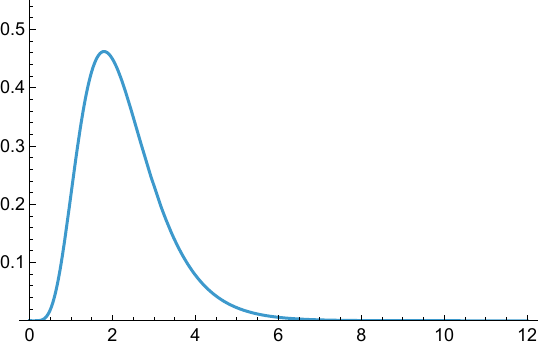}
      & \includegraphics[width=0.3\textwidth]{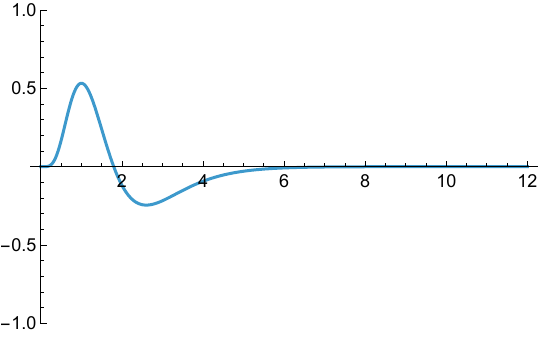}
      &
        \includegraphics[width=0.3\textwidth]{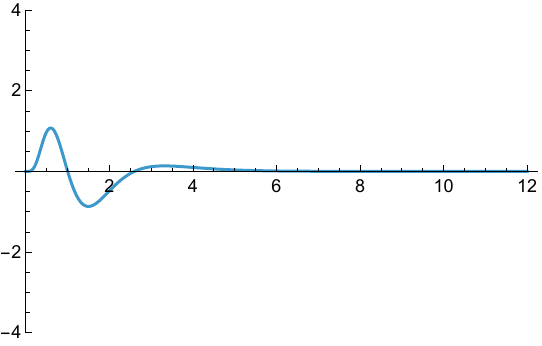}
      \\
      $n = 0$, $\sigma = 2$
      & $n = 1$, $\sigma = 2$
      & $n = 2$, $\sigma = 2$ \\
      \includegraphics[width=0.3\textwidth]{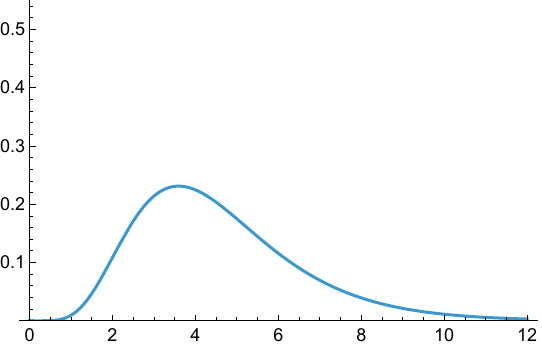}
      & \includegraphics[width=0.3\textwidth]{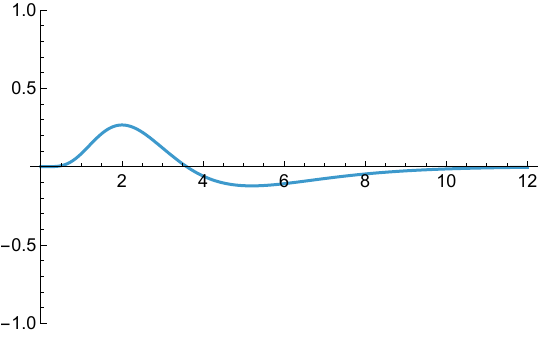}
      &
        \includegraphics[width=0.3\textwidth]{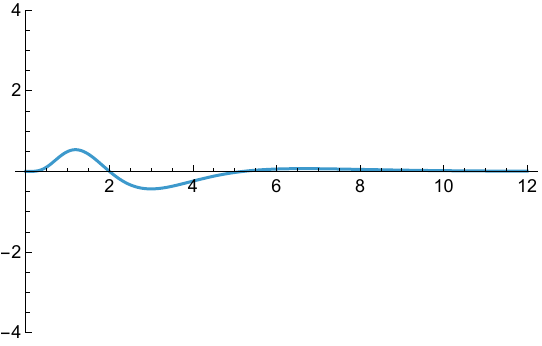}
      \\
    \end{tabular}      
  \end{center}
  \caption{The time-causal limit kernel $\Psi(t;\; \tau, c)$ according
    to (\ref{eq-def-time-caus-lim-kern-inf-conv}), truncated after the
    8 truncated exponential kernels having the longest time constants,
    with its scale-normalized temporal derivatives
    $\Psi_{\zeta^n}(t;\; \tau, c) = \tau^{n \, \gamma/2} \, \Psi_{t^n}(t;\; \tau, c)$
    up to order $n = 2$,
    with the scale normalization power $\gamma = 1$
    corresponding to $L_1$-normalization across scales,
    for different combinations of the temporal scales
    $\sigma = \sqrt{\tau} \in \{ 1, 2 \}$ and
    different values to the distribution parameter
    $c \in \{ \sqrt{2}, 2 \}$. (Horizontal axes: time $t \in [0, 12]$.
    Vertical axes: kernel values with different ranges
    $[0, 0.50]$, $[-1.0, 1.0]$ or $[-4.0, 4.0]$, depending on
    the order $n$ of temporal differentiation.)}
  \label{fig-limitkern-graphs}
\end{figure*}

\subsubsection{The time-causal limit kernel}

While the above theory applies to very general classes of mother
wavelets, we will in this treatment focus solely on mother wavelets that are
obtained from temporal derivatives of a special type of time-causal
kernel, denoted the time-causal limit kernel
(Lindeberg \citeyear{Lin16-JMIV} Section~5;
Lindeberg \citeyear{Lin23-BICY} Section~3)
 \begin{equation}
  \label{eq-time-caus-lim-kern}
  h(t;\; \tau) = \Psi(t;\; \tau, c),
\end{equation}
characterized by having a Fourier transform of the form
\begin{equation}
  \label{eq-FT-comp-kern-log-distr-limit}
     \hat{\Psi}(\omega;\; \tau, c) 
     = \prod_{k=1}^{\infty} \frac{1}{1 + i \, c^{-k} \sqrt{c^2-1} \, \sqrt{\tau} \, \omega}.
\end{equation}
This time-causal kernel has been derived from criteria concerning
the notion of temporal scale space representation, where any continuous
signal $f \colon \bbbr \rightarrow \bbbr$ is to be represented by a temporal
scale-space representation $L \colon \bbbr \times \bbbr_+ \rightarrow \bbbr$
(see Lindeberg \citeyear{Lin23-BICY} Equation~(26))
\begin{equation}
  \label{eq-def-time-caus-scsp}
  L(t;\; \tau, c) = \int_{\xi \in \bbbr}  \Psi(\xi;\; \tau, c) \, f(t - \xi) \, d\xi
\end{equation}
over multiple temporal scale levels $\tau \in \bbbr_+$ in such a way
that no new local extrema or zero-crossings may be created from finer
to coarser levels of temporal scales.
The parameter $c$ is specifically a distribution parameter with
$c > 1$, that determines the discrete scale levels in units of the
variance of the time-causal limit kernel according to%
\footnote{This logarithmic spacing of the temporal scale levels is
  similar to the spacing generated by a uniform spacing with
  respect to the canonical logarithmic
  parameterization of the scale parameter into an effective scale
  parameter $\tau_{\text{eff}}$ according to
  $\tau_{\text{eff}} = A + B \log \tau$, as can be derived from
  axiomatic arguments for a regular spatial scale-space representation
  (Lindeberg \citeyear{Lin92-PAMI}).
  Compared to a uniform spacing of the scale levels relative to the scale
  parameter measured in units of the variance $\tau$ of the temporal
  smoothing kernel, the logarithmic distribution according to
  (\ref{eq-temp-sc-levels}) also leads to significantly shorter delays,
  as further described in Lindeberg (\citeyear{Lin16-JMIV}) Section~4
  and Appendix~1.}
(Lindeberg \citeyear{Lin16-JMIV} Equation~(18))
\begin{equation}
  \label{eq-temp-sc-levels}
  \tau_k = \tau_0 \, c^{2k},
\end{equation}
where $\tau_0 > 0$ represents a reference scale, and
where for common purposes of implementation one often chooses
$\tau_0 = 1$ and either $c = \sqrt{2}$ or $c = 2$.
Alternatively, the time-causal limit kernel can also be parameterized
in terms of the standard deviation $\sigma = \sqrt{\tau}$ of the
time-causal limit kernel, which is of dimension $[\mbox{time}]$.

Figure~\ref{fig-limitkern-graphs} shows graphs of the time-causal limit kernel for a few
values of the temporal scale parameter $\sigma = \sqrt{\tau}$, together with its
first- and second-order temporal derivatives.

\subsubsection{Characterization of continuous smoothing kernels}
\label{sec-char-1d-smooth-kernels}

A theorem by Schoenberg (\citeyear{Sch50})
characterizes the families of continuous smoothing kernels, that
are variation-diminishing in the sense that the number of
zero-crossings in a smoothed signal must never exceed the number of
zero-crossings in the original signal, to having a a bilateral
Laplace-Stieltjes transform of the form
\begin{equation}
  \label{eq-char-var-dim-kernels-cont-case-Laplace}
  \int_{\xi = - \infty}^{\infty} e^{-s \xi} \, h(\xi) \, d\xi =
  C \, e^{\gamma s^2 + \delta s}
  \prod_{i = 1}^{\infty} \frac{e^{a_i s}}{1 + a_i s}
  \quad
\end{equation}
for $-c < \mbox{Re}(s) < c$ and some $c > 0$,
where $C \neq 0$, $\gamma \geq 0$, $\delta$ and $a_i$ are real and
$\sum_{i=1}^{\infty} a_i^2$ is convergent.

From this result,%
\footnote{In
  Equation~(\protect\ref{eq-char-var-dim-kernels-cont-case-Laplace}),
  (i)~the factor $e^{\gamma s^2}$ corresponds to the Laplace-Stieltjes
  transform of the Gaussian kernel, (ii)~the factor $1/(1 + a_i s)$
  corresponds to the Laplace-Stieltjes transform of a truncated
  exponential kernel, while (iii)~the factors $e^{\delta s}$ and
  $e^{a_i s}$ correspond to the Laplace-Stieltjes transforms
  of shift operations. The overall product form of this expression
  does furthermore correspond to convolutions between these 
  primitive smoothing transformations.}
it follows that any
continuous 1-D smoothing kernel can, beyond trivial rescaling
operations and shifts, be decomposed into the following
classes of primitive smoothing transformations:
\begin{itemize}
\item
  convolution with {\em Gaussian kernels\/}
  \begin{equation}
  \label{eq-Gauss-polya-comp}
  h(\xi) = e^{-\gamma \xi^2},
\end{equation}
\item
  convolution with {\em truncated exponential functions\/}
  \begin{equation}
    \label{eq-truncexp-polya-comp}
    h(\xi) =
    \left\{
      \begin{array}{lcl}
        e^{- |\lambda| \xi} & & \xi \geq 0, \\
        0                 & & \xi < 0,
      \end{array}
    \right.
      \quad\quad
    h(\xi) =
    \left\{
      \begin{array}{lcl}
        e^{|\lambda| \xi} & & \xi \leq 0, \\
        0                 & & \xi > 0,
      \end{array}
    \right.
  \end{equation}
   for some strictly positive $|\lambda|$.
\end{itemize}
With respect to smoothing of temporal signals over multiple temporal
scales, this result has the following important implications:
\begin{itemize}
\item
  Over a symmetric non-causal temporal domain, where the
  future is assumed to be accessible, as it can be for pre-recorded
  temporal signals, this result singles out the Gaussian kernel as the
  canonical semi-group of temporal smoothing kernels
  (Karlin \citeyear{Kar68} Theorem~5.2,
  Lindeberg \citeyear{Lin90-PAMI} Theorem~5)
  that obey
  \begin{equation}
    h(\cdot;\; \tau_1) * h(\cdot;\; \tau_2) = h(\cdot;\; \tau_1 + \tau_2)
  \end{equation}
  where $\tau_1 \geq 0$ and $\tau_2 \geq 0$ are the temporal scale
  parameters.
\item
  Over a time-causal temporal domain, where the future cannot be
  accessed, the only possible temporal
  smoothing kernels are truncated exponential kernels
  \begin{equation}
    \label{eq-trunc-exp-kern}
    h_{\exp}(t;\; \mu_k) 
    = \left\{
      \begin{array}{ll}
        \frac{1}{\mu_k} e^{-t/\mu_k} & t \geq 0, \\
        0         & t < 0,
      \end{array}
    \right.
  \end{equation}
  coupled in cascade
  (Lindeberg and Fagerstr{\"o}m \citeyear{LF96-ECCV},
  Lindeberg \citeyear{Lin23-BICY} Section~2.2).
\end{itemize}
In this way, we can obtain a complete characterization of what
idealized temporal smoothing kernels to use, when to smooth
one-dimensional signals to coarser levels of temporal scales
in the cases of either a time-causal or a non-causal temporal domain.

\subsubsection{Scale covariance property of the time-causal limit kernel}
\label{sec-temp-sc-cov-limit-kern}

The definition of the time-causal limit kernel according to
(\ref{eq-time-caus-lim-kern}), originally performed in
Lindeberg (\citeyear{Lin16-JMIV}) Section~5 and then
refined in Lindeberg (\citeyear{Lin23-BICY}) Section~3, starts from the above
characterization of time-causal temporal smoothing kernels as only
possible to express in terms of truncated exponential kernels coupled
in cascade, to choose the time constants $\mu_k$ in an infinite
cascade of convolution operations
\begin{equation}
  \label{eq-def-time-caus-lim-kern-inf-conv}
  \Psi(\cdot;\; \tau, c) = *_{k=1}^{\infty} h_{\exp}(\cdot;\; \mu_k)
\end{equation}
based on intermediate temporal scale levels $\tau_k$ according to
(\ref{eq-temp-sc-levels}) in such a way that the choice of the
temporal scale levels according to
\begin{equation}
  \mu_k = c^{-k} \sqrt{c^2-1} \sqrt{\tau}
\end{equation}
implies that the resulting time-causal limit kernel obeys temporal
scale covariance.

Temporal scale covariance here means that, if we for any temporal
scaling factor $S_t$, that is an integer power of the distribution
parameter $c > 1$ according to
\begin{equation}
  S_t = c^j \quad\quad \mbox{for} \quad\quad j \in \bbbz,
\end{equation}
consider a scaling
transformation of the original signal $f$ into a rescaled signal $f'$
defined by
\begin{equation}
  f'(t') = f(t) \quad\quad \mbox{for} \quad\quad t' = S_t \, t,
\end{equation}
then the resulting temporal scale-space representations
\begin{align}
  \begin{split}
    \label{eq-def-time-caus-temp-scsp-repr}
    L(\cdot;\; \tau, c) =  \Psi(\cdot;\; \tau, c) * f(\cdot)
  \end{split}\\
  \begin{split}
    \label{eq-def-time-caus-temp-scsp-repr-prim}    
    L'(\cdot;\; \tau', c) =  \Psi(\cdot;\; \tau', c) * f'(\cdot)
  \end{split}
\end{align}
are equal (Lindeberg \citeyear{Lin16-JMIV} Equation~(47))
\begin{equation}
  \label{eq-temp-sc-cov}
  L'(t';\; \tau', c) = L(t;\; \tau, c)
\end{equation}
for matching values of the temporal moments $t' = S_t \, t$
and the temporal scale parameters
\begin{equation}
  \tau' = S_t^2 \, \tau.
\end{equation}
This result thus means that the time-causal limit kernel is
self-similar under temporal scaling transformations, which also, with
appropriate complementary temporal scale normalization, carries over to temporal
derivatives of the time-causal limit kernel according to
\begin{equation}
  L_{t^n}(t;\; \tau) = \partial_t^n L(t;\; \tau),
\end{equation}
as will be described later in Section~\ref{sec-sc-cov-sc-norm-ders-scsp}.

Note, however, that since the temporal scale levels are inherently
discrete, as implied by the construction of the temporal scale-space
representation from multiple truncated exponential kernels coupled in
cascade, the temporal scale covariance property holds exactly only
when the temporal scaling factor $S_t$ is an integer power of the
distribution parameter $c$ according to $S_t = c^j$ for $j \in \bbbz$.
For other values of $S_t$, the corresponding result will only be approximate.

\subsubsection{Properties of the time-causal limit kernel}

\paragraph{Characterization of the temporal duration}

The variance of the time-causal limit kernel, which characterizes the
temporal duration, is given by
(Lindeberg \citeyear{Lin16-JMIV} Equation~(35))
\begin{equation}
  \label{eq-var-time-caus-limit-kern}
  V(\Psi(\cdot;\; \tau, c)) = \sum_{k=1}^{\infty} \mu_k^2 = \tau.
\end{equation}

\paragraph{Characterizations of the temporal delay}

The temporal mean of the time-causal limit kernel,
which constitutes one way of characterizing the temporal delay,
is (Lindeberg \citeyear{Lin16-JMIV} Equation~(34))
\begin{equation}
  \label{eq-mean-time-caus-limit-kern}
  M(\Psi(\cdot;\; \tau, c))  = \sum_{k=1}^{\infty} \mu_k
  = \sqrt{\frac{c+1}{c-1}} \, \sqrt{\tau}.
\end{equation}
An alternative characterization of the temporal delay is given by the
position of the temporal maximum of the time-causal limit kernel.
Unfortunately, it is hard to calculate the derivatives of the
time-causal limit kernel in compact closed form.
The maximum point can, however, be estimated from an approximation of
the time-causal limit kernel in terms of Koenderink's
(\citeyear{Koe88-BC}) scale-time model according
to (Lindeberg \citeyear{Lin23-BICY} Equation~(39))
\begin{equation}
  \label{eq-approx-temp-pos-time-caus-log-distr}
  t_{\text{max}}
  \approx \frac{(c+1)^2 \, \sqrt{\tau}}
                       {2 \sqrt{2} \sqrt{(c-1) \, c^3}} = \delta.
\end{equation}

\paragraph{Trade-off issue regarding the choice of the distribution
  parameter $c$}

From these estimates of the temporal delay, in combination with the
result (\ref{eq-temp-sc-cov}) concerning exact scale covariance
properties only for temporal scaling factors $S_t$ that are integer
powers of the distribution parameter $c$, we have a trade-off issue
in the respect that:
\begin{itemize}
\item
  A lower value of the distribution parameter $c$ will
  imply a denser set of temporal scale levels, with numerically better
  ability to approximate temporal scale covariance for temporal scaling
  factors $S_t$ that are not integer powers of the distribution
  parameter $c$.
\item
  A larger value of the distribution parameter $c$ will lead to shorter
  temporal delays, as estimated according to
  (\ref{eq-mean-time-caus-limit-kern})
  and (\ref{eq-approx-temp-pos-time-caus-log-distr}),
  with better ability to respond fast in real-time situations.
\end{itemize}

\paragraph{Differential equation formulation of the time-causal
  temporal scale-space representation}

In terms of differential equations,%
\footnote{The fact that the time-causal scale-space representation
$L(t;\; \tau, c)$ according to (\ref{eq-def-time-caus-scsp}) satisfies
the recurrence relation (\ref{eq-first-ord-int}) between adjacent
levels of temporal scales can be formally verified as follows:
Let us first rewrite the recurrence relation between the input signal
$f_{\text{in}}(t)$ and the output signal $f_{\text{out}}(t)$ between adjacent levels
of temporal scales $\tau_{k-1}$ and $\tau_k$ as
$\partial_t f_{\text{out}}(t) = -1/\mu_k \times ( f_{\text{out}}(t) -
f_{\text{in}}(t))$.
Taking the Laplace transform of this relation then gives
$s \, {\mathcal L}(f_{\text{out}})(s) = -1/\mu_k \times
({\mathcal L}(f_{\text{out}})(s) - {\mathcal L}(f_{\text{in}})(s))$.
Let us next assume that the Laplace transforms ${\mathcal L}(f_{\text{out}})(s)$ and
${\mathcal L}(f_{\text{in}})(s)$ of $f_{\text{out}}(t)$ and $f_{\text{in}}(t)$, respectively,
are related by a transfer function $H(s)$ according to
${\mathcal L}(f_{\text{out}})(s) = H(s) \, {\mathcal L}(f_{\text{in}})(s)$.
Inserting this expression into the Laplace transform of the recurrence
relation, then gives
$s \, H(s) \, {\mathcal L}(f_{\text{in}})(s) =
-1/\mu_k \times (H(s) \, {\mathcal L}(f_{\text{in}})(s) - {\mathcal
  L}(f_{\text{in}})(s))$,
from which we can solve for the transfer function as
$H(s) = 1/(1 + \mu_k \, s)$. From tables of the Laplace transforms of
standard functions, we have that the Laplace transform of the
truncated exponential kernel $e^{-\alpha \, t}$ for $t > 0$,
0 elsewhere, is $1/(s + \alpha)$.
From this, we can conclude that the transfer function $H(s)$
corresponds to the Laplace transform of the truncated exponential
kernel $h_{\exp}(t;\; \mu_k) = 1/\mu_k \times e^{-t/\mu_k}$
for $t > 0$, 0 elsewhere,
which proves the result.}
the relationships between adjacent
levels of temporal scales are given by
(Lindeberg \citeyear{Lin23-BICY} Equation~(11))
\begin{equation}
  \label{eq-first-ord-int}
   \partial_t L(t;\; \tau_k, c) 
   = - \frac{1}{\mu_k} \left( L(t;\; \tau_{k}, c) - L(t;\; \tau_{k-1}, c) \right)
\end{equation}
with the initial condition $L(t;\; 0, c) = f(t)$.

An important property of this type of temporal scale-space
representation is that it is {\em time-recursive\/}. The temporal
scale-space representations $L(t;\; \tau_k, c)$ over the
temporal scale channels with indices $k$ constitute a {\em sufficient
temporal memory of the past\/}, to compute the temporal scale-space
representation and the next temporal moment, given a new input in the
input signal $f(t)$.

The form of the first-order integrators (\ref{eq-first-ord-int}) also
explicitly implies that local perturbations in the signal are
gradually smoothed out from finer to coarser levels of temporal
scales, because of the negative feedback with
respect to the differences $L(t;\; \tau_{k}, c) - L(t;\; \tau_{k-1}, c)$
between the adjacent temporal scale channels.

\paragraph{Cascade smoothing property of the time-causal limit
  kernel}

Due to the definition of the time-causal limit kernel according to
(\ref{eq-def-time-caus-lim-kern-inf-conv}),
it follows that it obeys the following cascade smoothing
property between adjacent levels of temporal scales
$\tau$ and $\tau/c^2$
(Lindeberg \citeyear{Lin16-JMIV} Equation~(41)):
\begin{equation}
  \label{eq-recur-rel-limit-kernel}
  \Psi(\cdot;\; \tau, c)
  = h_{\exp}(\cdot;\; \tfrac{\sqrt{c^2-1}}{c} \sqrt{\tau}) * \Psi(\cdot;\; \tfrac{\tau}{c^2}, c).
\end{equation}
This property
means that, when computing the temporal scale-space
representation or a temporal derivative of the temporal scale-space
representation, it is not necessary to redo all the computations when
to compute those representations over multiple temporal scales.
Instead, it is sufficient to perform the major amount of temporal
smoothing once and for all, and then compute the representations at
the coarser levels of temporal scales, by applying a usually much
smaller number of truncated exponential kernels to the corresponding
representations at finer temporal scales.

Between adjacent temporal
scale levels of the smoothed temporal scale-space representations, we
therefore have
\begin{equation}
  \label{eq-recur-rel-limit-kernel-scsp}
  L(\cdot;\; \tau, c)
  = h_{\exp}(\cdot;\; \tfrac{\sqrt{c^2-1}}{c} \sqrt{\tau}) * L(\cdot;\; \tfrac{\tau}{c^2}, c),
\end{equation}
and regarding temporal derivatives
\begin{equation}
  \label{eq-recur-rel-limit-kernel-ders}
  L_{t^n}(\cdot;\; \tau, c)
  = h_{\exp}(\cdot;\; \tfrac{\sqrt{c^2-1}}{c} \sqrt{\tau}) * L_{t^n}(\cdot;\; \tfrac{\tau}{c^2}, c).
\end{equation}
Since each truncated exponential kernel is a simplifying kernel, that
guarantees non-creation of new local extrema or zero-crossings,
this result means that also the temporal derivatives obey a
simplifying property from finer to coarser levels of temporal scales.

Combined with the variance-based characterization of the time-causal limit
kernel in Equation~(\ref{eq-var-time-caus-limit-kern}),
this simplifying property from finer to coarser levels of temporal
scales means that convolution of a signal with the time-causal limit
kernel with scale parameter $\tau$ can be regarded as effectively
suppressing temporal structures at temporal scales finer than
$\sigma = \sqrt{\tau}$.

\paragraph{Truncating the infinite convolution kernel for
  implementation purposes}

For purposes of numerical or physical implementation, the time-causal
limit kernel can, because of the rapid convergence property of the
temporal scale levels, as corresponding to the convergence properties
of a geometric series, be truncated for a finite number $K$ of layers
of exponential kernels coupled in cascade.

For this purpose, we can choose the time-constants of the $K$
selected truncated exponential kernels coupled in cascade according to
(Lindeberg \citeyear{Lin16-JMIV} Equations~(19)-(20))%
\footnote{To derive these relations, we have made use of the facts that
  (i)~for non-negative convolution kernels, the variances are additive
  under the convolution operation, and (ii)~the variance of truncated
  exponential kernel $h_{\exp}(t;\; \mu_k)$ with time constant
  $\mu_k$ is $V(h_{\exp}(\cdot;\; \mu_k) = \mu_k^2$. These properties
  lead to the relationships in
  Equation~(\ref{eq-muk-log-distr}), given the logarithmic sampling
  of the temporal scale levels $\tau_k =  c^{2k}$ according to
  (\ref{eq-temp-sc-levels}). Concerning
  Equation~(\ref{eq-mu1-log-distr}), we have on the other hand
  determined the first time constant $\mu_1$ in such a way that the
  variance of the first truncated exponential kernel
  $V(h_{\exp}(\cdot;\; \mu_1) = \mu_1^2$ is equal to the sum of the
  variances of all the truncated exponential kernels in the infinite
  convolution operation (\ref{eq-def-time-caus-lim-kern-inf-conv})
  that have been truncated away. In this way, the variance of the
  resulting composed convolution kernel is equal to the same value
  $\tau$ as would be obtained with the original time-causal limit
  kernel $\Psi(t;\; \tau, c)$ with temporal scale parameter $\tau$
  according to (\ref{eq-var-time-caus-limit-kern}).}
\begin{align}
  \begin{split}
     \label{eq-mu1-log-distr}
     \mu_1 & = c^{1-K} \sqrt{\tau},
  \end{split}\\
  \begin{split}
     \label{eq-muk-log-distr}
     \mu_k & = \sqrt{\tau_k - \tau_{k-1}} = c^{k-K-1} \sqrt{c^2-1} \sqrt{\tau} \quad (2 \leq k \leq K),
  \end{split}
\end{align}
where the time constant $\mu_1$ of the first truncated exponential
kernel has been determined in such a way that the temporal variance of the
resulting composed kernel is exactly $\tau$:
\begin{equation}
  V(h(\cdot;\; \mu_1, \mu_2, \dots, \mu_K))
  = \sum_{k = 1}^K \mu_k^2 = \tau.
\end{equation}
In this way, the first truncated
exponential kernel approximates the temporal smoothing
effect of all the truncated exponential kernels that have been
truncated away, as quantified in terms of the temporal variances of
the involved temporal smoothing kernels.

\paragraph{Fulfilment of the admissibility criterion for a mother wavelet}

Based on a partial fraction expansion of the Laplace transform of the
time-causal limit kernel defined
according to  (\ref{eq-FT-comp-kern-log-distr-limit}):
\begin{equation}
  H_{\mbox{\scriptsize $\Psi$}}(q;\; \tau, c)
  = \prod_{k=1}^{\infty} \frac{1}{1 + \mu_k \, q}
  = \sum_{k=1}^{\infty} \frac{A_k}{1 + \mu_k \, q},
\end{equation}
with the time constants
$\mu_k$ as functions of $\tau$ and $c$ according to
(\ref{eq-muk-log-distr}),
we can determine the coefficients $A_k$ according to
(Lindeberg \citeyear{Lin23-BICY} Equation~(73))
\begin{equation}
  A_k = \prod_{i=1, i \neq k}^{\infty} \frac{1}{1 - \frac{\mu_i}{\mu_k}},
\end{equation}
implying that the $n$:th order temporal derivative of the time-causal limit
kernel will have the following series representation
(Lindeberg \citeyear{Lin23-BICY} Equation~(75)):
\begin{equation}
  \label{eq-ser-decomp-der-time-caus-limit-kern}
  (\partial_{t^n} \Psi)(t;\; \tau, c)
  = \sum_{k = 1}^{\infty} \left( \frac{-1}{\mu_k} \right)^n
  \frac{A_k}{\mu_k} \,  e^{-t/\mu_k} \quad (t \geq 0).
\end{equation}
Due to the exponential decrease of this expression as function of
$t$, it thereby follows that (\ref{eq-fast-decrease-mother-wavelet})
\begin{equation}
  \int_{t \in \bbbr} (1 + |t|) \, |\Psi(t;\; \tau, c)| \, dt < \infty,
\end{equation}
which formally verifies that the time-causal limit kernel fulfils
the admissibility condition (\ref{eq-adm-crit-mother-wavelet})
for being a mother wavelet.

\section{Results}
\label{sec-results}

Given the above treatment about special properties of the time-causal limit
kernel, we will in this section start by describing
(i)~how mother wavelets can be defined from temporal derivatives of
the time-causal limit kernel kernel, and how such mother wavelets can
be used for defining a continuous wavelet transform based on a
discrete temporal scale parameter. We will also describe
(ii)~how the normalization of this kernel over scales
relates to the notion of scale-normalized derivatives in scale-space
theory, including the special role of $L_1$-normalization across scales
and scale selection properties for local characteristic
model signals, as well as (iii)~how that scale normalization relates
to the normalization over scales in more traditional wavelet
representations.

Based on these foundations, we will then describe
(iv)~how the first- and second-order temporal derivatives of the time-causal
limit kernel can be regarded as a time-causal analogues of the Ricker
wavelet, also referred to as the Mexican hat wavelet, and
corresponding to the second-order derivative of the Gaussian kernel,
as well as also analogues of the first-order derivatives of the
non-causal Gaussian kernel.

The intention behind this treatment is to lay out the theoretical
foundations for using the proposed time-causal wavelets as
computational primitives in more composed signal processing and
learning approaches.
Specifically, since the time-causal limit kernel, that we will use as
a fundamental basis for defining the time-causal wavelet
representations in terms of derivatives thereof,
does not have any compact closed-form expression, we
will in this section devote special emphasis on describing its theoretical properties,
so as to be able to infer properties of the derived time-causal
wavelet representations, and in this way mediate a better theoretical
understanding.

An effectively more compact description of some of the main concepts
and results will then be given in
Section~\ref{sec-time-caus-wavelets},
with extensions to discrete temporal wavelets in
Sections~\ref{sec-disc-approx}--\ref{sec-disc-wavelet-repr},
discrete approximations of scaling properties in
Section~\ref{sec-char-scaling-props} and
experimental results of computing the proposed types of time-causal
wavelet representations for different types of temporal signals in Section~\ref{sec-exps}.

\subsection{Time-causal and time-recursive mother wavelets and wavelet
representations}

Our proposal is that the temporal derivatives of the time-causal limit
kernel can with appropriate complementary normalization be used as
mother wavelets
\begin{equation}
  \chi(t;\; c) = \frac{\partial_{t^n} \Psi(t;\; 1, c)}
                              {\| \partial_{t^n} \Psi(t;\; 1, c) \|_p}
\end{equation}
for computing time-causal and time-recursive wavelet
representations according to (\ref{eq-def-wavelet-transf}):
\begin{equation}
  \label{eq-def-wavelet-transf-time-caus}
  ({\mathcal W}_{\chi} f)(u, v;\; c) =
    \int_{t \in \bbbr} f(t) \, e^{-u/2} \, \chi(e^{-u} \, t - v;\; c) \, dt.
\end{equation}
A conceptual difference between this formulation of a time-causal
wavelet transform and the more traditional formulation
(\ref{eq-def-wavelet-transf}), however, is that because of the
discrete nature of the temporal scale levels $\tau_k = c^{2k}$
according to (\ref{eq-temp-sc-levels}), the logarithmic variable $u$
in the time-causal wavelet transform
(\ref{eq-def-wavelet-transf-time-caus})
will only assume discrete values $u_j$ according to
\begin{equation}
  e^{u_j} = c^j,
\end{equation}
that is of the form
\begin{equation}
  u_j = j \, \log c \quad\quad \mbox{for} \quad\quad j \in \bbbz.
\end{equation}
According to traditional wavelet theory, one usually makes
use of $L_2$-norms when normalizing the mother wavelet.
For purposes of computing derivatives
over multiple scales in scale-space theory, however, other choices of
$L_p$-norms are more common for relating image structures over scales
(Lindeberg \citeyear{Lin97-IJCV} Table~3),
as will be detailed further below.

\subsubsection{Relations to scale-normalized derivatives in
  scale-space theory}

For purposes in scale-space theory, the notion of
scale-normal\-ized temporal derivatives can be introduced according to
(Lindeberg \citeyear{Lin97-IJCV} Equation~(6);
Lindeberg \citeyear{Lin17-JMIV} Equation~(6))
\begin{equation}
  \label{eq-temp-sc-norm-ders}
  \partial_{\zeta} = \tau^{\gamma/2} \partial_t,
\end{equation}
where $\gamma \in \bbbr_+$ is a scale normalization parameter.

\subsubsection{Scale covariance property of scale-normalized temporal
  derivatives}
\label{sec-sc-cov-sc-norm-ders-scsp}

As shown in Lindeberg (\citeyear{Lin16-JMIV}) Appendix~3,
under a temporal scaling transformation of the form
\begin{equation}
  f'(t') = f(t) \quad\quad \mbox{for} \quad\quad t' = c^{j'-j} \, t,
\end{equation}
with the corresponding transformation between corresponding
temporal scale levels of the form
\begin{equation}
  \tau' = c^{2(j'-j)} \tau,
\end{equation}
the scale-normalized temporal derivatives defined from
temporal differentiation of convolutions with the time-causal
limit kernel $\Psi(t;\; \tau, c)$ do
for any temporal input signal relate according to
\begin{align}
  \begin{split}
    \label{eq-transf-prop-sc-norm-temp-ders-limit-kern}
    L'_{\zeta'^n}(t';\, \tau', c) 
    & = c^{(j'-j)n (\gamma-1)} \, L_{\zeta^n}(t;\, \tau, c).
  \end{split}
\end{align}
Specifically, local temporal
scale estimates $\hat{\tau}$ and $\hat{\tau}'$, as determined from local extrema 
over temporal scales in the two temporal domains, will be assumed at corresponding temporal scale
levels and will thus be transformed in a scale-covariant way for any
temporal scaling transformation of the form $t' = c^{j'-j} \, t$,
according to
\begin{equation}
  \hat{\tau}' = c^{2(j'-j)} \hat{\tau}.
\end{equation}
Furthermore, in the special case when $\gamma = 1$,
the scale-normalized derivatives are equal
\begin{equation}
  \label{eq-transf-prop-sc-norm-temp-ders-limit-kern-perfect-sc-inv}
  L'_{\zeta'^n}(t';\, \tau', c) = L_{\zeta^n}(t;\, \tau, c).
\end{equation}
This means that we can then readily compare the magnitudes of
scale-normalized derivatives between matching temporal moments
$\{ t \leftrightarrow t' \}$
and temporal scales
$\{ \tau \leftrightarrow \tau' \}$.


\subsubsection{Scale selection properties for characteristic temporal
  signals}
\label{sec-scsel-theory-gauss-scsp}

By studying the evolution properties of such temporal scale-normalized
derivatives over scales for the {\em non-causal\/} temporal scale-space
representation defined from convolution with Gaussian kernels
according to
\begin{equation}
  L(\cdot;\; \tau, c) =  g(\cdot;\; \tau) * f(\cdot),
\end{equation}
where the non-causal temporal Gaussian kernel is given by
\begin{equation}
  \label{eq-1D-gauss-kernel}
  g(t;\; \tau) = \frac{1}{\sqrt{2 \pi \, \tau}} \, e^{-t^2/2\tau},
\end{equation}
it specifically follows that:
\begin{itemize}
\item
  For a 1-D Gaussian blob 
  \begin{equation}
    f(t) = g(t;\; \tau_0)
  \end{equation}
  the strongest magnitude response of the scale-normalized
  second-order derivative operator
  \begin{equation}
    L_{\zeta\zeta}(t;\; \tau) = \tau \, L_{tt}(t;\; \tau) 
  \end{equation}
  is assumed for the time moment $t = 0$ and the temporal scale level
  (Lindeberg \citeyear{Lin17-JMIV} Equation~(13))  
  \begin{equation}
    \label{eq-sc-sel-peak-1D-general-gamma}
    \hat{\tau} = \frac{2\gamma}{3 - 2\gamma} \, \tau_0 
  \end{equation}
  proportional to the temporal scale $\tau_0$ in the input signal.
  For the specific choice of
  \begin{equation}
    \label{eq-gamma-0p75-2nd-der-gauss-temp-scsp}
    \gamma = \gamma_2 = \frac{3}{4}
  \end{equation}
  we have a perfect match
  to the characteristic temporal scale in the input signal according to
  \begin{equation}
    \label{eq-sigma-hat-spec-blob}
    \hat{\tau} = \tau_0.
  \end{equation}
\item
  For a 1-D Gaussian edge
  \begin{equation}
    f(t) = \int_{\xi = -\infty}^{t} g(\xi;\; \tau_0) \, d\xi
  \end{equation}
  the strongest maximum response of the scale-normalized
  first-order derivative operator
  \begin{equation}
    L_{\zeta}(t;\; \tau) = \tau^{1/2} \, L_{t}(t;\; \tau) 
  \end{equation}
  is assumed for the time moment $t = 0$ and the temporal scale level
  (Lindeberg \citeyear{Lin17-JMIV} Equation~(24))
  \begin{equation}
    \label{eq-sc-sel-ramp-1D-general-gamma}
    \hat{\tau} = \frac{\gamma}{1 - \gamma} \, \tau_0
  \end{equation}
  proportional to the temporal scale $\tau_0$ in the input signal.
  For the specific choice of
  \begin{equation}
    \label{eq-gamma-0p5-1st-der-gauss-temp-scsp}
    \gamma = \gamma_1 = \frac{1}{2}    
  \end{equation}
  we have a perfect match
  to the characteristic temporal scale in the input signal according to
  \begin{equation}
    \label{eq-sigma-hat-spec-edge}
    \hat{\tau} = \tau_0.
  \end{equation}
\end{itemize}
In this way, the scale selection mechanism based on local extrema over
temporal scales leads to scale estimates $\hat{\tau}$ that reflect the
inherent temporal scales $\tau$ in the input signal.
Notably, these results bear close relationships to the matched filter theorem
(Woodward \citeyear{Woo53-book}, Turin
\citeyear{Tur60-matchedfilter}),
in that the scale selection mechanism will choose filters for detecting the
different types of temporal structures in the input signal, that as well as
possible match their size.

\subsubsection{Interpretations in terms of normalization over scale
  with respect to $L_p$-norms of Gaussian derivative kernels}
\label{sec-interpret-sc-norm-Lp-norms}

In relation to normalization of the filters with respect to
$L_p$-norms, it can specifically be shown 
(Lindeberg \citeyear{Lin97-IJCV} Section~9.1) 
that the notion of $\gamma$-normalized derivatives corresponds to 
normalizing the $n$:th-order scale-normalized Gaussian derivative kernels
$g_{\zeta^n}(t;\; \tau)$ over a one-dimensional domain
to constant $L_p$-norms over scale 
\begin{equation}
  \label{eq-Lp-norm-gauss-ders}
  \| g_{\zeta^n}(\cdot;\; \tau) \|_p 
  = \left( 
        \int_{t \in \bbbr} |g_{\zeta^n}(t;\; \tau)|^p \, dx
      \right)^{1/p} 
   = G_{n,\gamma}
\end{equation}
with
\begin{equation}
 \label{eq-sc-norm-p-from-gamma}
  p = \frac{1}{1 + n(1 - \gamma)},
\end{equation}
where the perfectly scale invariant case
\begin{equation}
  \gamma = 1
\end{equation}
corresponds to
$L_1$-normalization for all orders $n$ of temporal differentiation.
Such default $L_1$-normalization corresponds to scale-normalized
temporal derivatives that are dimensionless in units of
$[\mbox{time}]$ and will therefore treat temporal structures at all
temporal scales in a similar manner.

Compare also with the transformation property of scale-normalized
temporal derivatives in the scale covariance property
(\ref{eq-transf-prop-sc-norm-temp-ders-limit-kern}), where the choice
of $\gamma = 1$ corresponding to $p = 1$ makes the scale-normalized temporal
derivatives equal at corresponding temporal moments $\{ t \leftrightarrow t' \}$
and temporal scales $\{ \tau \leftrightarrow \tau' \}$.

With regard to the above detection of the scale of a 1-D Gaussian blob
from extrema over scale in the magnitude of the scale-normalized second-order
derivative using the scale normalization power $\gamma_2 = 3/4$,
or the detection of the scale of a 1-D Gaussian edge from extrema
over scale in the magnitude of the scale-normalized first-order
derivative using the scale normalization power $\gamma_1 = 1/2$, it can
be noted that both of these scale normalization methods correspond to
normalizing the corresponding Gaussian derivative kernels to constant
$L_p$-norms over scale for the choice of $p = 2/3$.

\subsubsection{Normalization over scales in wavelet representations}
\label{sec-sc-norm-wavelet-repr}

For a rescaled mother wavelet of the form
(\ref{eq-rescaled-mother-wavelet})
\begin{equation}
  \label{eq-rescaled-mother-wavelet-again}
  \chi_a(t) = \frac{1}{a^b} \, \chi(\tfrac{t}{a}),
\end{equation}
where $a \in \bbbr_+$ and $b \in \bbbr$, its $L_p$-norm is given by
\begin{equation}
  \| \chi_a(\cdot) \|_p = \frac{a^{1/p}}{a^b} \| \chi(\cdot) \|_p,
\end{equation}
implying that the $L_p$-norm of the rescaled mother wavelet
(\ref{eq-rescaled-mother-wavelet-again}) is constant over scales if
and only if
\begin{equation}
  b = \frac{1}{p}.
\end{equation}
In regular wavelet representations, the choice $p = 2$ corresponding
to $b = 1/2$ constitutes the default choice.
In scale-space theory, the theoretical results described in
Sections~\ref{sec-sc-cov-sc-norm-ders-scsp}
and~\ref{sec-interpret-sc-norm-Lp-norms} imply
that $p = 1$ corresponding to $b = 1$ does, on the other hand,
constitutes the default choice.

\begin{figure*}[hbtp]
  \begin{center}
    \begin{tabular}{ccc}
      $n = 0$, $\sigma = 1$
      & $n = 1$, $\sigma = 1$
      & $n = 2$, $\sigma = 1$ \\
      \includegraphics[width=0.3\textwidth]{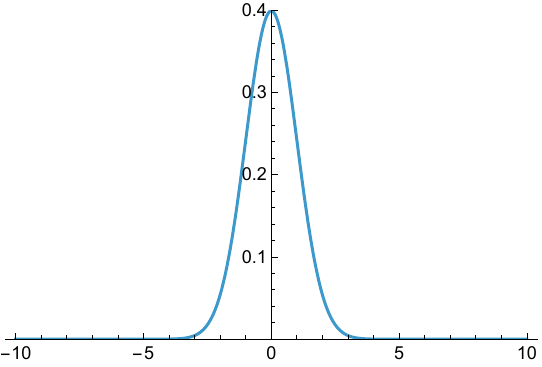}
      & \includegraphics[width=0.3\textwidth]{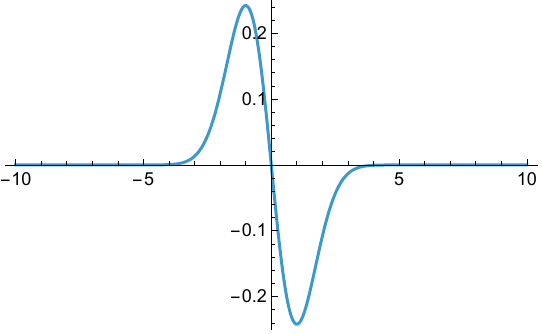}
      &
        \includegraphics[width=0.3\textwidth]{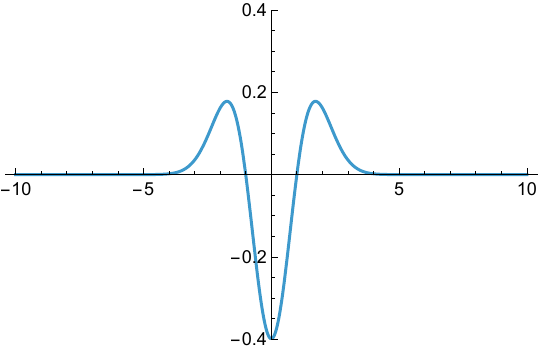}
      \\
    \end{tabular}      

    \bigskip
    
    \begin{tabular}{ccc}
      
      $n = 0$, $\sigma = 2$
      & $n = 1$, $\sigma = 2$
      & $n = 2$, $\sigma = 2$ \\
      \includegraphics[width=0.3\textwidth]{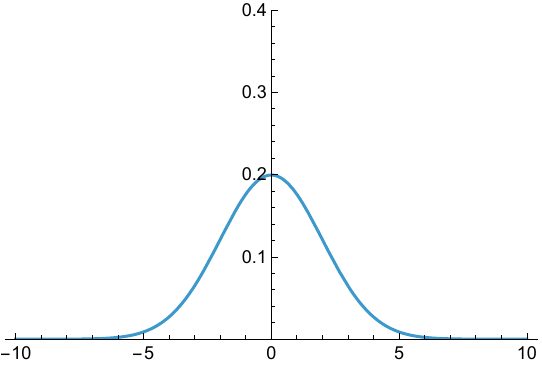}
      & \includegraphics[width=0.3\textwidth]{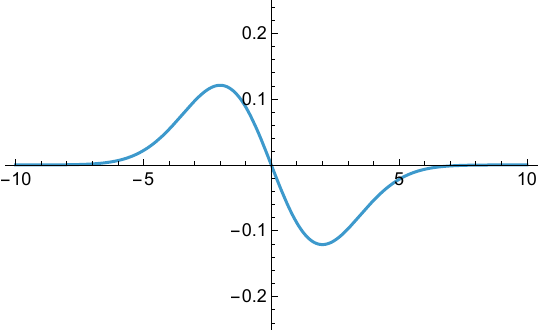}
      &
        \includegraphics[width=0.3\textwidth]{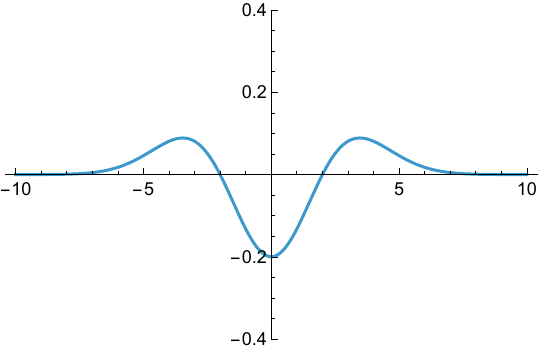}
      \\
    \end{tabular}      
  \end{center}
  \caption{The Gaussian kernel $g(t;\; \tau)$
    according
    to (\ref{eq-1D-gauss-kernel})
    with its scale-normalized temporal derivatives
    $g_{\zeta^n}(t;\; \tau) = \tau^{n \, \gamma/2} \, g_{t^n}(t;\; \tau)$
    up to order $n = 2$,
    with the scale normalization power $\gamma = 1$
    corresponding to $L_1$-normalization across scales,
    for different temporal scales $\sigma = \sqrt{\tau} \in \{ 1, 2 \}$.
    (Horizontal axes: time $t \in [-10, 10]$.
    Vertical axes: kernel values with different ranges, depending on
    the order $n$ of temporal differentiation.)}
  \label{fig-gaussder-graphs}
\end{figure*}

\subsubsection{The first- and second-order derivatives of the time-causal limit
  kernel as time-causal analogues of the Ricker wavelet / the Mexican
  hat wavelet and the wavelet based on the first-order derivative of
  the Gaussian kernel}

In regular wavelet theory, the Ricker wavelet, also known as the
Mexican hat wavelet, and corresponding to the second-order derivative
of the Gaussian kernel
\begin{equation}
  g_{tt}(t;\; \tau)
  = \frac{(t^2 - \tau)}{\tau^2} \, g(t;\; \tau)
  = \frac{(t^2 - \tau)}{\tau^2} \, \frac{1}{\sqrt{2 \pi \tau}} \, e^{-t^2/2\tau}
\end{equation}
constitutes a natural choice for defining a mother wavelet.
With normalization to unit $L_2$-norm over scales for
\begin{equation}
  \| g_{tt}(\cdot;\; \tau) \|_2
  = \frac{\sqrt{\frac{3}{2}}}{2 \sqrt[4]{\pi} \, \tau ^{5/4}},
\end{equation}
we obtain the following mother wavelet
\begin{equation}
  \chi_{g_{tt},L_2}(t;\; \tau) 
  = \frac{g_{tt}(t;\; \tau)}{\| g_{tt}(\cdot;\; \tau) \|_2}
  = \frac{2 \left(t^2-\tau \right) \, e^{-\frac{t^2}{2 \tau }}}
             {\sqrt{3} \sqrt[4]{\pi } \, \tau^{5/4}},
\end{equation}
and with normalization to unit $L_1$-norm over scales for
\begin{equation}
  \| g_{tt}(\cdot;\; \tau) \|_1 = \sqrt{\frac{8}{e \pi }} \, \frac{1}{\tau },
\end{equation}
we obtain
\begin{equation}
  \chi_{g_{tt},L_1}(t;\; \tau)
  = \frac{g_{tt}(t;\; \tau)}{\| g_{tt}(\cdot;\; \tau) \|_1}
  = \frac{\sqrt{e} \left(t^2-\tau \right) \, e^{-\frac{t^2}{2 \tau }}}
             {4 \tau ^{3/2}}.
           \end{equation}
In a corresponding manner, we can also start from the first-order derivative
of the Gaussian kernel
\begin{equation}
  g_t(t;\; \tau)
  = -\frac{t}{\tau} \, g(t;\; \tau)
  = -\frac{t}{\tau} \, \frac{1}{\sqrt{2 \pi \tau}} \, e^{-t^2/2\tau}
\end{equation}
to define a mother wavelet.
With normalization to unit $L_2$-norm over scales for
\begin{equation}
  \| g_t(\cdot;\; \tau) \|_2
  = \frac{1}{2 \sqrt[4]{\pi} \, \tau ^{3/4}},
\end{equation}
we obtain the following mother wavelet
\begin{equation}
  \chi_{g_t,L_2}(t;\; \tau) 
  = \frac{g_t(t;\; \tau)}{\| g_t(\cdot;\; \tau) \|_2}
  = -\frac{\sqrt{2} \, t \, e^{-\frac{t^2}{2 \tau }}}
               {\sqrt[4]{\pi} \, \tau ^{3/4}},
\end{equation}
and with normalization to unit $L_1$-norm over scales for
\begin{equation}
  \| g_t(\cdot;\; \tau) \|_1 = \sqrt{\frac{2}{\pi}} \, \frac{1}{\sqrt{\tau}},
\end{equation}
we obtain
\begin{equation}
  \chi_{g_t,L_1}(t;\; \tau)
  = \frac{g_t(t;\; \tau)}{\| g_t(\cdot;\; \tau) \|_1}
  = -\frac{t \, e^{-\frac{t^2}{2 \tau }}}{2 \, \tau }.
\end{equation}
Figure~\ref{fig-gaussder-graphs} shows graphs of the Gaussian
derivative operators underlying these constructions, here visualized
in terms of scale-normalized derivatives for scale normalization
power $\gamma = 1$ and corresponding to $L_1$-normalization over
scales.

Notably, given the uniqueness of the Gaussian kernel for constructing
a scale-space representation over a symmetric one-dimensional domain,
as based on the characterization of one-dimensional smoothing kernels
in Section~\ref{sec-char-1d-smooth-kernels}, and the uniqueness of
temporal smoothing with truncated exponential kernels over a
time-causal temporal domain, as also resulting from the same
characterization of one-dimensional smoothing kernels,
with the special role of the time-causal limit kernel as leading to
temporal scale covariance, as described in
Section~\ref{sec-temp-sc-cov-limit-kern},
the Gaussian kernel can be regarded as the canonical choice of
smoothing kernel over a non-causal temporal domain, whereas the
time-causal limit kernel can be regarded as a canonical choice of
smoothing kernel over a time-causal temporal domain.

Given this theoretical and conceptual background, we propose to
consider the first- and second-order temporal derivatives of the time-causal limit
kernel according (\ref{eq-time-caus-lim-kern}) as the time-casual
analogues of the first- and second-order temporal derivatives of the Gaussian
kernel.

With complementary normalization, we then obtain the following
first-order time-causal mother wavelet based on $L_2$-normalization
\begin{equation}
  \chi_{\Psi_t,L_2}(t;\; \tau, c) 
  = \frac{\Psi_t(t;\; \tau, c)}{\| \Psi_t(\cdot;\; \tau, c) \|_2}
\end{equation}
and the following first-order time-causal mother wavelet based on $L_1$-normalization
\begin{equation}
  \chi_{\Psi_t,L_1}(t;\; \tau, c) 
  = \frac{\Psi_t(t;\; \tau, c)}{\| \Psi_t(\cdot;\; \tau, c) \|_1},
\end{equation}
as well as the following
second-order time-causal mother wavelet based on $L_2$-normal\-ization
\begin{equation}
  \chi_{\Psi_{tt},L_2}(t;\; \tau, c) 
  = \frac{\Psi_{tt}(t;\; \tau, c)}{\| \Psi_{tt}(\cdot;\; \tau, c) \|_2}
\end{equation}
and the following second-order time-causal mother wavelet based on $L_1$-normalization
\begin{equation}
  \chi_{\Psi_{tt},L_1}(t;\; \tau, c) 
  = \frac{\Psi_{tt}(t;\; \tau, c)}{\| \Psi_{tt}(\cdot;\; \tau, c) \|_1}.
\end{equation}
In a corresponding manner, time-causal analogues of
Gaussian-derivative-based mother wavelets for other orders $n$ of temporal
differentiation as well as other exponents $p$
\begin{equation}
  \chi_{g_{t^n},L_p}(t;\; \tau) 
  = \frac{g_{t^n}(t;\; \tau)}{\| g_{t^n}(\cdot;\; \tau) \|_p}
\end{equation}
can also be formulated according to
\begin{equation}
  \chi_{\Psi_{t^n},L_p}(t;\; \tau, c)
  = \frac{\Psi_{t^n}(t;\; \tau, c)}{\| \Psi_{t^n}(\cdot;\; \tau, c) \|_p}.
\end{equation}
Unfortunately, the norms $\| \Psi_{t^n}(\cdot;\; \tau, c) \|_p$ are
hard to compute in closed form
for the temporal derivatives of the time-causal limit kernel.
Therefore, for implementation purposes, these norms can either be computed
numerically, or be moved outside of the details of the actual numerical
implementation of the wavelet analysis of the signal,
by instead basing the corresponding computations on
temporal derivatives of the signal that are not explicitly normalized
in magnitude with respect to the $L_p$-norms of the mother wavelets,
such as by using scale-normal\-ized
derivatives according to (\ref{eq-temp-sc-norm-ders}).

For such purposes, we provide the following numerical values of the
$L_2$-norms and $L_1$-norms of the temporal derivatives of the
time-causal limit kernel up to order 2, computed based on a truncation of the
time-causal limit kernel to the 8 truncated exponential kernels having
the longest time constants $\mu_k$ for the case of the temporal scale
parameter being $\tau = 1$ and the distribution
parameter set to $c = 2$:
\begin{align}
  \begin{split}
    \label{l2-norm-psi-t-c2}
    \| \Psi_t(\cdot;\; 1, 2) \|_2 \approx  0.635,
  \end{split}\\
  \begin{split}
    \label{l1-norm-psi-t-c2}
    \| \Psi_t(\cdot;\; 1, 2) \|_1 \approx 0.995,
  \end{split}\\
  \begin{split}
    \label{l2-norm-psi-tt-c2}
    \| \Psi_{tt}(\cdot;\; 1, 2) \|_2 \approx 2.084,
  \end{split}\\
  \begin{split}
    \label{l1-norm-psi-tt-c2}
    \| \Psi_{tt}(\cdot;\; 1, 2) \|_1 \approx 2.385,
  \end{split}
\end{align}
and for the case of the distribution
parameter $c = \sqrt{2}$:
\begin{align}
  \begin{split}
    \label{l2-norm-psi-t-csqrt2}
    \| \Psi_t(\cdot;\; 1, \sqrt{2}) \|_2 \approx 0.513,
  \end{split}\\
  \begin{split}
    \label{l1-norm-psi-t-csqrt2}
    \| \Psi_t(\cdot;\; 1, \sqrt{2}) \|_1 \approx 0.924,
  \end{split}\\
  \begin{split}
    \label{l2-norm-psi-tt-csqrt2}
    \| \Psi_{tt}(\cdot;\; 1, \sqrt{2}) \|_2 \approx 0.983,
  \end{split}\\
  \begin{split}
    \label{l1-norm-psi-tt-csqrt2}
    \| \Psi_{tt}(\cdot;\; 1, \sqrt{2}) \|_1 \approx 1.555.
  \end{split}
\end{align}
To compute the values of the $L_2$-norms and the $L_1$-norms
of the temporal derivatives
of the time-causal limit kernels for other values of the temporal
scale parameter $\tau$, we can then make use of the following
recursive relationship (Lindeberg \citeyear{Lin16-JMIV} Equation~(177))
\begin{equation}
  \| \Psi_{t^n}(\cdot;\; c^{2j}, c) \|_p 
  = c^{-j(n+1)+j/p} \,
      \| \Psi_{t^n}(\cdot;\; 1, c) \|_p
\end{equation}
to express the $L_2$-norms and the $L_1$-norms at the scale levels
\begin{equation}
  \tau_j = c^{2j},
\end{equation}
which are related to the base level $\tau_0 = 1$ according to integer
powers of the distribution parameter squared $c^2$, in units of the
variance of the time-causal limit kernel.

\subsection{Time-causal wavelet representations for continuous signals}
\label{sec-time-caus-wavelets}

\subsubsection{Time-causal wavelet representations based on temporal
  derivatives of the time-causal limit kernel}

Given the theory presented above, for any input signal $f \colon \bbbr \rightarrow \bbbr$
and given any order $n > 0$ of temporal differentiation of the
time-causal limit kernel $\Psi(\cdot;\; \tau_k, c)$,
we can for a suitable choice of the
scale normalization factor $\gamma$ define the continuous temporal
derivative of the time-causal scale-space representation
$L_{\zeta^n} \colon \bbbr \times \bbbz \rightarrow \bbbr$
according to
\begin{align}
  L_{\zeta^n}(\cdot;\; \tau_k, c)
  = \Psi_{\zeta^n}(\cdot;\; \tau_k, c) * f(\cdot) = \nonumber\\
 = \tau^{n \gamma/2} \, \partial_{t^n} \, \Psi(\cdot;\; \tau_k, c) * f(\cdot)
\end{align}
for the temporal scale levels chosen as
(Lindeberg \citeyear{Lin16-JMIV} Equation~(18))
\begin{equation}
  \tau_k = \tau_0 \, c^{2k} \quad\quad \mbox{for $k \in \bbbz$}.
\end{equation}
With this parameterization, the discrete parameter $k \in \bbbz$ has a similar
role as the continuous variable $u$ has in the parameterization of the scaling
group under the exponential map $e^{u}$ in the formulation of the
continuous wavelet transform according
to~(\ref{eq-def-wavelet-transf}).

If desired, we can then complement this representation by an
additional normalization with respect to a suitable $L_p$-norm of the
$n$:th order temporal derivative of the time-causal limit kernel
according to the theory above,
then, however, also taking into account the scaling properties of the
scale-normalized temporal derivatives.

In the following analysis to be performed, we will, however, not
explicitly normalize the mother wavelets to constant $L_p$-norm,
instead fully relying on the up to a uniform scaling factor
normalization over scales in terms of scale-normalized temporal
derivatives according to (\ref{eq-temp-sc-norm-ders}).

\subsubsection{Time-causal bandpass wavelet representations}
\label{eq-temp-band-pass-wavelets}

If one from an input signal $f(t)$
constructs a set of successively smoothed temporal representations
$L(t; \tau_k, c)$ by convolution with the time-causal limit kernel
according to (\ref{eq-def-time-caus-temp-scsp-repr}),
then the differences between these representations
\begin{equation}
  \label{eq-bandpass-wavelet-repr}
  \Delta L_{\text{DoT}}(t;\; \tau_k, c) = L(t; \tau_k) - L(t; \tau_{k-1}, c)
  \quad \mbox{for $k   \in [2, K]$},
\end{equation}
with the specifically finest bandpass channel given by
\begin{equation}
   \label{eq-bandpass-wavelet-repr-finest-scale}
   \Delta L_{\text{DoT}}(t;\; \tau_1, c) =  L(t; \tau_1) - L(t; \tau_0) = L(t; \tau_1) - f(t),
\end{equation}
will constitute a bandpass representation of the original signal
$f(t)$, with very close structural relationships to a
difference-of-Gaussians representation in the area of image
processing 
(Burt and Adelson \citeyear{BA83-COM}, Crowley and Stern \citeyear{Cro84-dolp}).

In terms of convolution kernels, such a representation can
equivalently, although less efficiently, be computed with
a time-causal convolution kernel of the form
\begin{equation}
  \Delta L_{\text{DoT}}(\cdot;\; \tau_k, c)
  = \Delta \Psi(\cdot;\; \tau_k, \tau_{k-1}, c) * f(\cdot),
\end{equation}
where the difference-of-the-time-causal-limit-kernels (DoT) kernel
$\Delta \Psi$ is given by
\begin{equation}
  \label{eq-diff-of-time-caus-limit-kern}
  \Delta \Psi(t;\; \tau_k, \tau_{k-1}, c) = \Psi(t; \tau_k, c) - \Psi(t; \tau_{k-1}, c)
\end{equation}
for (Lindeberg \citeyear{Lin16-JMIV} Equation~(18))
\begin{equation}
  \tau_k = c^2 \, \tau_{k-1}.
\end{equation}
Specifically, if such a representation is computed between
a finest minimum temporal scale level $\tau_1$, for the
specific choice of defining $L(t; \tau_0, c) = f(t)$ in the boundary case,
and a coarsest maximum temporal scale level $\tau_K$,
then reconstruction of the original signal or the representation at
any other temporal scale level
is straightforward from such a representation,
by merely adding the corresponding bandpass representations over
the temporal scale channels for all the temporal scale levels:
\begin{equation}
  \label{eq-reconstr-temp-scsp-time-caus}
  L(t; \tau_j, c)
  = L(t; \tau_K, c) - \sum_{k = j+1}^K \Delta L_{\text{DoT}}(t;\; \tau_k, c),
\end{equation}
which for the special choice of $j = 0$ gives
\begin{equation}
  \label{eq-reconstr-orig-sign-from-bandpass}
  f(t) = L(t; \tau_K, c) - \sum_{k = 1}^K \Delta L_{\text{DoT}}(t;\; \tau_k, c).
\end{equation}

\subsubsection{Exact reconstruction from the time-causal bandpass representation}

Notably, provided that
\begin{itemize}
\item
  the bandpass channels $\Delta L_{\text{DoT}}(t;\; \tau_k, c)$ for $k \in [1, K]$
  are computed from the temporal scale
  channels according to (\ref{eq-bandpass-wavelet-repr})
  and (\ref{eq-bandpass-wavelet-repr-finest-scale}) and that
\item
  the reconstruction is performed from the bandpass channels
  $\Delta L_{\text{DoT}}(t;\; \tau_k, c)$ according to
  (\ref{eq-reconstr-orig-sign-from-bandpass}),
\end{itemize}
then the reconstruction will be exact, if we can assume that the influence of
numerical errors can be disregarded and provided that the temporal
scale-space representation $L(t; \tau_K, c)$ at the coarsest temporal
scale $\tau_K$ is also provided as input to the reconstruction algorithm.

This property follows directly from inserting the expressions
(\ref{eq-bandpass-wavelet-repr}) and
(\ref{eq-bandpass-wavelet-repr-finest-scale})
for the bandpass representations into the expression
(\ref{eq-reconstr-orig-sign-from-bandpass}) for the reconstruction,
which gives
\begin{align}
   & L(t; \tau_K, c) - \sum_{k = 1}^K \Delta L_{\text{DoT}}(t;\; \tau_k, c) = \nonumber\\
   & = L(t; \tau_K, c)
  - (L(t; \tau_K) - L(t; \tau_{K-1}, c))
     - ... \nonumber\\
  & \quad -  (L(t; \tau_2) - L(t; \tau_{1}, c))
    -  (L(t; \tau_1) - f(t)) = \nonumber\\
  \label{eq-proof-exact-reconstr-from-bandpass-repr}
  &  = f(t).
\end{align}
Of course, when initiating the first-order temporal integrators for
the temporal smoothing operations with truncated exponential kernels
coupled in cascade, there will be transition effects in the first-order
integrators, as originating from the first temporal moment when the
time-recursive analysis is initiated. From the way that the bandpass representations are
defined from explicit differences between the temporal scale channels,
such transitions effects will, however, not affect the exactness of
the reconstruction.%
\footnote{By corresponding arguments, a similar exactness of reconstruction
  applies also to the corresponding discrete bandpass representations
  to be described in Section~\ref{sec-disc-wavelet-repr},
  as obtained from differences between temporal scale channels
  obtained from discrete temporal smoothing based on
  first-order recursive filters in cascade.
  The algebraic form of the proof for that result is similar to the
  above result (\ref{eq-proof-exact-reconstr-from-bandpass-repr}) for
  proving the result about exact reconstruction from
  the continuous bandpass representation.}

\subsubsection{Non-causal bandpass representations based on the second-order
  derivatives of the Gaussian kernel}

Notably, there is a close structural relationship between the above
notion of bandpass wavelets and to the Ricker wavelet, also known as the
Mexican hat wavelet, in that the difference-of-Gaussians operator
can be seen as an approximation to the Laplacian operator in the 2-D
case or the second-order derivative of the Gaussian in the 1-D case.

Since the non-causal Gaussian representation of a 1-D
signal defined by convolution with Gaussian kernels
\begin{equation}
  L(\cdot;\; \tau) = g(\cdot;\; \tau) * f(\cdot)
\end{equation}
satisfies the 1-D diffusion equation
\begin{equation}
  \label{eq-1D-diff-eq}
  \partial_{\tau} L = \frac{1}{2} \, L_{tt},
\end{equation}
it follows from the fact that the difference between the non-causal Gaussian-based temporal
scale-space representations at the scales $\tau$ and
$\tau + \Delta \tau$ will constitute an approximation of the
derivative of the scale-space representation in the scale direction
\begin{equation}
  \partial_{\tau} L(t;\; \tau)
  \approx \frac{L(t;\; \tau + \Delta \tau) - L(t;\; \tau)}{\Delta \tau}.
\end{equation}
This difference in the Gaussian-based temporal scale-space
representations can also be approximated by convolving the
input image with the second-order derivative of the Gaussian kernel
\begin{equation}
  \label{eq-diff-temp-sc-from-2nd-der-gauss}
  \frac{L(t;\; \tau + \Delta \tau) - L(t;\; \tau)}{\Delta \tau}
  \approx \frac{1}{2} (g_{tt}(\cdot;\; \tau) * f(\cdot))(t;\; \tau).
\end{equation}
Thus, we have that, up to a scalar normalization factor and a
normalization with respect to the scale parameter, the Ricker wavelet,
also known as the Mexican hat wavelet, approximately encodes differences in
a Gaussian-based scale-space representation between temporal scales.

The analogous finite-difference-based
relationship to (\ref{eq-diff-temp-sc-from-2nd-der-gauss}) 
becomes exact, if we replace the second-derivative of the Gaussian
kernel with the corresponding difference-of-Gaussian kernel
\begin{equation}
  \operatorname{DoG}(t;\; \tau,  \Delta \tau)
  = g(t;\; \tau + \Delta \tau) - g(t;\; \tau),
\end{equation}
leading to the following relationship for differences-of-Gaussians bandpass representations
\begin{align}
  \label{eq-diff-temp-sc-from-DoG}
  \Delta L_{\text{DoG}}(t;\; \tau) & = L(t;\; \tau + \Delta \tau) -
  L(t;\; \tau) = \nonumber\\
  & =  (\operatorname{DoG}(\cdot;\; \tau,  \Delta \tau)* f(\cdot))(t;\; \tau),
\end{align}
and corresponding to following relationships between the
diff\-erence-of-Gaussians kernel and the second-derivative of the
Gaussian kernel according to
\begin{equation}
  \frac{\operatorname{DoG}(t;\; \tau,  \Delta \tau)}{\Delta t}
  \approx \frac{1}{2} \, g_{tt}(t;\; \tau),
\end{equation}
with the approximation becoming gradually better for decreasing
values of the scale difference $\Delta \tau$.

\subsubsection{Exact reconstruction of the original signal from
  non-causal Laplacian or differences-of-Gaussians responses}

Regarding the second-order derivative of the non-causal
Gaussian kernel, it follows from the diffusion equation
interpretation of the non-causal 1-D scale-space representation
(\ref{eq-1D-diff-eq}), that provided that second-order
Gaussian-derivative-based responses are computed
over a continuum of scale levels over some range $[t_1, t_2]$,
then the signal at the finer scale level $t_2$ can be reconstructed
from the signal at the coarser scale level $t_2$ in combination with
the second-order derivative responses over this scale range according to
\begin{multline}
  \label{eq-reconstr-ricker-wavelet}
  L(t;\; \tau_2) - L(t;\; \tau_1)
  = \int_{t = t_1}^{t_2} \partial_{\tau}   L(t;\; \tau) \, d\tau = \\
  = \frac{1}{2} \int_{t = t_1}^{t_2} L_{tt}(t;\; \tau) \, d\tau
  = \frac{1}{2} \int_{t = t_1}^{t_2}
         (g_{tt}(\cdot;\; \tau) * f(\cdot))(t;\; \tau) \, d\tau,
\end{multline}
which implies the following explicit relationship regarding non-causal
reconstruction from a continuum of Laplacian-of-Gaussian responses:
\begin{equation}
  L(t;\; \tau_1)
  = L(t;\; \tau_2)
  - \frac{1}{2} \int_{t = t_1}^{t_2}
         (g_{tt}(\cdot;\; \tau) * f(\cdot))(t;\; \tau) \, d\tau
\end{equation}
with the following limit case for $\tau_1 = 0$ corresponding to
$L(t;\; 0) = f(t)$:
\begin{equation}
  f(t)
  = L(t;\; \tau_2)
  - \frac{1}{2} \int_{t = }^{t_2}
         (g_{tt}(\cdot;\; \tau) * f(\cdot))(t;\; \tau) \, d\tau.
\end{equation}
Concerning non-causal reconstruction of the original signal from a discrete set of
differences-of-Gaussians bandpass responses $\Delta L_{\text{DoG}}(\cdot;\; \tau_k)$,
we correspondingly have 
\begin{equation}
  f(t)
  = L(t;\; \tau_K)
  - \sum_{k = 0}^{K-1}
         \Delta L_{\text{DoG}}(t;\; \tau_k) 
\end{equation}
with the differences-of-Gaussians response $L_{\text{DoG}}(\cdot;\; \tau_0)$
at the finest scale level $\tau_0 = 0$ defined according to
\begin{equation}
  \Delta L_{\text{DoG}}(\cdot;\; \tau_0) = L(t;\; \tau_1) - f(t)
\end{equation}
and with the differences-of-Gaussians responses $L_{\text{DoG}}(\cdot;\; \tau_k)$
at the coarser scales $\tau_k$ defined as
\begin{equation}
  \Delta L_{\text{DoG}}(\cdot;\; \tau_k)  = L(t;\; \tau_{k+1}) - L(t;\; \tau_{k}).
\end{equation}
These expressions can be seen as non-causal analogues of the
corresponding exact reconstruction (\ref{eq-reconstr-temp-scsp-time-caus})
from a time-causal bandpass representation, again demonstrating a
close structural relationship to regard the temporal derivatives of the
time-causal limit kernel as time-causal analogues of the derivatives
of the regular Gaussian kernel, alternatively differences thereof.

\subsubsection{Conceptual difference between the bandpass representation
  based on the time-causal limit kernel in relation to the bandpass
  representation based on the non-causal Gaussian kernel}

Regarding the difference-of-the-time-causal-limit 
kernel (\ref{eq-diff-of-time-caus-limit-kern}) used in the temporal
bandpass wavelets representation in
Section~\ref{eq-temp-band-pass-wavelets},
it follows from the differential equation interpretation
(\ref{eq-first-ord-int}) in terms of first-order integrators
coupled in cascade, that this difference kernel does instead
represent the {\em first-order\/} temporal derivative of the
time-causal limit kernel according to
\begin{equation}
  \label{eq-rel-bandpass-kernel-1st-temp-der}
  \Delta \Psi(t;\; \tau_k, \tau_{k-1}, c)
  = - \mu_k \, \Psi_{t}(t;\; \tau_k, c),
\end{equation}
and also with the notable difference that the difference between the
temporal scale levels $\tau_k$ and $\tau_{k-1}$ now has to be macroscopic
of the form 
\begin{equation}
  \tau_k = c^2 \, \tau_{k-1}.
\end{equation}

\subsection{Discrete approximations of temporal derivatives of the
  time-causal limit kernel}
\label{sec-disc-approx}

To transfer the above theory over a continuous time-causal temporal
domain to a corresponding discrete time-causal domain, there is indeed
a corresponding discrete temporal scale-space theory to build upon,
which can be used for constructing both theoretically well-founded and numerically
well-conditioned discrete implementations.

\subsubsection{Discretization of the first-order integrators}

A basic result is that the canonical discretization of each layer in the
set of first-order integrators coupled in cascade
(\ref{eq-first-ord-int})
\begin{equation}
  \label{eq-diff-eq-adj-sc-cont-time-caus-scsp}
  \partial_t f_{\text{out}}(t)
  = -\frac{1}{\mu_{k,\text{cont}}} \, ( f_{\text{out}}(t) - f_{\text{in}}(t)) 
\end{equation}
is a first-order recursive filter of the form
\begin{equation}
  \label{eq-norm-update}
  f_{\text{out}}(t) - f_{\text{out}}(t-1)
  = - \frac{1}{1 + \mu_{k,\text{disc}}} \, (f_{\text{out}}(t-1)) - f_{\text{in}}(t)),
\end{equation}
where we here assume that the discrete signals in
Equation~(\ref{eq-norm-update}) are sampled with a grid spacing
$\Delta t$ relative to the continuous signals in
Equation~(\ref{eq-diff-eq-adj-sc-cont-time-caus-scsp}).
Then, it can be shown that (Lindeberg \citeyear{Lin25-TIT} Appendix~B.3)
that the discrete recurrence relation
(\ref{eq-norm-update}) constitutes a true discretization of
the continuous differential equation
(\ref{eq-diff-eq-adj-sc-cont-time-caus-scsp})
provided that
\begin{equation}
  \mu_{k,\text{disc}} = \frac{\mu_{k,\text{cont}}}{\Delta t}.
\end{equation}

\subsubsection{Classification of discrete temporal smoothing kernels}
\label{sec-char-1d-disc-smooth-kernels}
The result that the time-causal smoothing operation should be 
discretized on the form (\ref{eq-norm-update}) also builds upon a classification of discrete
variation-diminishing convolution transformations
by Schoenberg (\citeyear{Sch48}), which states that a discrete kernel
guarantees non-creation of new zero-crossings in the convolved signal
if and only if its generating function is of the form
\begin{equation}
  \label{eq-char-pf}
  \varphi(z) = c \; z^k \; e^{(q_{-1}z^{-1} + q_1z)}
  \prod_{i=1}^{\infty} \frac{(1+\alpha_i z)(1+\delta_i
    z^{-1})} {(1-\beta_i z)(1-\gamma_i z^{-1})},
\end{equation}
where
$c > 0$, $k \in \bbbz$, 
$q_{-1}, q_1, \alpha_i, \beta_i, \gamma_i, \delta_i \geq 0$ and
$\sum_{i=1}^{\infty}(\alpha_i + \beta_i + \gamma_i + \delta_i) <
\infty$,
from which we can recognize the $z$-transform of the discrete
recurrence relation (\ref{eq-norm-update}) as one of the members of
the class of possible functions, as occurring in the denominator of
the product expression in (\ref{eq-char-pf}):
\begin{equation}
  \label{eq-first-order-rec-filters}
  f_{\text{out}}(x)
  =
  f_{\text{in}}(x) + \beta_i \, f_{\text{out}}(x - 1)
  \quad\quad
  (0 \leq \beta_i < 1).
\end{equation}

\subsubsection{Discrete analogue of the time-causal limit kernel}
\label{sec-disc-anal-time-caus-limit-kern}

For formulating a discrete analogue of the time-causal limit kernel,
one special fact to consider is that while the temporal mean $M$ of
the first-order recursive filter is structurally similar as for the
truncated exponential filter $M = \mu_k$, the expression for the
variance of the discrete recursive filter is structurally different
\begin{equation}
  \Delta \tau_k = \mu_k^2 + \mu_k.
\end{equation}
Thus, to determine the time constants $\mu_k$ from the desired scale
increments $\Delta \tau_k$ between adjacent temporal scale levels
according to
\begin{equation}
  \label{eq-Delta-tau}
  \Delta \tau_k
  = \tau_k - \tau_{k-1}
  =\tau_0 \left( c^{2k} - c^{2(k-1)} \right),
\end{equation}
when $k > 1$, for some suitably sufficiently low%
\footnote{In order to compute an actual discrete approximation of the
  convolution of a discrete signal at a certain level, a sufficient
  number of initial scale levels needs to be implemented, to
  approximate the truncation of the infinite convolution sufficiently
  well. For our purposes, we often use a minimum of 4 to 8
  temporal scale levels.}
selected first temporal level $\tau_k > 0$,
and the first amount of incremental smoothing
emulating a discrete approximation of all the truncated exponential
kernels that have been truncated away
according to
\begin{equation}
  \Delta \tau_1 =\tau_0 \, c^{2},
\end{equation}
we should use the following time-constants in the recursive filters
that implement the discrete analogue of the time-causal limit kernel
(Lindeberg \citeyear{Lin23-BICY} Equation~(55))
\begin{equation}
  \label{eq-disc-time-constant}
  \mu_k = \frac{\sqrt{1 + 4 \Delta \tau_k}-1}{2}.
\end{equation}

\begin{figure*}[hbtp]
  \begin{center}
    $\mathbf{c = 2}$
    
    \medskip
    
    \begin{tabular}{ccc}
      $n = 0$, $\sigma = 1$
      & $n = 1$, $\sigma = 1$
      & $n = 2$, $\sigma = 1$ \\
      \includegraphics[width=0.18\textwidth]{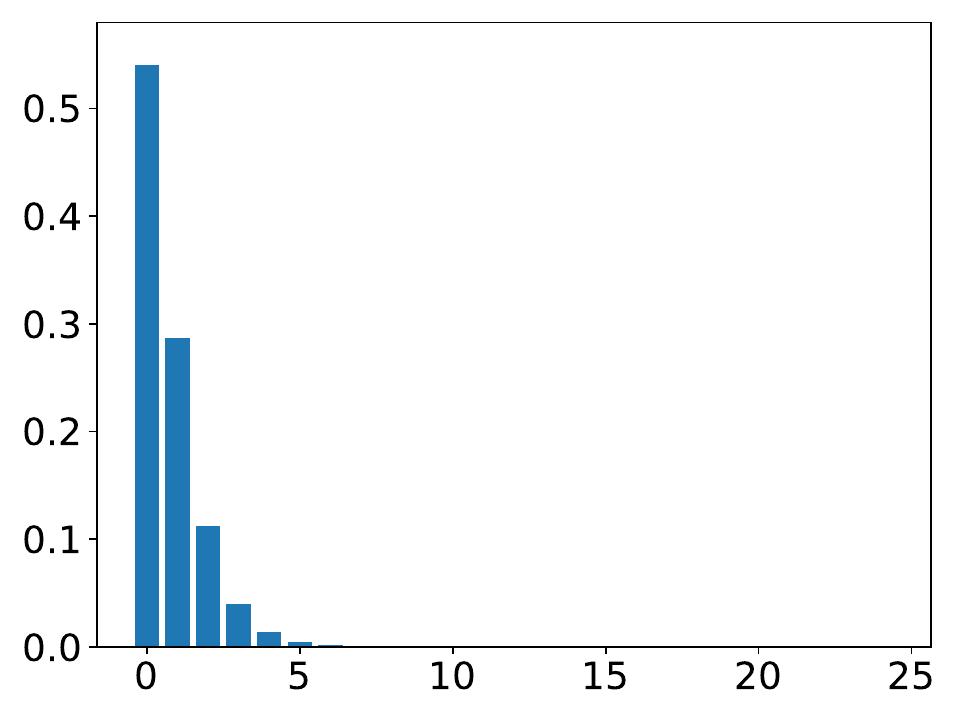}
      & \includegraphics[width=0.18\textwidth]{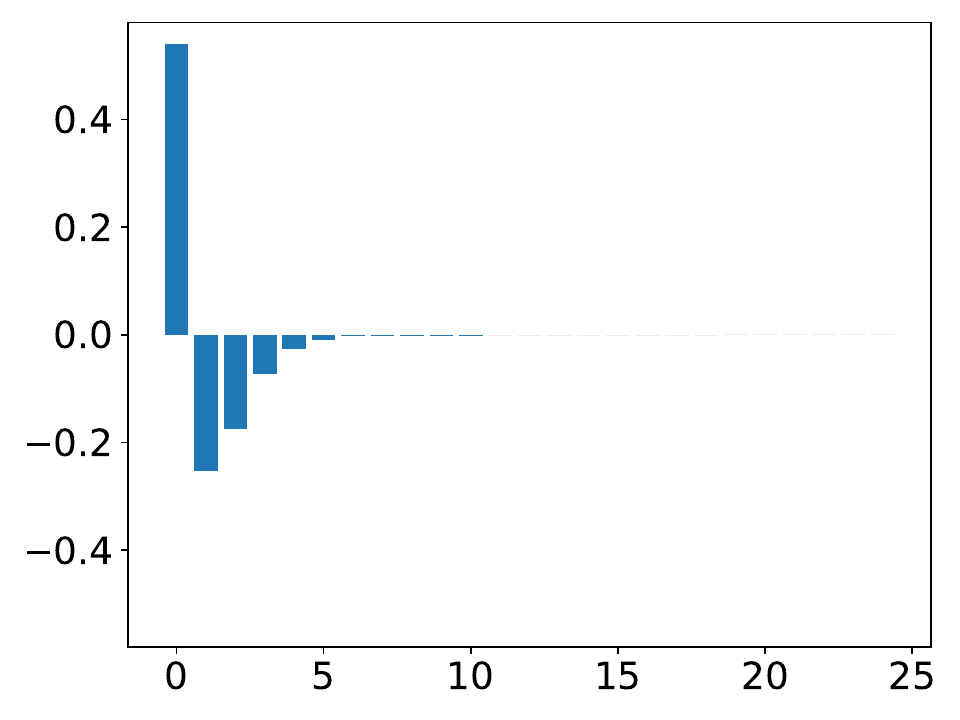}
      &
        \includegraphics[width=0.18\textwidth]{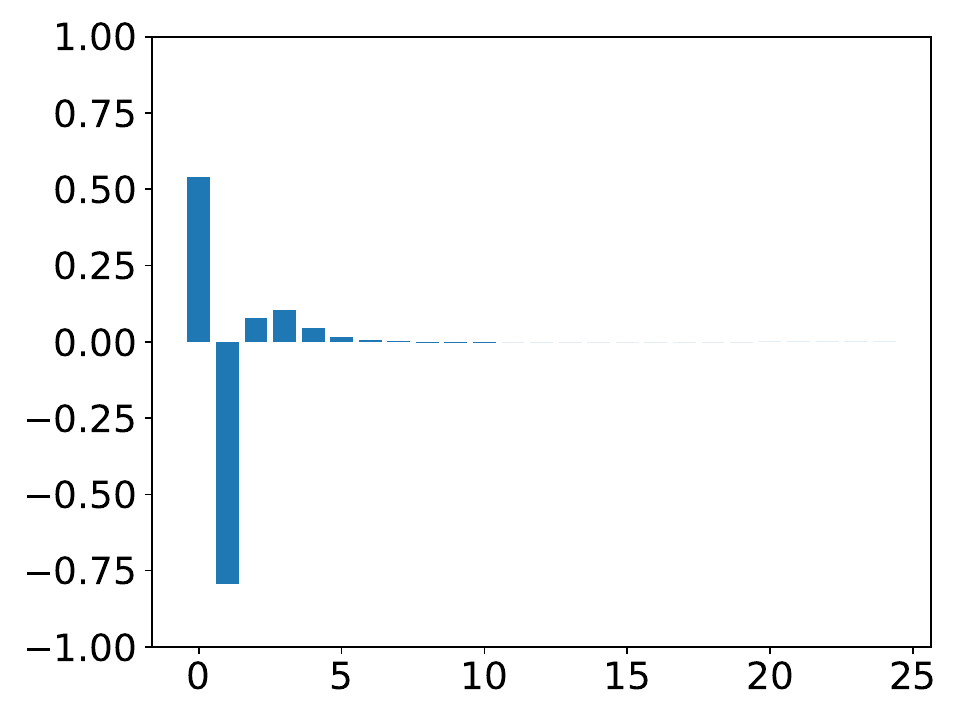}
      \\
      $n = 0$, $\sigma = 2$
      & $n = 1$, $\sigma = 2$
      & $n = 2$, $\sigma = 2$ \\
      \includegraphics[width=0.18\textwidth]{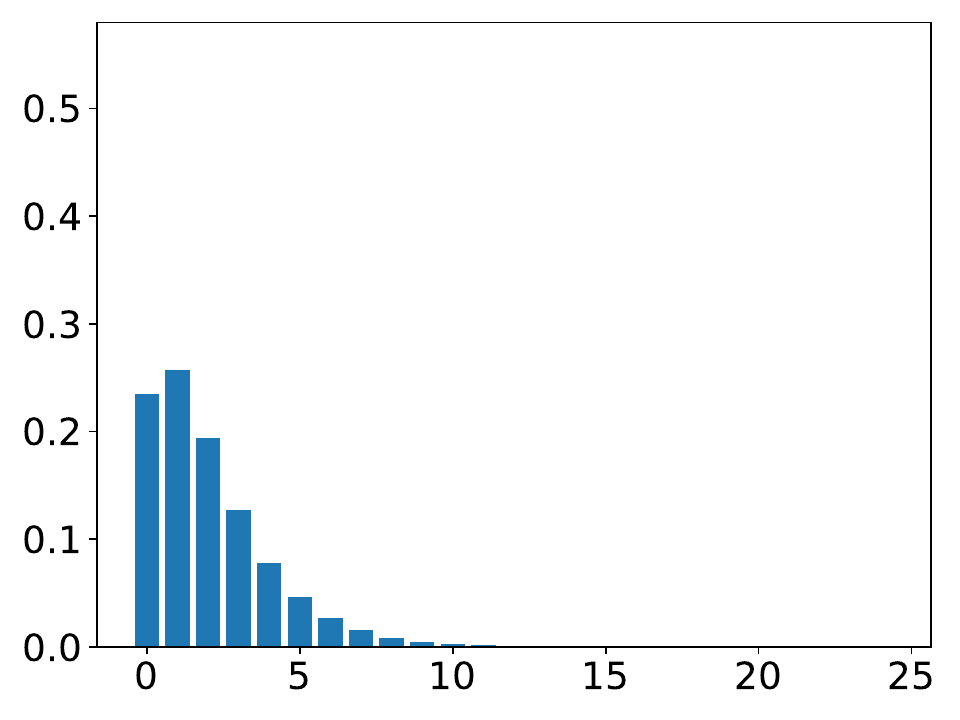}
      & \includegraphics[width=0.18\textwidth]{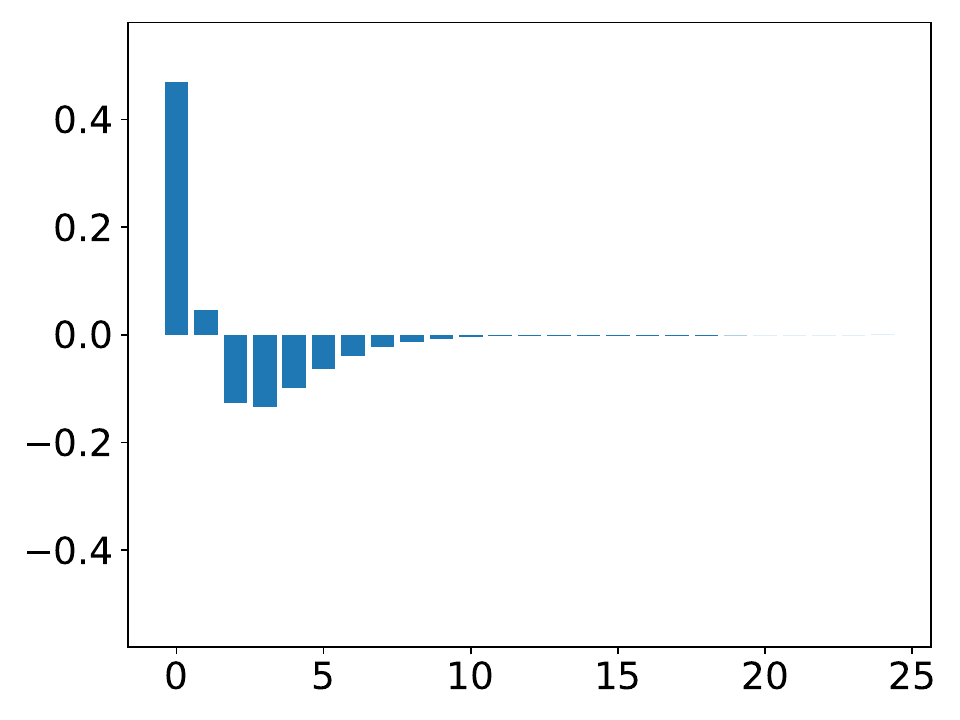}
      &
        \includegraphics[width=0.18\textwidth]{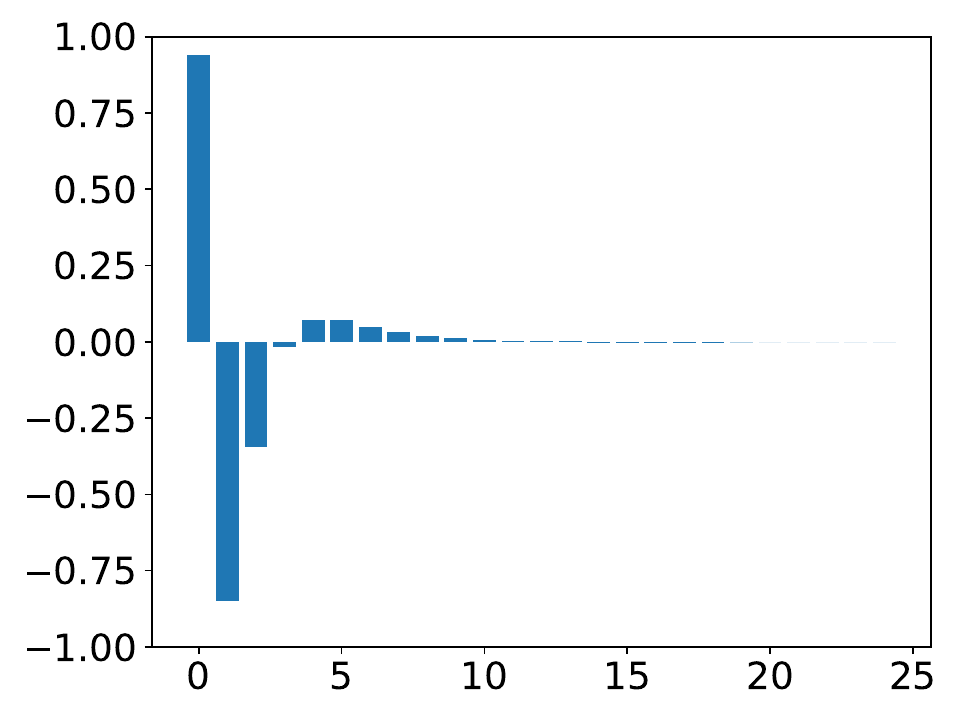}
       \\
      $n = 0$, $\sigma = 4$
      & $n = 1$, $\sigma = 4$
      & $n = 2$, $\sigma = 4$ \\
      \includegraphics[width=0.18\textwidth]{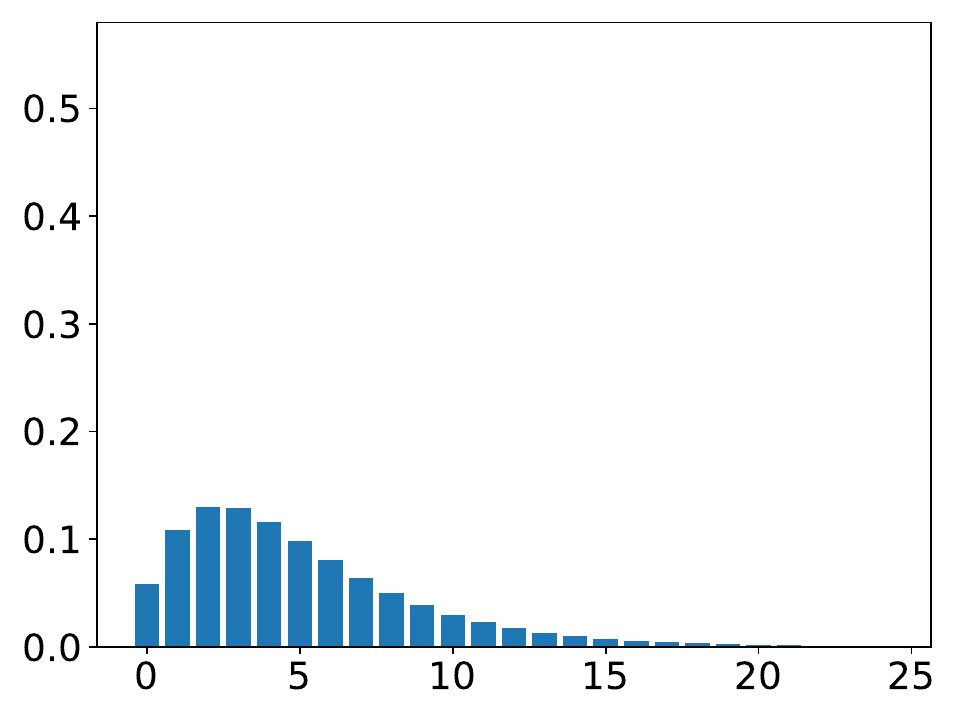}
      & \includegraphics[width=0.18\textwidth]{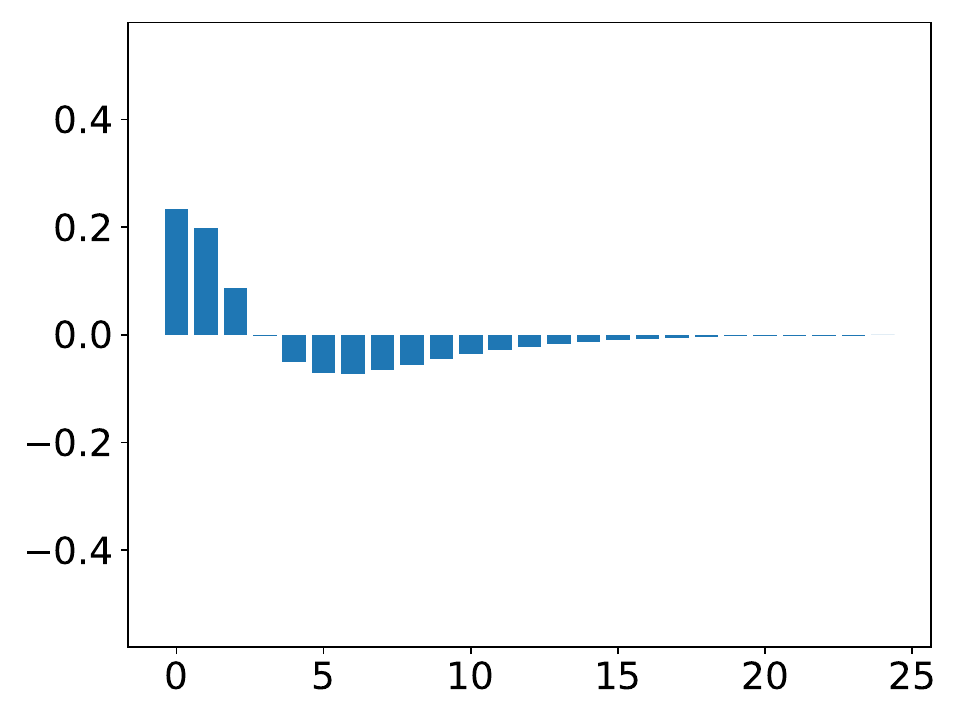}
      &
        \includegraphics[width=0.18\textwidth]{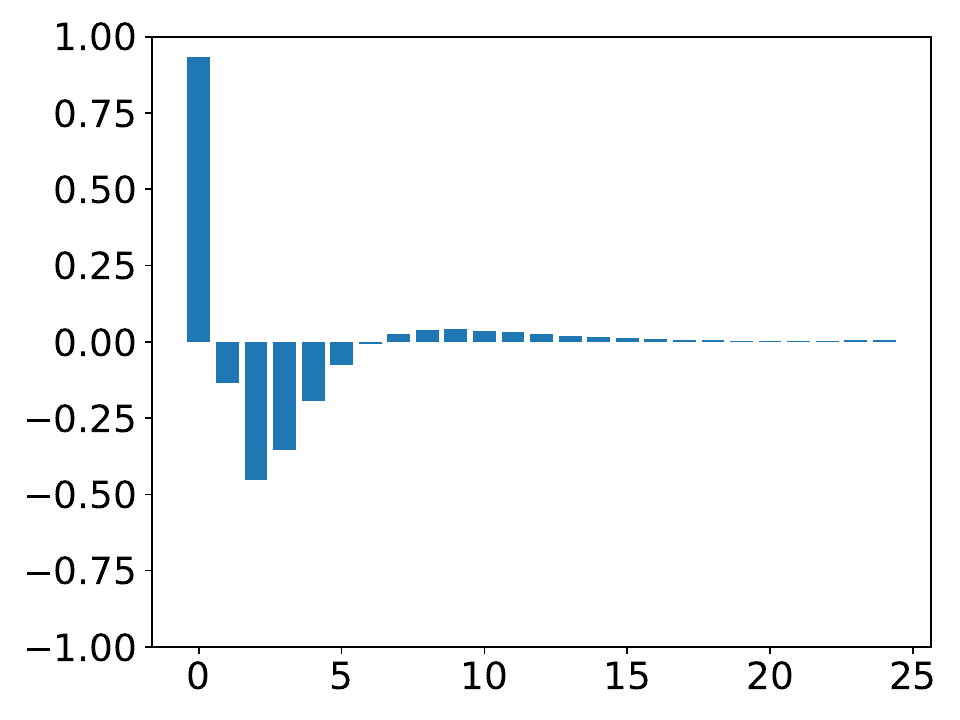}
     \\
    \end{tabular}      

    \bigskip
    \bigskip    

    $\mathbf{c = \sqrt{2}}$

    \medskip
  
    \begin{tabular}{ccc}
      $n = 0$, $\sigma = 1$
      & $n = 1$, $\sigma = 1$
      & $n = 2$, $\sigma = 1$ \\
      \includegraphics[width=0.18\textwidth]{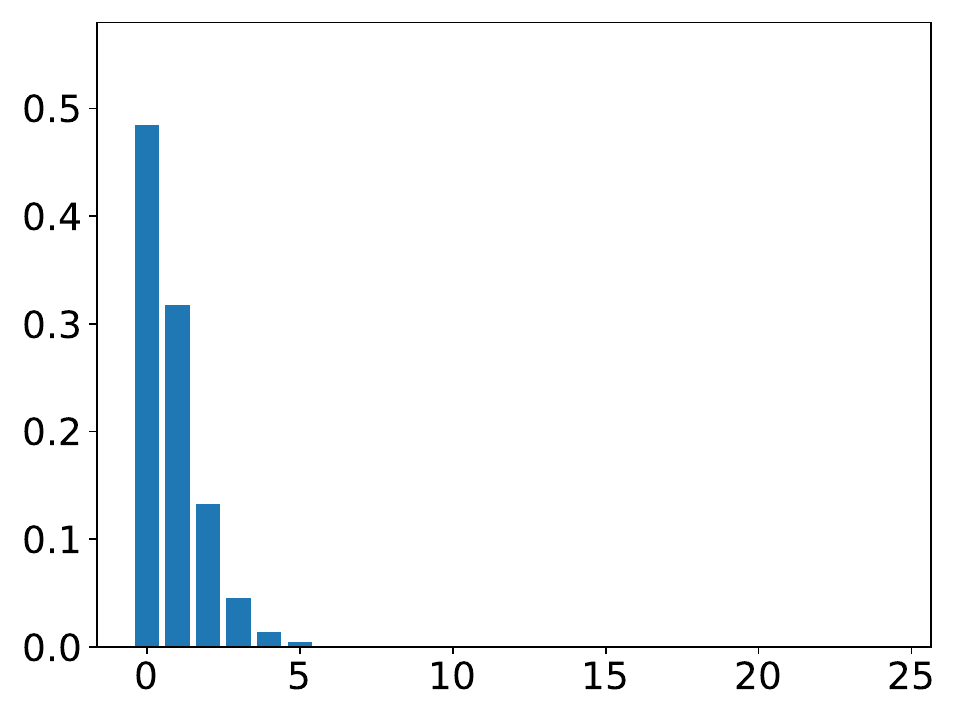}
      & \includegraphics[width=0.18\textwidth]{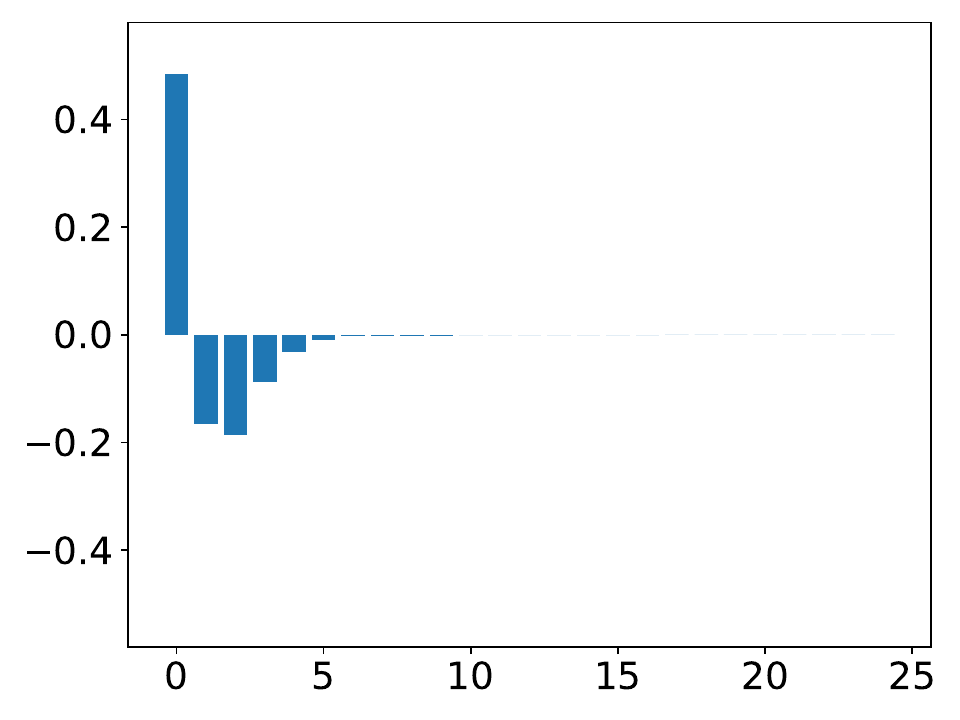}
      &
        \includegraphics[width=0.18\textwidth]{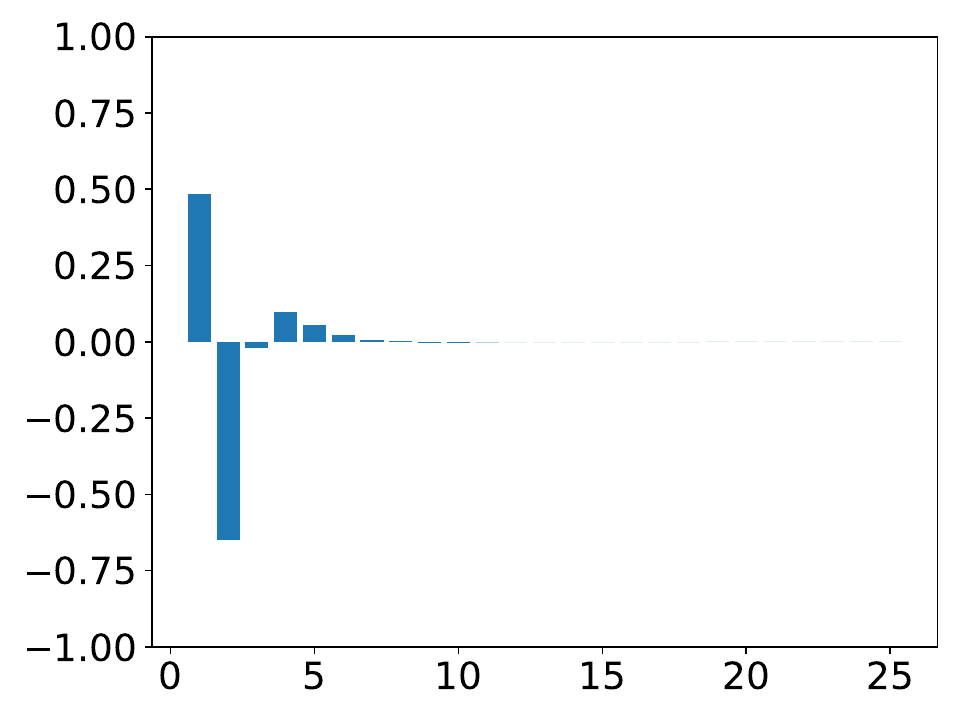}
      \\
      $n = 0$, $\sigma = 2$
      & $n = 1$, $\sigma = 2$
      & $n = 2$, $\sigma = 2$ \\
      \includegraphics[width=0.18\textwidth]{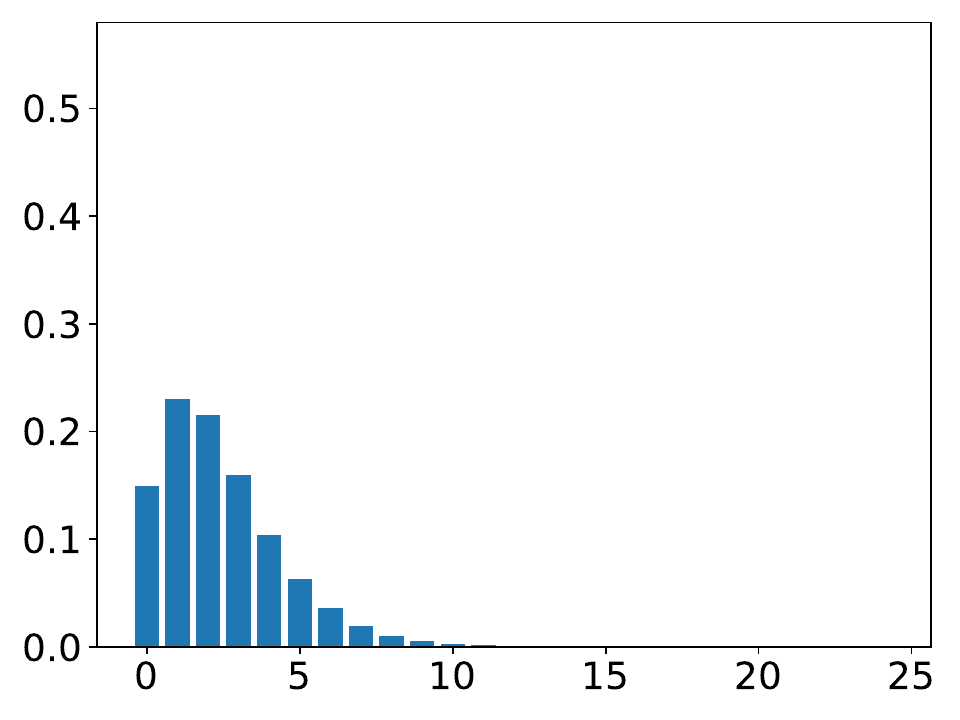}
      & \includegraphics[width=0.18\textwidth]{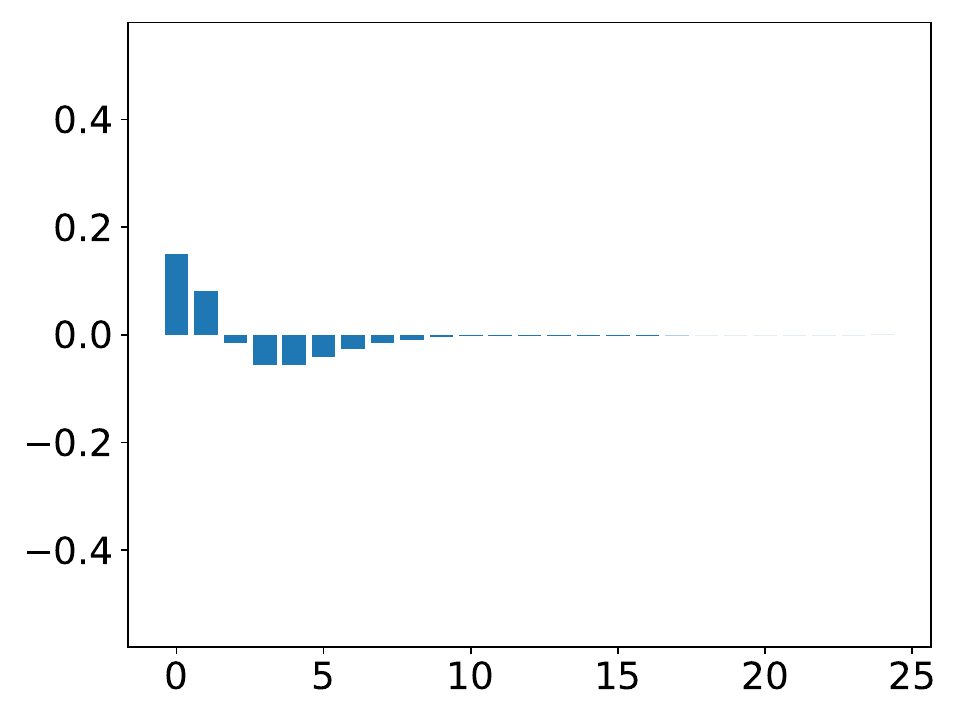}
      &
        \includegraphics[width=0.18\textwidth]{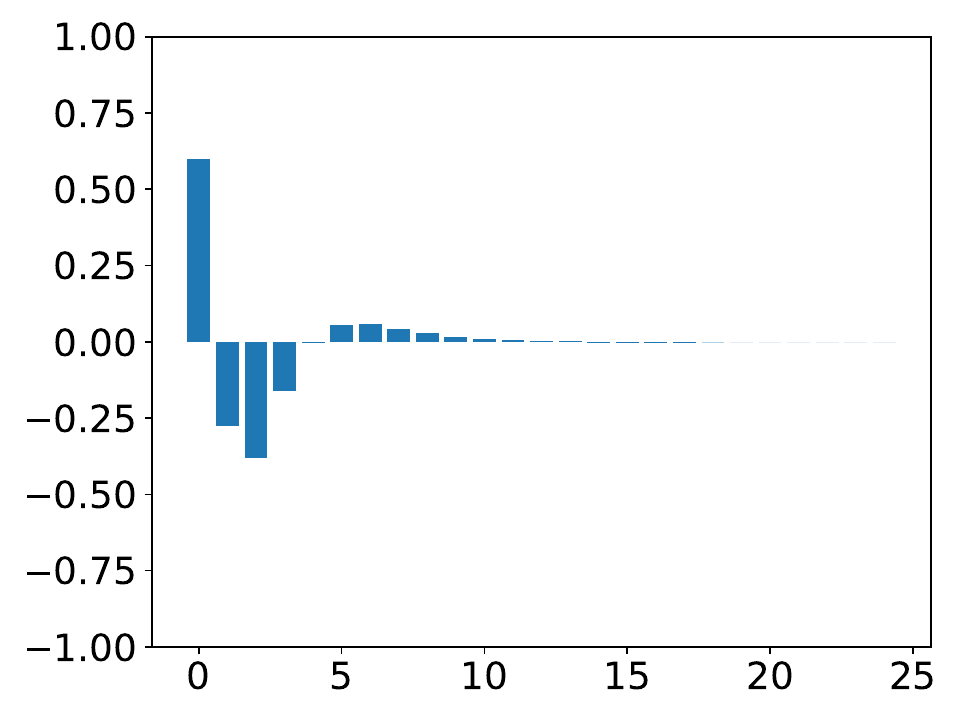}
     \\
      $n = 0$, $\sigma = 4$
      & $n = 1$, $\sigma = 4$
      & $n = 2$, $\sigma = 4$ \\
      \includegraphics[width=0.18\textwidth]{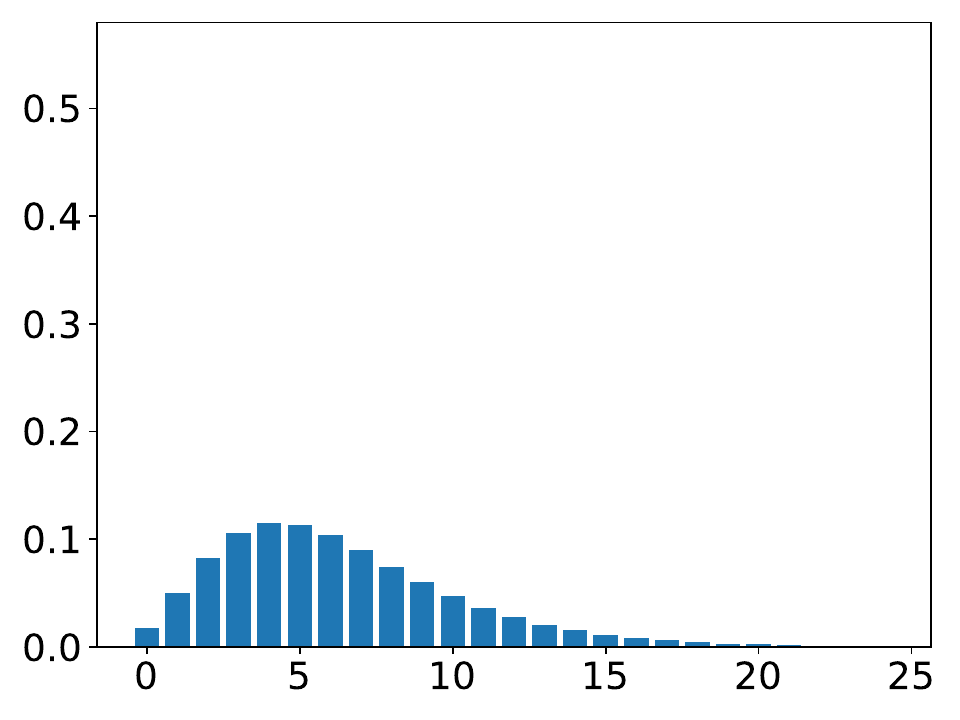}
      & \includegraphics[width=0.18\textwidth]{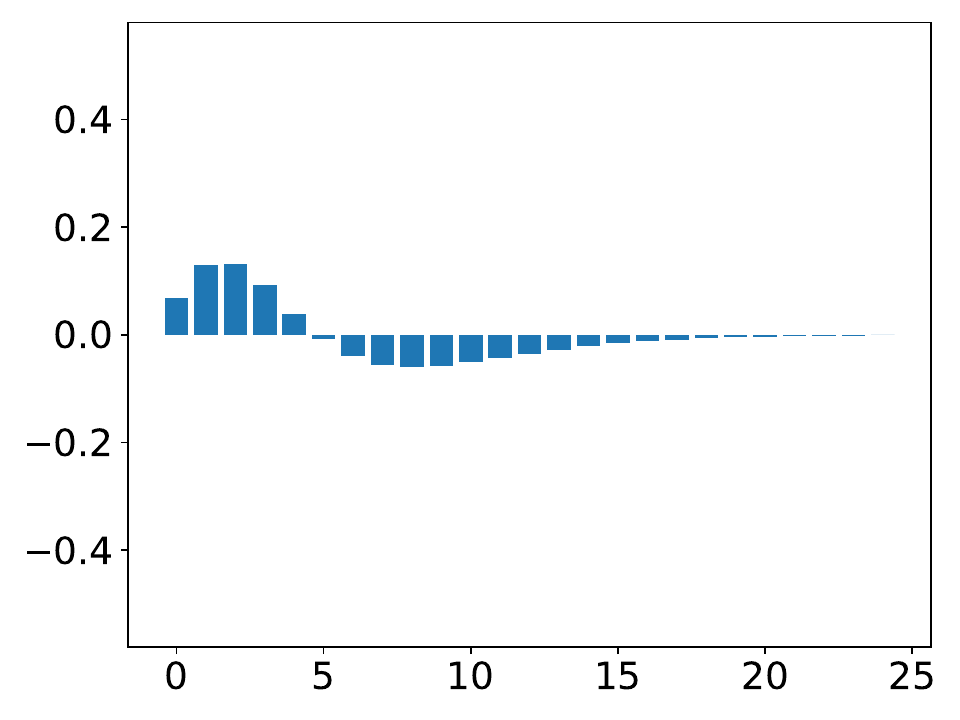}
      &
        \includegraphics[width=0.18\textwidth]{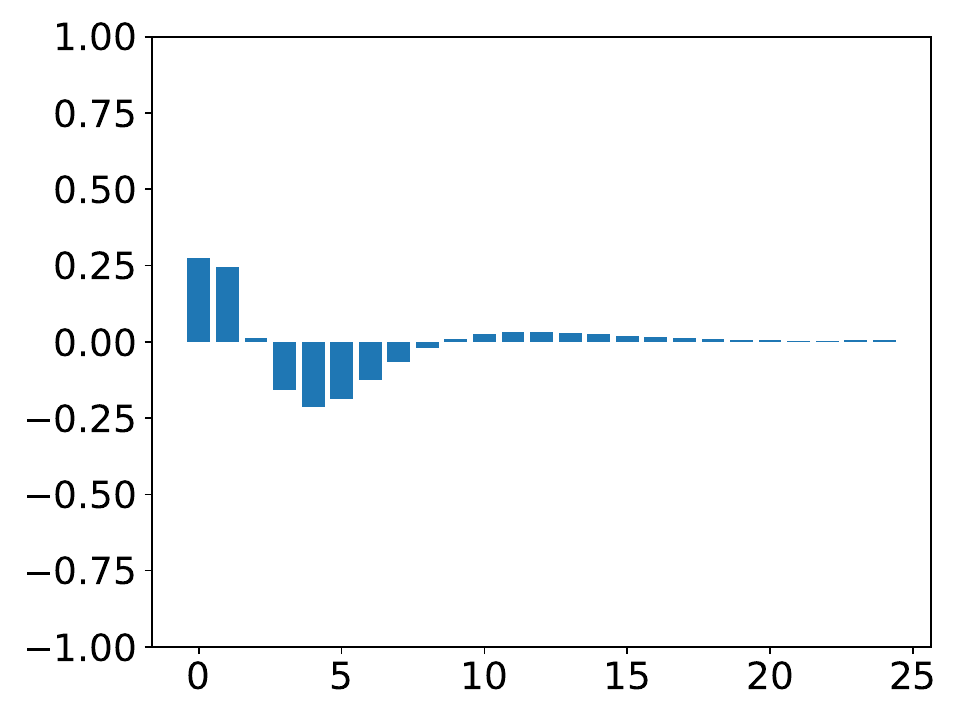}
      \\
    \end{tabular}      
  \end{center}
  \caption{Equivalent discrete approximations of the  time-causal limit kernel
    $\Psi(t;\; \tau, c)$ as obtained from a set of first-order
    recursive filters coupled in cascade, with the time constants
    determined according to (\ref{eq-disc-time-constant}), and with
    the infinite cascade truncated after the
    8 recursive filters having the longest time constants,
    together with discrete approximations of
    its scale-normalized temporal derivatives
    $\Psi_{\zeta^n}(t;\; \tau, c) = \tau^{n \, \gamma/2} \, \Psi_{t^n}(t;\; \tau, c)$
    up to order $n = 2$, obtained by applying first- and second-order
    temporal difference operators $\delta_t$ and $\delta_{tt}$
    according to (\ref{eq-temp-der-approx-molecules}) to the discrete
    approximations of time-causal limit kernel,
    with the scale normalization power $\gamma = 1$
    corresponding to $L_1$-normalization across scales,
    for different combinations of the temporal scales
    $\sigma = \sqrt{\tau} \in \{ 1, 2, 4 \}$ and
    different values to the distribution parameter
    $c \in \{ \sqrt{2}, 2 \}$. (Horizontal axes: time $t \in [0, 25]$.
    Vertical axes: kernel values with different ranges
    $[0, 0.50]$, $[-0.40, 0.40]$ or $[-1.0, 1.0]$, depending on
    the order $n$ of temporal differentiation.)}
  \label{fig-disc-limitkern-graphs}
\end{figure*}

\subsubsection{Discrete approximations of temporal derivatives of the
  time-causal limit kernel}

Given that the input signal has been smoothed to a set of temporal
scales in this way, discrete approximations of temporal derivatives
can then be computed with temporal difference operators of the form
\begin{equation}
  \label{eq-temp-der-approx-molecules}
     \delta_t = (-1, +1), \quad\quad
     \delta_{tt} = (1, -2, 1),
\end{equation}
and scale-normalized derivatives according to
(\ref{eq-temp-sc-norm-ders}) in turn be computed by multiplying
the regular discrete derivative approximations by an appropriate power
of the scale parameter $\tau$ in units of the variance of the
underlying the equivalent discrete temporal smoothing kernel.

Figure~\ref{fig-disc-limitkern-graphs} shows examples of such
equivalent discrete
derivative approximation kernels, obtained by filtering a discrete
delta function with the proposed methodology for computing
discrete approximation of temporal derivatives at multiple temporal
scales.
As can be seen from these graphs, the discrete nature of the
smoothing kernels and their discrete derivative approximations is
significant for lower temporal scale levels, while the shapes of the
kernels approach their continuous counterparts in
Figure~\ref{fig-limitkern-graphs} with increasing values
of the temporal scale parameter $\tau$.

\begin{algorithm*}[hbtp]
  \begin{algorithmic}[0]
    \Procedure{disc-time-caus-wavelets}{$f, \mu$}\Comment{$f$
      input stream, $\mu$ of size $K$}
      \State $level \gets 0$ \Comment{of size $K$}
      \State $level\_prev \gets 0$ \Comment{of size $K$}
      \State $level\_prev2 \gets 0$ \Comment{of size $K$}
      \State $n \gets 0$ \Comment{time counter}

      \Repeat
         \State $signal \gets f(n \, \Delta t)$
            \Comment{read the input stream with time increment $\Delta t$}
            \For{$k \gets 1, K$}
              \If{$k = 1$}
                 \State $level_{k} \gets level\_prev_{k} + (signal - level\_prev_{k})/(1 + \mu_{k})$
                    \Comment{first smoothed layer}
                 \State $bandpass_k \gets level_{k} - signal$
                 \Comment{lowest bandpass representation}
              \Else
                 \State $level_{k} \gets level\_prev_{k} +
                    (level_{k-1} - level\_prev_{k})/(1 + \mu_{k})$
                  \Comment{higher smoothed layers}
                  \State $bandpass_k \gets level_{k} - level_{k-1}$
                  \Comment{higher bandpass representations}
              \EndIf
           \EndFor
           \State $firstder \gets level - level\_prev$
              \Comment{first-order temporal derivatives}
           \State $secondder \gets level - 2*level\_prev + level\_prev2$
              \Comment{second-order temporal derivatives}
           \State $level\_prev2 \gets level\_prev$
              \Comment{update buffer for the time frame 2 time steps ago}
           \State $level\_prev \gets level$
              \Comment{update buffer for the previous time frame}
           \State $n \gets n+1$
              \Comment{prepare for the next time frame}
        \Until{interrupt}
   \EndProcedure
  \end{algorithmic}

  \caption{Pseudocode for combining discrete time-causal temporal filtering
    with the computation of bandpass wavelet representations
    and the computation of first- and second-order temporal
    derivatives, based on a set of
    first-order recursive filters of the form (\ref{eq-norm-update}) coupled in cascade,
    which implements convolution with a discrete approximation of the
    time-causal limit kernel $\Psi(t;\; \tau, c)$ according to (\ref{eq-time-caus-lim-kern}).
    Here, it is assumed that the time constants $\mu_{k} > 0$
    have been already computed according to (\ref{eq-disc-time-constant}).
    The variable $level\_prev$ 
    represents a memory from the previous frame, necessary to compute
    the temporal differences that drive the recursive filters, 
    also serving as a temporal buffer for computing first-order temporal
    derivatives. The variable $level\_prev2$ 
    represents a memory from 2 time steps ago, used for computing
    second-order temporal derivatives. In this respect, the
    algorithm is strictly time-recursive, since it only makes use of
    information from the present moment and a very short-term
    memory from the previous 1 or 2 frames.}
  \label{fig-pseudo-code-rec-filt-casc}
\end{algorithm*}

\begin{algorithm*}[hbtp]
  \begin{algorithmic}[0]
    \Procedure{disc-decode-bandpass}{$b$, $l$}
       \Comment{$b$ input stream of size $K$, $l$ input stream of size 1}
       \State $n \gets 0$ \Comment{time counter}
     
       \Repeat
          \State $bandpass \gets b(n \, \Delta t)$
           \Comment{read stream with bandpass representations with time increment $\Delta t$}
          \State $lastlevel \gets l(n \, \Delta t)$
            \Comment{read stream at coarsest temporal scale with time increment $\Delta t$}
           \State $reconstr \gets lastlevel$ \Comment{of size $1$}
           \For{$k \gets K, 1$}
              \State $reconstr \gets reconstr - bandpass_k$
                 \Comment{accumulate bandpass representations over scales}
           \EndFor
           \State $n \gets n+1$
              \Comment{prepare for the next time frame}
        \Until{interrupt}
   \EndProcedure
  \end{algorithmic}

  \caption{Pseudocode for reconstructing a signal from its bandpass
    wavelet representation $b$ combined with the time-causal temporal
    scale-space representation $l$ at the coarsest temporal scale.}
  \label{fig-pseudo-code-decode-bandpass}
\end{algorithm*}

\subsubsection{Streaming pipelines for the encoding and decoding stages}

Algorithm~\ref{fig-pseudo-code-rec-filt-casc} summarizes the different
steps involved in implementing a discrete approximation of convolution
with the time-causal limit kernel in this way, as well as the
computation of temporal bandpass representations and the computation
of first- and second-order temporal derivatives.%
\footnote{While one would in an actual implementation typically only make use of
  either bandpass representations or temporal derivatives,
  we here describe both approaches in the same pseudocode to save space.}
Algorithm~\ref{fig-pseudo-code-decode-bandpass} gives corresponding
pseudocode for reconstructing the original temporal signal from its
bandpass representation, given complementary access to the time-causal
scale-space representation at the coarsest temporal scale.

\subsection{Time-causal wavelet representations for discrete signals}
\label{sec-disc-wavelet-repr}

Given that we use the scheme in
Algorithm~\ref{fig-pseudo-code-rec-filt-casc} for smoothing a discrete
signal stream $f \colon \bbbz \rightarrow \bbbr$ with an approximation
of the discrete analogue of the time-causal limit kernel
$\Psi_{\text{disc}}(t;\; \tau, c)$, and then computing discrete
approximations of temporal
derivatives using the time-causal discrete temporal derivative operators
$\delta_t$ and $\delta_{tt}$ according to
(\ref{eq-temp-der-approx-molecules}), complemented with scale
normalization according to
\begin{equation}
  \delta_{\zeta^n} = \delta_{t^n,\text{norm}} = \tau^{n \gamma/2} \, \delta_{t^n}
\end{equation}
for some suitably selected value of the scale normalization parameter
$\gamma > 0$,
we thereby obtain numerical approximations
of the discrete analogue
$L_{\zeta^n} \colon \bbbz \times \bbbz \rightarrow \bbbr$
of the scale-normalized temporal derivative of
order $n$ of the time-causal
scale-space representation
of the signal $f$ according to
\begin{align}
   \label{eq-def-time-caus-temp-der-disc-wavelet}
  L_{\zeta^n}(\cdot;\; \tau_k, c)
  & = \Psi_{\zeta^n,\text{disc}}(\cdot;\; \tau_k, c) * f(\cdot) = \nonumber\\
  & = \tau^{n \gamma/2} \, \delta_{t^n} \, \Psi_{\text{disc}}(\cdot;\; \tau_k, c) * f(\cdot)
\end{align}
for the temporal scale levels chosen as
\begin{equation}
  \tau_k = \tau_0 \, c^{2k} \quad\quad \mbox{for $k \in \bbbz$}.
\end{equation}
With this parameterization, the discrete parameter $k \in \bbbz$ again
has a similar
role as the continuous variable $u$ has in the parameterization of the scaling
group under the exponential map $e^{u}$ in the formulation of the
continuous wavelet transform according to~(\ref{eq-def-wavelet-transf}).

Alternatively, given the raw undifferentiated temporal scale-space
representation equivalently obtained by convolution with the discrete
analogue $\Psi_{\text{disc}}(t;\; \tau_k, c)$ of the time-causal limit kernel
according to
\begin{equation}
  L(\cdot;\; \tau_k, c)
  = \Psi_{\text{disc}}(\cdot;\; \tau_k, c) * f(\cdot),
\end{equation}
we can compute a bandpass wavelet representation from
\begin{equation}
  \label{eq-disc-time-caus-bandpass-repr}
  \Delta L_{\text{DoT}}(t;\; \tau_k, c) = L(t; \tau_k) - L(t; \tau_{k-1}, c),
\end{equation}
which is computationally equivalent to convolving the input signal $f$
with the bandpass wavelet filter
\begin{equation}
  \Delta \Psi_{\text{disc}}(t;\; \tau_k, \tau_{k-1}, c)
  = \Psi_{\text{disc}} (t; \tau_k) - \Psi_{\text{disc}} (t; \tau_{k-1}, c)
\end{equation}
between adjacent levels of temporal scales
\begin{equation}
  \tau_k = c^2 \, \tau_{k-1}.
\end{equation}
A technical difference compared to the continuous case is, however,
the following:
Whereas for the continuous time-causal wavelets, the bandpass wavelet
representation is directly related to the first-order temporal
derivative of the time-causal scale-space representation according to
(\ref{eq-first-ord-int})
\begin{align}
  \Delta L_{\text{DoT}}(t;\; \tau_{k}, c)
  & = L(t;\; \tau_{k}, c) - L(t;\; \tau_{k-1}, c) = \nonumber\\
  & = - \mu_k \, \partial_t L(t;\; \tau_k, c),
\end{align}
for the discrete time-causal wavelets, the bandpass wavelet representation
is only related to the discrete approximation of the first-order
temporal derivative of the time-causal scale-space representation,
if we additionally include a temporal shift when
defining the bandpass representation according to
the recurrence relation (\ref{eq-norm-update}):
\begin{multline}
  L(t-1;\; \tau_{k}, c) - L(t;\; \tau_{k-1}, c) = \\
  = - (1 + \mu_k) \, (L(t;\; \tau_{k}, c) - L(t-1;\; \tau_{k}, c) = \\
  = - (1 + \mu_k) \, \delta_t L(t;\; \tau_{k}, c).
\end{multline}
Therefore, if a main purpose is to perform reconstruction from the
time-causal wavelet responses, it is more straightforward to make use
of the explicit bandpass representation as opposed to using wavelet
responses in terms of discrete approximations of temporal derivatives
obtained from temporal difference operators.

\subsection{Characterizations of scaling properties in discretizations of
  the continuous theory}
\label{sec-char-scaling-props}

The theory for scale-covariant time-causal and time-recursive wavelets
in Section~\ref{sec-cont-theory} is developed for fully
continuous signals. In this section, we will characterize
how well the continuous scaling properties carry over to a discrete implementation,
as based on the methodology proposed in
Sections~\ref{sec-disc-approx} and~\ref{sec-disc-wavelet-repr},
and depending on the relationship between the temporal scale levels in
relation to the sampling distance in the signal.

\begin{figure}[hbtp]
  \begin{center}
    \begin{tabular}{c}
      {\footnotesize $\| \Psi_{\zeta^n,\text{disc}}(\cdot;\, \tau, c) \|_1$ for order $n = 1$} \\
      \includegraphics[width=0.45\textwidth]{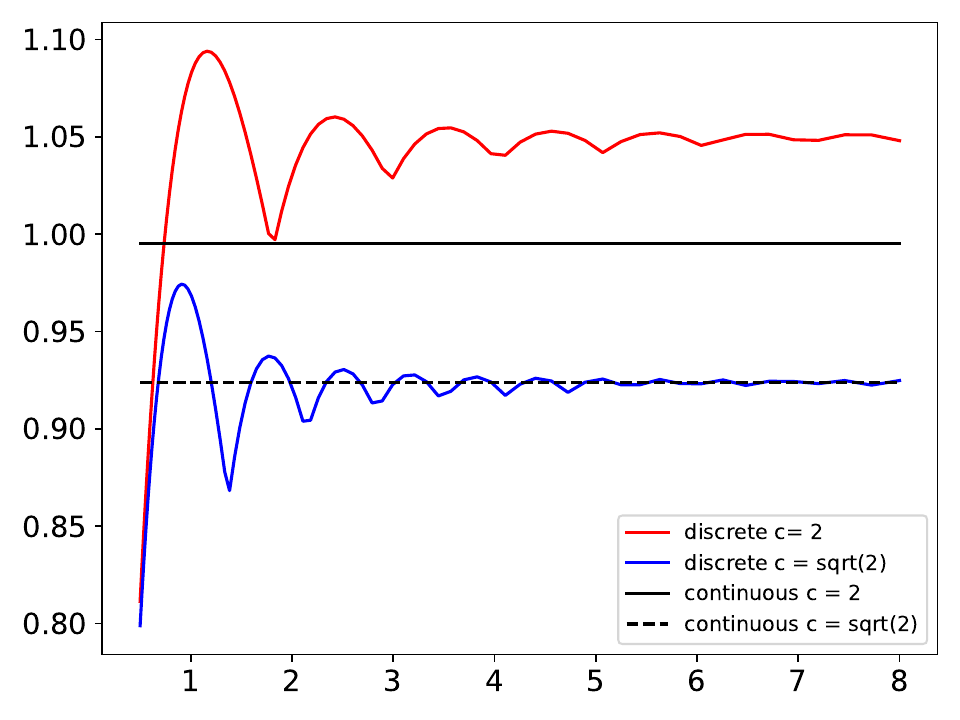}
      \\
      {\footnotesize $\| \Psi_{\zeta^n,\text{disc}}(\cdot;\, \tau, c) \|_1$ for order $n = 2$} \\
      \includegraphics[width=0.45\textwidth]{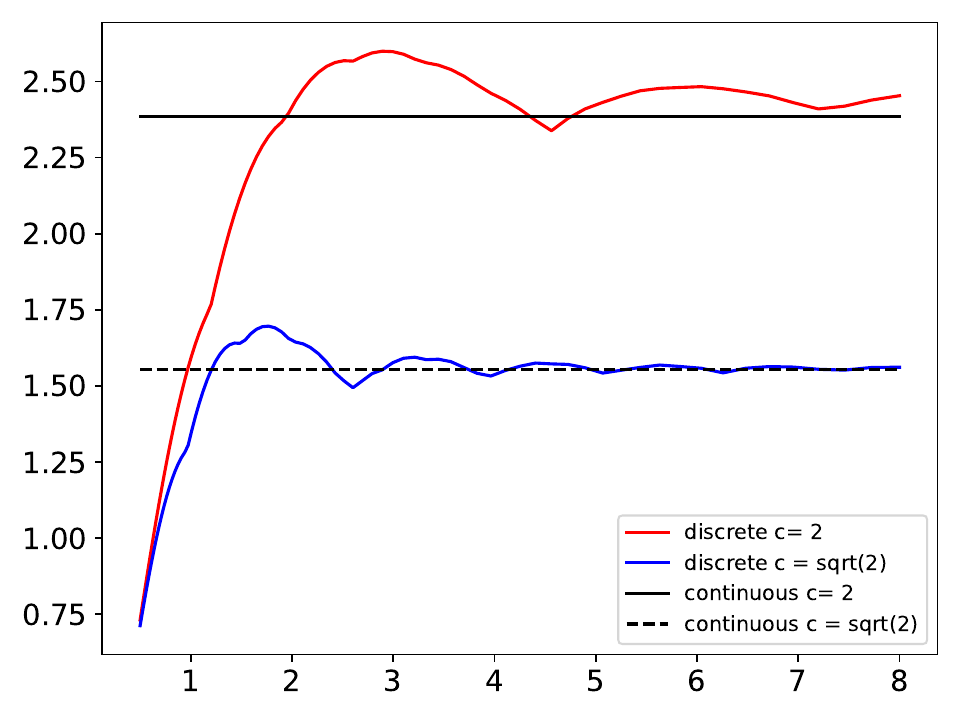}
    \end{tabular}
  \end{center}
  \caption{Graphs over the dependency on the scale parameter
    $\sigma = \sqrt{\tau}$ for the discrete $l_1$-norms according to
    (\ref{eq-lpnorms-limitkernders}) of the equivalent
    discrete kernels illustrated in Figure~\ref{fig-disc-limitkern-graphs},
    that correspond to the discrete approximations of the derivatives
    of the time-causal limit kernel, as obtained with the
    discretization methodology described in
    Section~\ref{sec-disc-approx} for the choice of the scale
    normalization power $\gamma = 1$.
    For reference, the graphs also show the evolution properties over
    scales of
    the $L_1$-norms of the temporal derivatives of the continuous
    time-causal limit kernel, according to
    (\ref{l1-norm-psi-t-c2}), (\ref{l1-norm-psi-tt-c2}),
    (\ref{l1-norm-psi-t-csqrt2}) and (\ref{l1-norm-psi-tt-csqrt2}), 
    which are constant over scales for $\gamma = 1$.
    As can be seen from these graphs, there are some transient phenomena
    at finer levels of scales, where the influence of the discretization
    effects is strongest. Towards coarser levels of scales, however, the
    discrete $l_p$-norms asymptotically approach constant values.
    When the distribution parameter $c = \sqrt{2}$, there are very good
    matches between the asymptotic values and the corresponding
    expressions for the fully continuous theory.
    When the distribution parameter $c =2$, there are, however,
    certain deviations.
  (Horizontal axes: temporal scale parameter in units of $\sigma =
  \sqrt{\tau}$. Vertical values: magnitudes of the discrete
  $l_p$-norms, with red curves showing the results for the
  distribution parameter $c = 2$ and blue curves showing the results
  for the distribution parameter $c = \sqrt{2}$.)}
  \label{fig-lpnorms}
\end{figure}

\subsubsection{Scale dependencies of the discrete $l_p$-norms of the
  equivalent discrete derivative approximation kernels}

Figure~\ref{fig-lpnorms} shows a first characterization in this respect, by
showing how the discrete $l_1$-norms of the equivalent convolution
kernels corresponding to discrete approximations of the time-causal
limit kernel
\begin{equation}
  \label{eq-lpnorms-limitkernders}
  \| \Psi_{\zeta^n,\text{disc}}(t;\; \tau, c) \|_p
  = \| \tau^{n \, \gamma/2} \, \delta_{t^n} \Psi_{\text{disc}}(t;\; \tau, c) \|_p
\end{equation}
for the scale normalization power $\gamma = 1$
vary as function of the scale parameter $\tau$, in the graphs notably
parameterized in terms of the standard deviation $\sigma = \sqrt{\tau}$
instead of the variance $\tau$.

For reference, the graphs also show the evolution properties of
the $L_1$-norms of the temporal derivatives of the continuous
time-causal limit kernel, according to
(\ref{l1-norm-psi-t-c2}), (\ref{l1-norm-psi-tt-c2}),  
(\ref{l1-norm-psi-t-csqrt2}) and (\ref{l1-norm-psi-tt-csqrt2}),
which are constant over scales for $\gamma = 1$.

As can be seen from these graphs, there are some transient phenomena
at finer levels of scales, where the influence of the discretization
effects is strongest. Towards coarser levels of scales, however, the
discrete $l_p$-norms asymptotically approach constant values.

When the distribution parameter is $c = \sqrt{2}$, we obtain very good
matches between the asymptotic values and the corresponding
expressions for the fully continuous time-causal model.
When the distribution parameter is $c =2$, there are, however,
certain deviations, of the order of 5~\% relative to the continuous
theory.
This quantification thereby demonstrates that the
discretization effects due to the discrete nature of the temporal
scale levels may be notable when the distribution parameter
$c = 2$, while there is a much better match between the continuous and
discrete theories when the distribution parameter is $c = \sqrt{2}$.

Concerning how to interpret these curves, it should be noted that here
we have treated the scale parameter as a continuous variable.
Given that the temporal scale levels are to be inherently discrete in
a temporal scale-space representation, this corresponds to emulating
the effect of using possibly different initial scale values $\tau_0$
when initiating the temporal scale levels according to
(\ref{eq-temp-sc-levels}).
In a practical implementation, a fixed value of the initial scale
value $\tau_0$ should, however, always be used, thus implying that the
resulting graphs will only assume values for the values of the scale
parameter $\sigma = \sqrt{\tau}$ that are integer multiples of the
distribution parameter $c$.

The reason why there are certain oscillations in these curves is
because of the positive and the negative lobes in the discrete
approximations of the derivatives in the time-causal limit kernel
interact with the grid spacing, which thereby affects the discrete
$l_1$-norms.

\begin{figure*}[hbtp]
  \begin{center}
    \begin{tabular}{cc}
      {\em discrete blob model $f_{\text{blob}}(t;\;
      \tau_{\text{ref}})$
      for $\sigma_{\text{ref}} = \sqrt{\tau_{\text{ref}}} = 1$}
      & {\em discrete edge model $f_{\text{edge}}(t;\;
        \tau_{\text{ref}})$
        for $\sigma_{\text{ref}} = \sqrt{\tau_{\text{ref}}} = 1$} \\
      \includegraphics[width=0.32\textwidth]{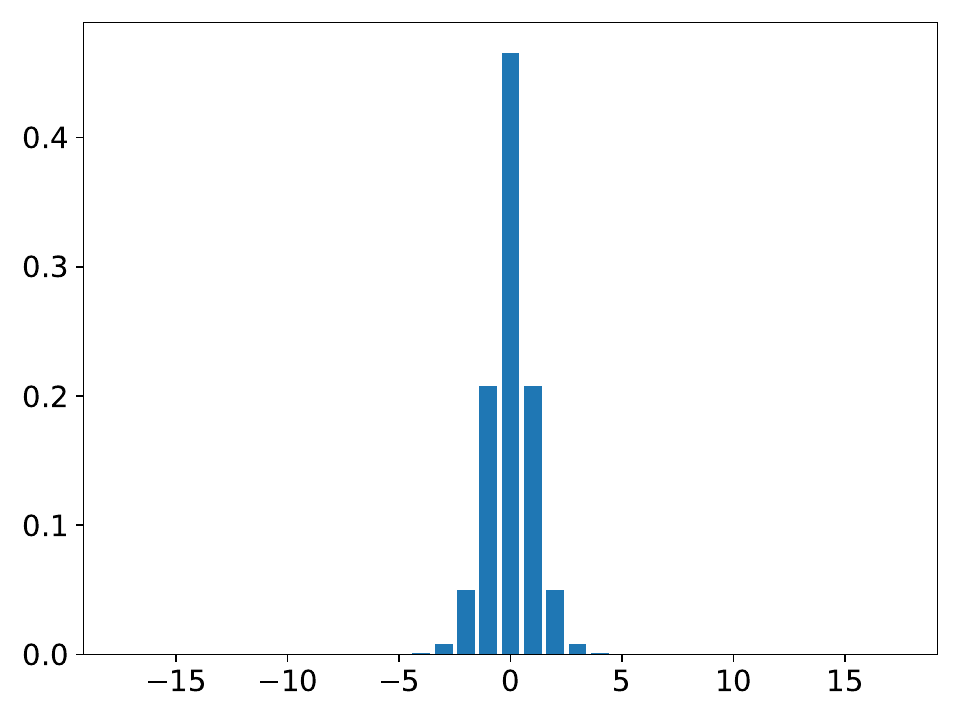}
      &
        \includegraphics[width=0.32\textwidth]{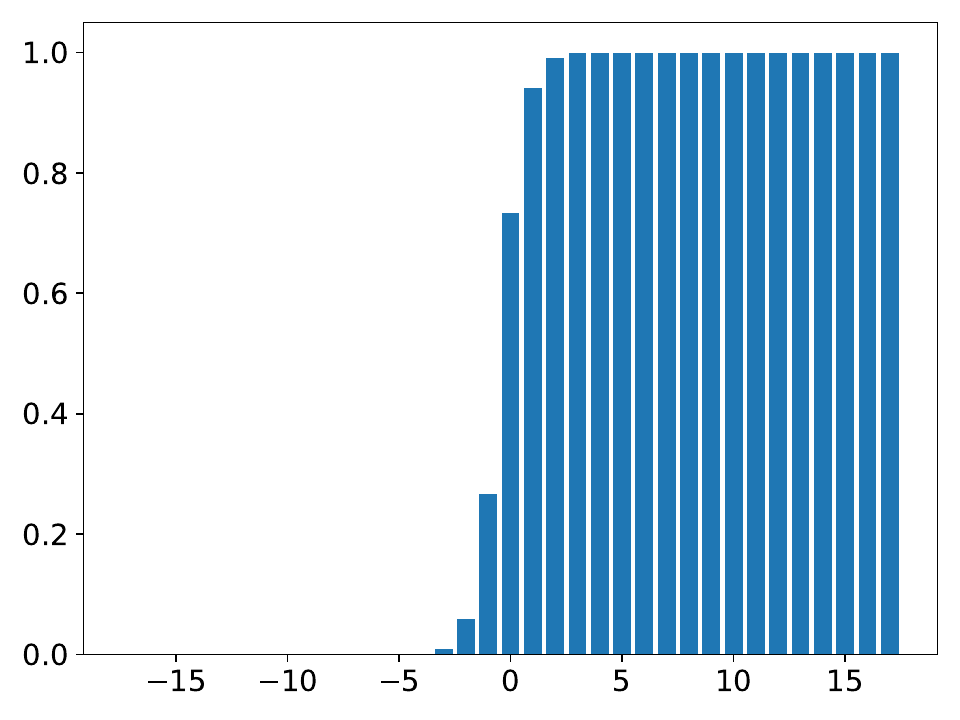}
      \\ $\,$ \\ 
      {\em scale-normalized second-order derivatives for $\gamma = 3/4$}
      & {\em scale-normalized first-order derivatives for $\gamma = 1/2$} \\
      \vspace{-8mm}
      \includegraphics[width=0.32\textwidth]{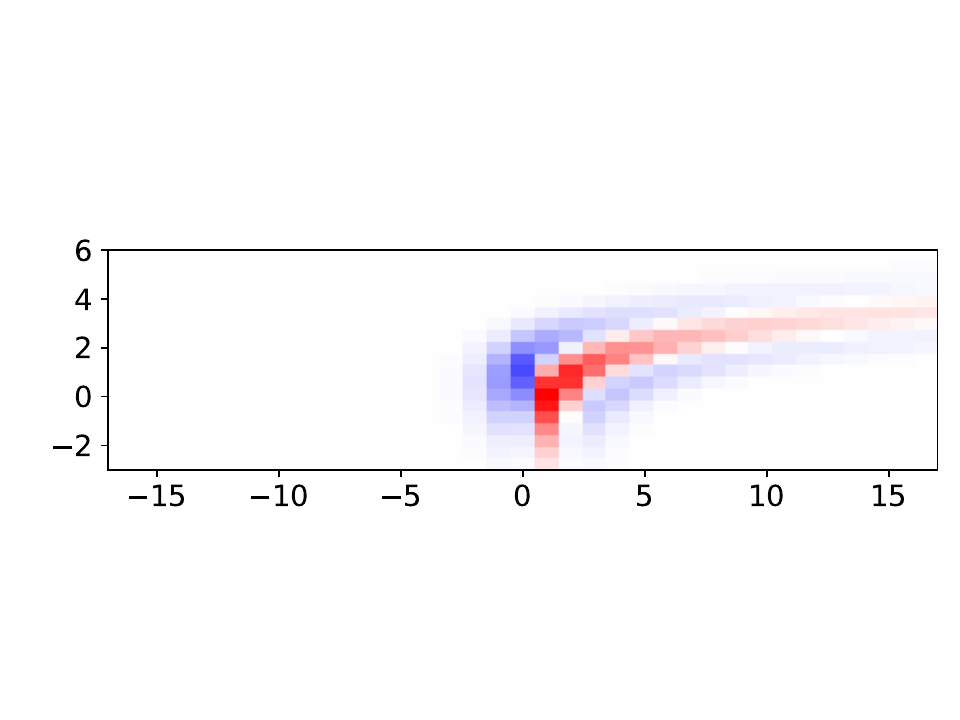}
      &
        \includegraphics[width=0.32\textwidth]{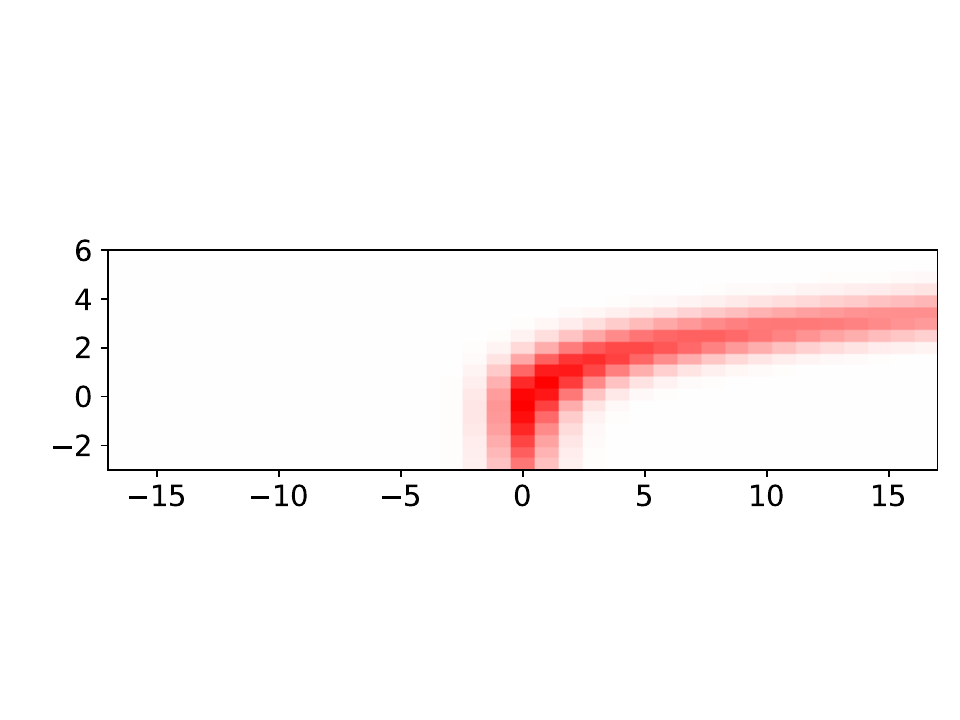}
      \\
      {\em discrete blob model $f_{\text{blob}}(t;\;
      \tau_{\text{ref}})$
      for $\sigma_{\text{ref}} = \sqrt{\tau_{\text{ref}}} = 4$}
      & {\em discrete edge model $f_{\text{edge}}(t;\;
        \tau_{\text{ref}})$
        for $\sigma_{\text{ref}} = \sqrt{\tau_{\text{ref}}} = 4$} \\
      \includegraphics[width=0.32\textwidth]{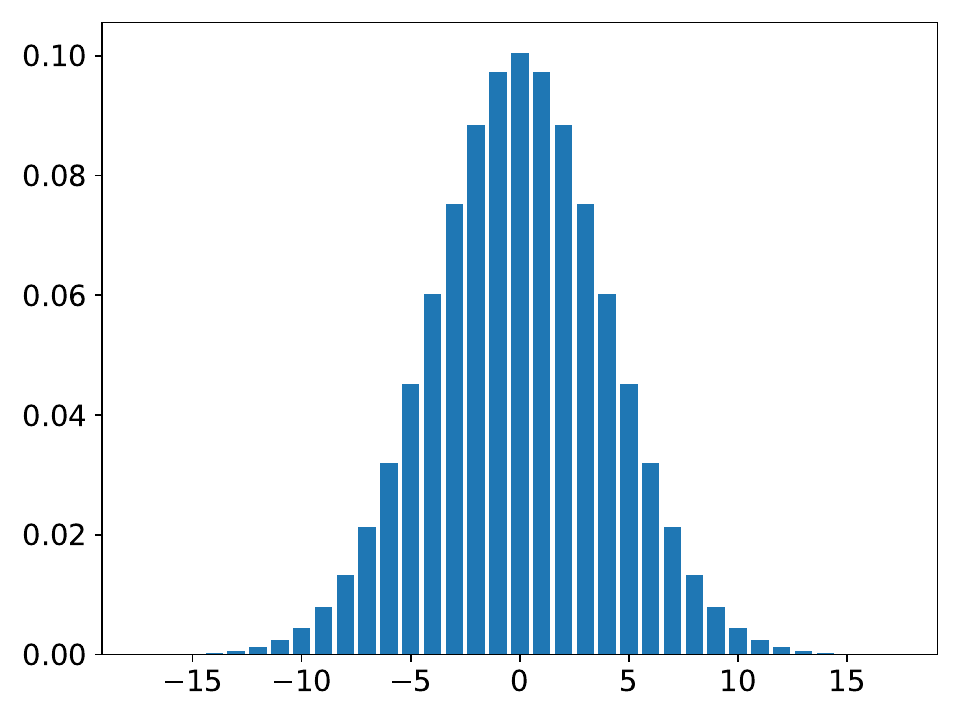}
      &
        \includegraphics[width=0.32\textwidth]{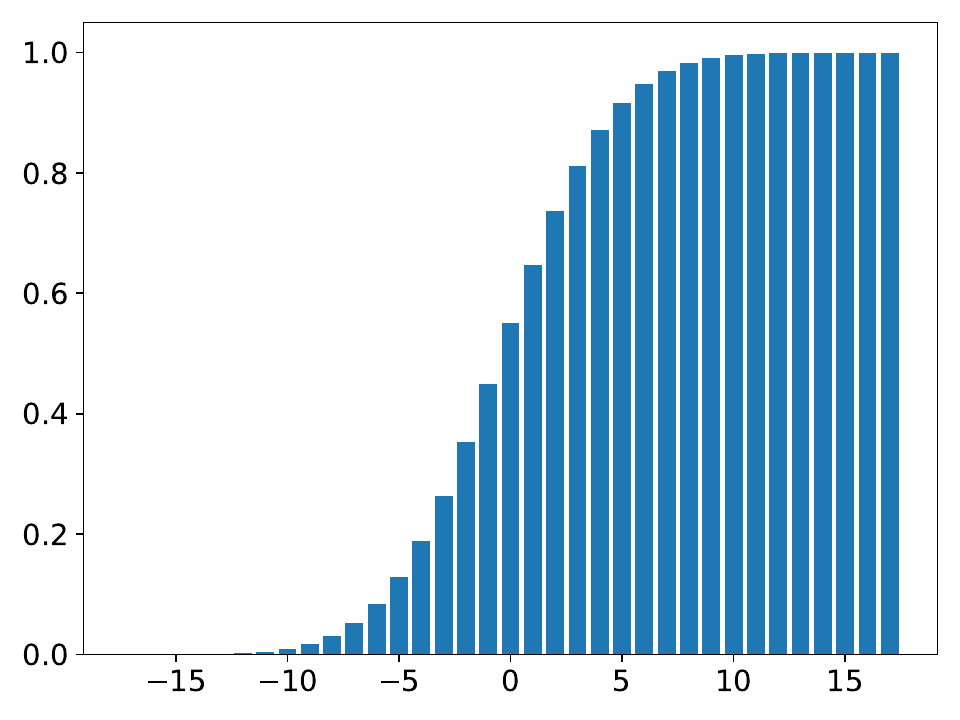}
      \\ $\,$ \\
      {\em scale-normalized second-order derivatives for $\gamma = 3/4$}
      & {\em scale-normalized first-order derivatives for $\gamma = 1/2$} \\
      \includegraphics[width=0.32\textwidth]{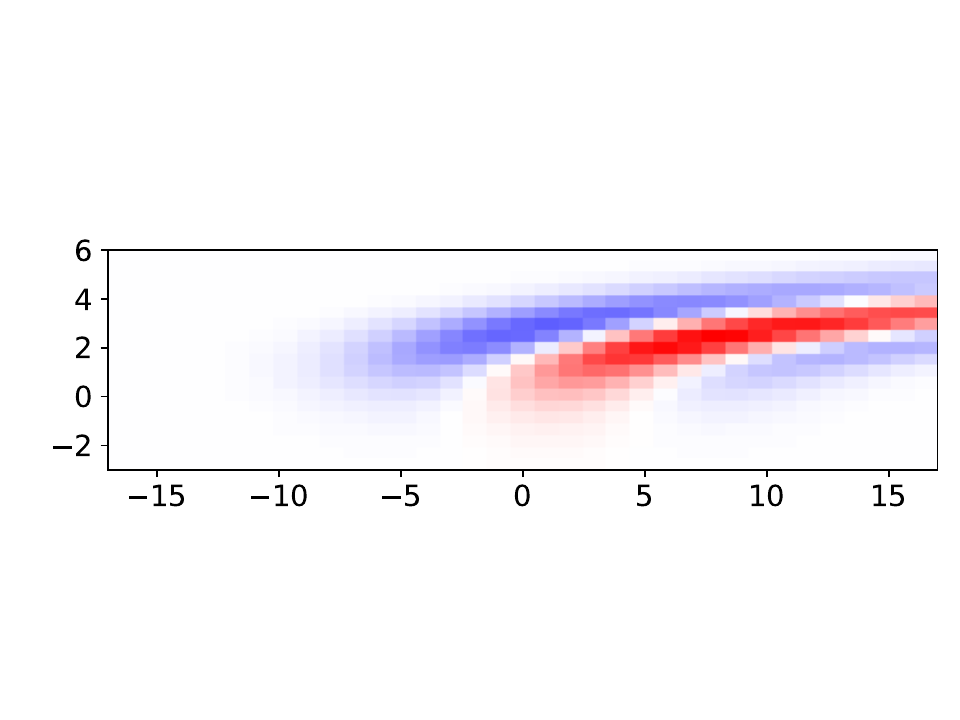}
      &
        \includegraphics[width=0.32\textwidth]{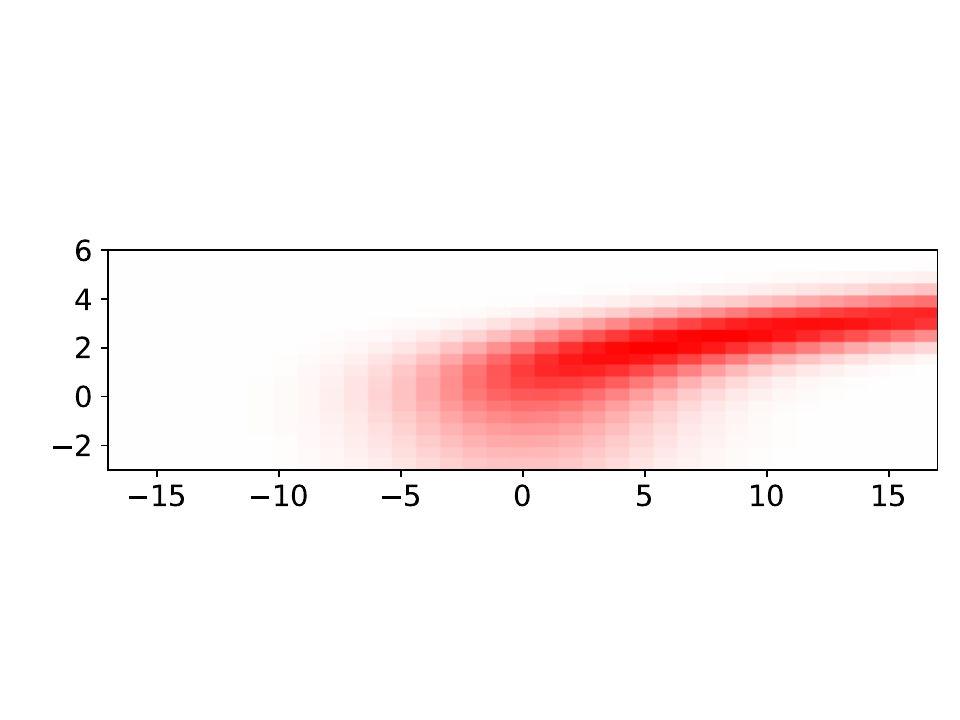} \vspace{-8mm}
    \end{tabular}      
  \end{center}
  \caption{Illustration of the conceptual stages involved in
    characterizing the scaling properties of the scale-normalized
    temporal derivatives of the time-causal limit kernel.
    The left column show two discrete blob-like
    model signals based on the discrete analogue of the Gaussian
    kernel according to (\ref{eq-blob-model}) for two different
    reference scales $\sigma_{\text{ref}} = \sqrt{\tau_{\text{ref}}} \in \{1, 4 \}$, together with
    the scale-normalized derivatives of order 2 for the scale
    normalization power $\gamma = 3/4$
    over the scale range $\sigma = \sqrt{\tau} \in [1/8, 64]$.
    The right column shows two edge-like model signals according to
    (\ref{eq-edge-model}) together with the scale-normalized
    derivatives of order 1 for $\gamma = 1/2$ over the same scale
    ranges.
    All results have been computed with the distribution parameter
    $c = \sqrt{2}$ for the discrete analogue of the time-causal
    limit kernel.
    (The reason why the slopes of the red and blue stripes are not
    vertical, but oblique, is because of the different amounts of
    temporal delay at different temporal scales.)
    (Horizontal axes: time $t \in [-17, 17]$. Vertical axes in the
    temporal scale-space representations: Effective scale = $\log_2
    \sqrt{\tau}$.) (In the time-causal wavelet representations, red
    denotes positive values and blue denotes negative values.)}
  \label{fig-scsel-ideal-blob-edge}
\end{figure*}

\subsubsection{Scale selection properties for local blob-like and
  edge-like structures in the signal}
\label{sec-char-scsel-props-blob-edge}

To investigate the scale selective properties of the proposed
methodology for defining time-causal wavelets, let us next focus on
the ability of the time-causal wavelets in terms of scale-normalized
derivatives of the time-causal limit kernel
to reflect the inherent characteristic scales in the input
data.

For this purpose, we generated blob-like and edge-like temporal
structures in terms of discrete approximations of the Gaussian kernel
in terms of the discrete analogue of the Gaussian kernel defined by
(Lindeberg \citeyear{Lin90-PAMI} Equation~(19))
\begin{equation}
  \label{eq-disc-anal-gauss-def}
  T(m;\; \tau) = e^{-\tau} \, I_m(\tau),
\end{equation}
where $I_m(\tau)$ denotes the modified Bessel functions of integer
order (Abramowitz and Stegun \citeyear{AS64}).
More, specifically blob-like temporal structures were generated as
(see Figure~\ref{fig-scsel-ideal-blob-edge}(left))
\begin{equation}
  \label{eq-blob-model}
  f_{\text{blob}}(t;\; \tau_{\text{ref}}) = T(t;\; \tau_{\text{ref}}) 
\end{equation}
and edge-like structures according to
(see Figure~\ref{fig-scsel-ideal-blob-edge}(right))
\begin{equation}
    \label{eq-edge-model}
  f_{\text{edge}}(t;\; \tau_{\text{ref}})
  = \sum_{m = -\infty}^t T(m;\; \tau_{\text{ref}}).
\end{equation}
Then, we computed the discrete time-causal wavelet
representations of these signals over the range of temporal scale
levels $\sigma = \sqrt{\tau} \in [1/8, 64]$ (see
Figure~\ref{fig-scsel-ideal-blob-edge}),
by computing
\begin{itemize}
\item
  discrete approximations of
  the second-order scale-normal\-ized temporal derivatives
  for the scale normalization power $\gamma = 3/4$
  of the time-causal wavelet representation of the idealized
  blob models $f_{\text{blob}}(t;\; \tau_{\text{ref}})$
  according to (\ref{eq-blob-model}), as well as
\item
  discrete approximations
  of the first-order scale-normalized temporal derivatives
  for the scale normalization power $\gamma = 1/2$
  of the time-causal wavelet representation of the idealized
  edge models $f_{\text{edge}}(t;\; \tau_{\text{ref}})$
  according to (\ref{eq-edge-model}).
\end{itemize}

\begin{figure}[hbtp]
  \begin{center}
    \begin{tabular}{c}
      {\em Selected scale levels $\hat{\sigma}_{\text{blob}}$ for blob-like model signals} \\
      \includegraphics[width=0.45\textwidth]{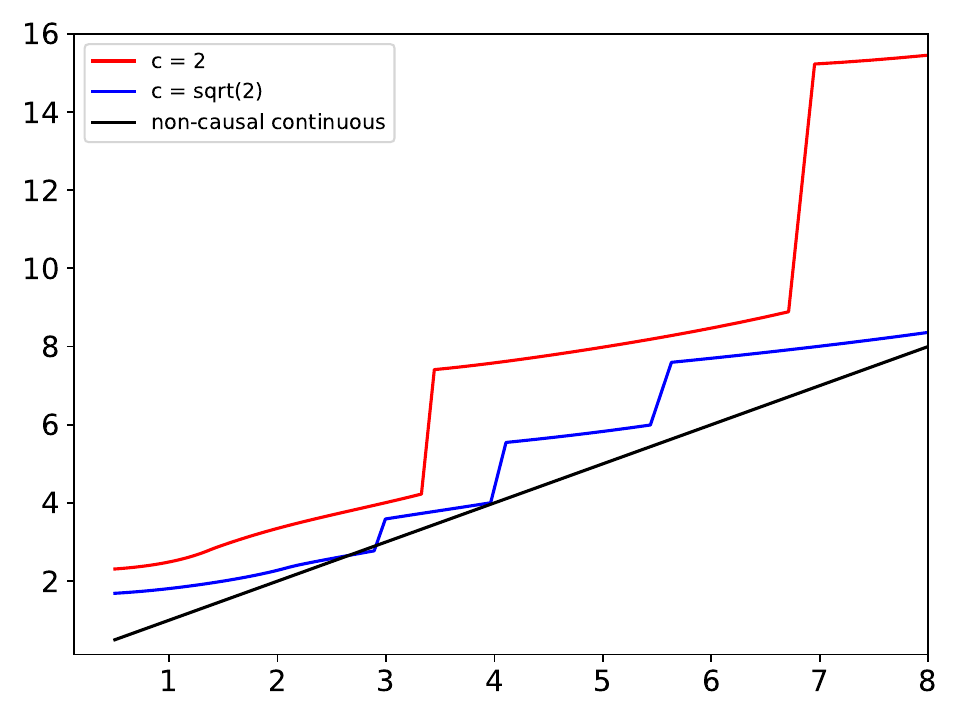}
      \\
      {\em Selected scale levels $\hat{\sigma}_{\text{edge}}$ for edge-like model signals} \\
      \includegraphics[width=0.45\textwidth]{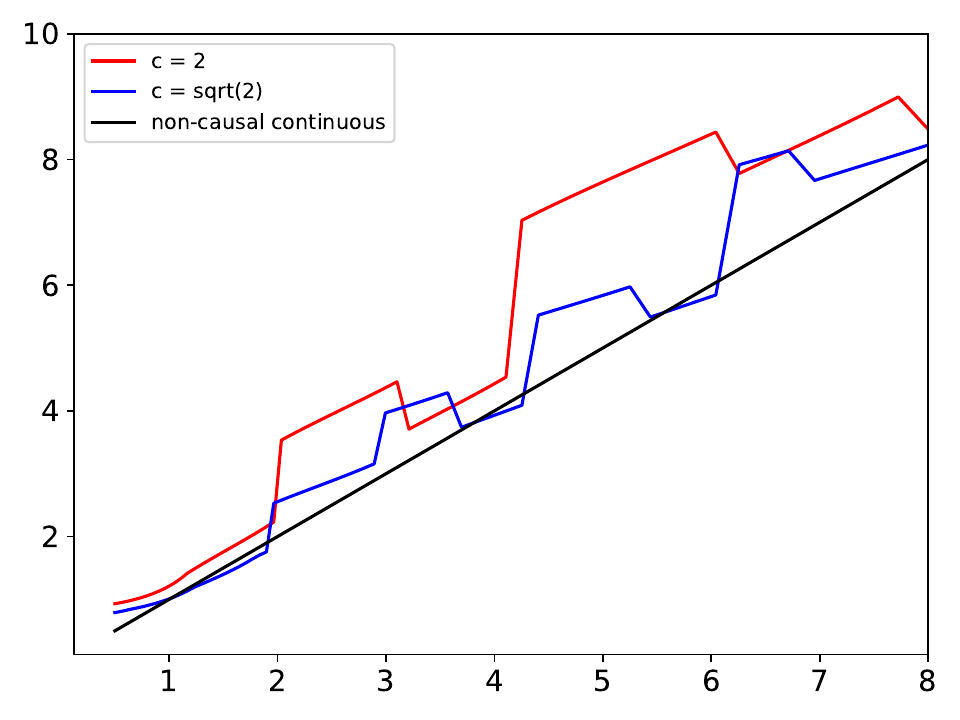}
    \end{tabular}
  \end{center}
  \caption{The selected scale levels
    $\hat{\sigma}_{\text{blob}} = \sqrt{\hat{\tau}_{\text{blob}}}$ and
    $\hat{\sigma}_{\text{edge}} = \sqrt{\hat{\tau}_{\text{edge}}}$
    from maxima over scales of
    scale-normalized temporal derivatives
    for blob-like structures according to (\ref{eq-sc-est-blob}) when using
    the scale normalization power $\gamma = 3/4$
    according to (\ref{eq-gamma-0p75-2nd-der-gauss-temp-scsp})
    and for edge-like structures according to (\ref{eq-sc-est-edge}) when using
    the scale normalization power $\gamma = 1/2$
    according to (\ref{eq-gamma-0p5-1st-der-gauss-temp-scsp}),
    computed from the
    time-causal wavelet representations of
    (top) ideal blob-like model signals of the form
    (\ref{eq-blob-model}) and
    (bottom) ideal edge-like model signals of the form
    (\ref{eq-edge-model}) under variations of the reference scale
    $\tau_{\text{ref}}$ in units of the standard deviation
    $\sigma_{\text{ref}} = \sqrt{\tau_{\text{ref}}}$ of the underlying model
    functions, and for different values of the distribution parameter
    $c = \sqrt{2}$ (blue curves) or $c = 2$ (red curves).
    For comparison, the corresponding scale estimates
    $\hat{\sigma} = \sigma_{\text{ref}}$ according to
    (\ref{eq-sigma-hat-spec-blob}) and (\ref{eq-sigma-hat-spec-edge})
    that would be
    obtained for a non-causal fully continuous Gaussian-derivative-based
    wavelet representation are
    also shown (black curves).
    As can be seen from the graphs, there is largely a linear dependency
    between the reference scales $\sigma_{\text{ref}}$ and the resulting scale
    estimates $\hat{\sigma}_{\text{ref}}$. There are, however, also
    clear discontinuities in the graphs, caused by the discrete nature
    of the temporal scale levels.
    Notably, the discretization effects decrease significantly when
    decreasing the distribution parameter from $c = 2$ to $c =
    \sqrt{2}$. For time-critical applications, however, a larger value
    of the distribution parameter leads to shorter temporal delays.
    (Horizontal axes: reference scale $\sigma_{\text{ref}}$ in units of
    the standard deviations of the underlying kernels.
    Vertical axes: scale estimates $\hat{\sigma}_{\text{ref}}$, also in
    units of the standard deviation of the underlying kernels.)}
  \label{fig-scsel-graphs-blob-edge}
\end{figure}

\noindent
Thereafter, we
determined the time moments and the temporal scales
at which these discrete approximations of the scale-normalized
temporal derivative responses assumed their maximum values over time
and temporal scales
\begin{multline}
    \label{eq-sc-est-blob}
    (\hat{t}_{\text{blob}}, \hat{\tau}_{\text{blob}})
    = \\
    = \argmax_{t, \tau}
    | (\Psi_{\zeta\zeta,\text{disc}}(\cdot;\; \tau, c) * f_{\text{blob}}(\cdot;\;
    \tau_{\text{ref}}))(t) |,
\end{multline}
\begin{multline}
    \label{eq-sc-est-edge}
    (\hat{t}_{\text{edge}}, \hat{\tau} _{\text{edge}})
    = \\
    = \argmax_{t, \tau}
    | (\Psi_{\zeta,\text{disc}}(\cdot;\; \tau, c) * f_{\text{edge}}(\cdot;\;
    \tau_{\text{ref}}))(t) |.
\end{multline}
Such points computed in a temporal scale-space representation are
referred to as scale-space extrema, and constitute a generalization of a
corresponding methodology developed for a spatial scale-space
representation (Lindeberg \citeyear{Lin97-IJCV,Lin98-IJCV}).
Specifically, based on the theory described in
Section~\ref{sec-scsel-theory-gauss-scsp}, in case the time-causal
limit kernel would be replaced by a non-causal Gaussian kernel, this
methodology would in the ideal continuous case determine the inherent
temporal duration of the corresponding idealized temporal blob and
edge models according to (\ref{eq-sigma-hat-spec-blob}) and
(\ref{eq-sigma-hat-spec-edge}).

Figure~\ref{fig-scsel-graphs-blob-edge} shows graphs of the dependency of
the resulting scale estimates 
for a range of reference scale levels $\tau_{\text{ref}}$,
when detecting scale-space extrema in this way for
the reference blob and edge signals $f_{\text{blob}}(t;\; \tau_{\text{ref}})$
and $f_{\text{edge}}(t;\; \tau_{\text{ref}})$ according to
(\ref{eq-blob-model}) and (\ref{eq-edge-model}), when
using time-causal wavelet representations for a fixed reference
scale $\tau_0 = 1$ in (\ref{eq-temp-sc-levels}) using 
the distribution parameter set to either $c = 2$ or $c= \sqrt{2}$.
Given the discrete nature of both the temporal scale levels and the
temporal sampling, we used local parabolic interpolation around
each extremum over temporal scales to increase the
accuracy in the resulting estimates of the temporal scales $\hat{\tau}$.

As can be seen from the results, the resulting scale estimates 
largely follow a linear increase in the selected scale levels as
function of the reference scales $\tau_{\text{ref}}$ of the model
signals. Due to the discrete nature of the temporal scale levels,
and the lack of provable scale covariance properties for temporal
scaling factors $S_t$ that are not integer powers of the distribution
parameter $c$, we cannot expect the graphs of the selected scale
levels to be smooth curves across the discontinuities over the
temporal scale levels.

Notably, the quantization effects decrease substantially when
decreasing the distribution parameter $c$ from $c = 2$ to $c =
\sqrt{2}$.
In this respect, there is a trade-off issue in that decreasing the
value of the distribution parameter $c$ leads to more accurate
scale estimates from the underlying signal, whereas increasing the
distribution parameter $c$ leads to shorter temporal delays;
compare with Equations~(\ref{eq-mean-time-caus-limit-kern})
and~(\ref{eq-approx-temp-pos-time-caus-log-distr}), and
see also the following section.

\begin{figure}[hbtp]
  \begin{center}
    \begin{tabular}{c}
      {\em Temporal delays in terms of mean values
              $M(\Psi_{\text{disc}}(\cdot;\; \tau, c))$} \\
      \includegraphics[width=0.45\textwidth]{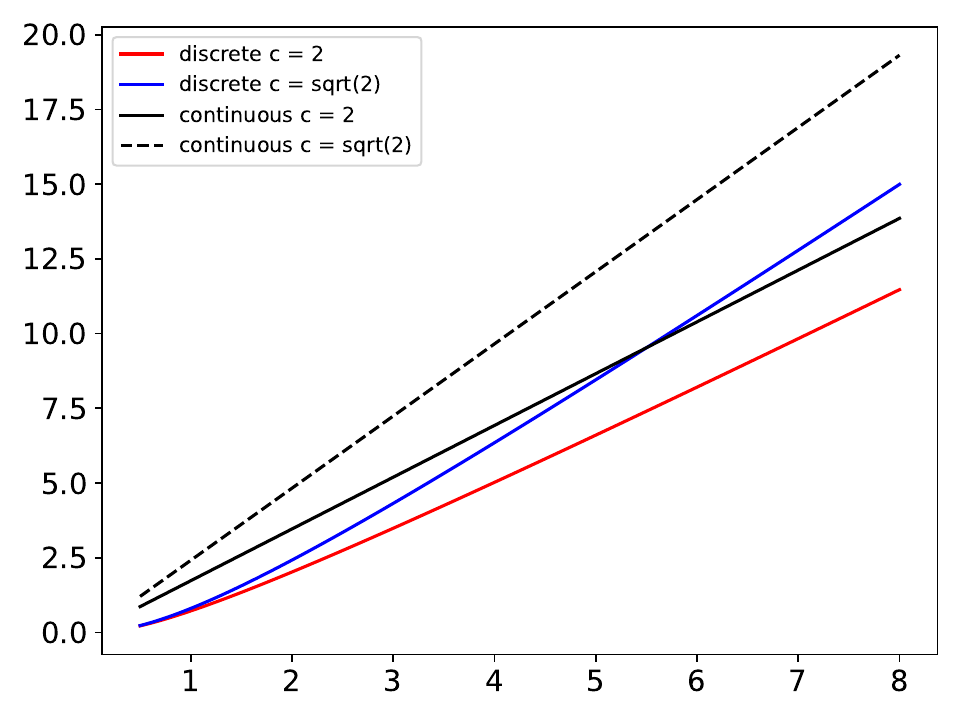}
      \\
      {\em Temporal delays in terms of
      $t_{\text{max}} = \argmax_t \Psi_{\text{disc}}(\cdot;\; \tau, c)$} \\
      \includegraphics[width=0.45\textwidth]{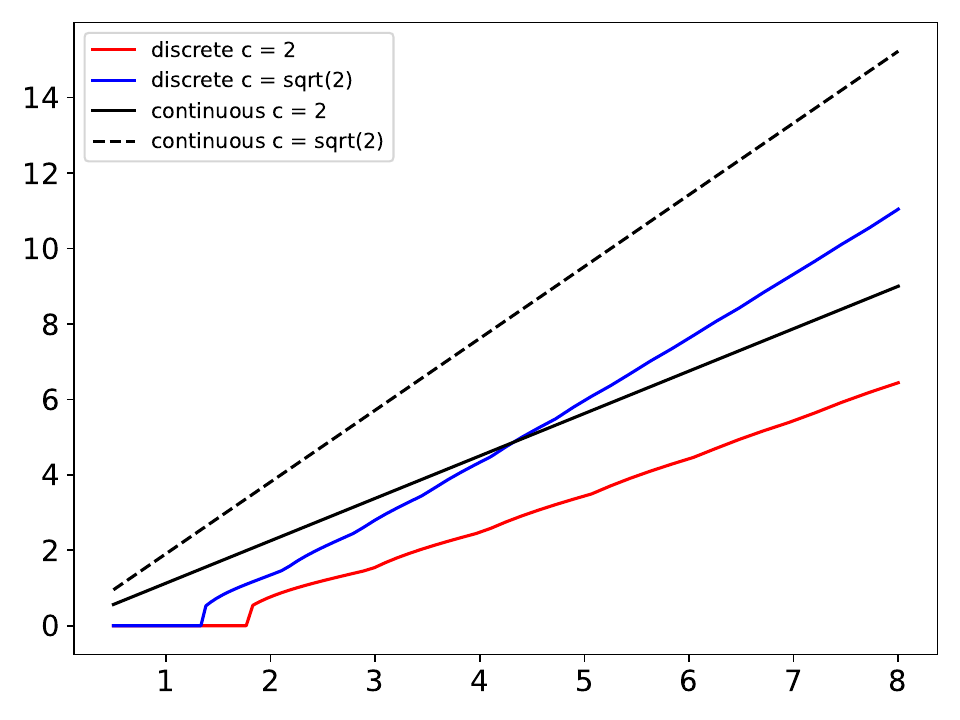}
    \end{tabular}
  \end{center}
  \caption{Measures of the temporal delays of the discrete
    approximations of the time-causal limit kernel in terms of the
    temporal mean according to
    (\ref{eq-mean-time-caus-limit-kern-disc}) and in terms of the
    temporal maximum point according to
    (\ref{eq-tmax-time-caus-limit-kern-disc}) for the distribution
    parameter $c = \sqrt{2}$ (blue curves) and $c = 2$ (red curves).
    For comparison, the results from the corresponding expressions
    (\ref{eq-mean-time-caus-limit-kern}) and
    (\ref{eq-approx-temp-pos-time-caus-log-distr}) for
    the continuous kernels are also shown (black curves).
    (Horizontal axes in the temporal scale-space representations:
    scale parameter $\sigma$ in units of
    the standard deviation of the underlying kernels.
    Vertical axes: delay estimates in units of $[\mbox{time}]$.}
  \label{fig-delay-graphs}
\end{figure}

\subsubsection{Temporal delays of the discrete approximations of the
  time-causal limit kernel}

To characterize the temporal delays for the discrete approximations of
the time-causal limit kernel, we can compute corresponding discrete
analogues of the temporal delay estimates for the continuous
time-causal limit kernel in terms of the temporal mean according
to Equation~(\ref{eq-mean-time-caus-limit-kern})
\begin{equation}
  \label{eq-mean-time-caus-limit-kern-disc}
  M(\Psi_{\text{disc}}(\cdot;\; \tau, c))
  =  \frac{\sum_{t=0}^{\infty} t \, \Psi_{\text{disc}}(t;\; \tau, c)}
              {\sum_{t=0}^{\infty} \Psi_{\text{disc}}(t;\; \tau, c)}
  = \sum_{k=1}^{\infty} \mu_k,
\end{equation}
with the discrete time constants $\mu_k$ according to
(\ref{eq-disc-time-constant}) and (\ref{eq-Delta-tau}),
and the time value at which the discrete approximation of the
time-causal limit kernel assumes its maximum value over scales
\begin{equation}
  \label{eq-tmax-time-caus-limit-kern-disc}
  t_{\text{max}} = \argmax_t \Psi_{\text{disc}}(t;\; \tau, c).
\end{equation}
Figure~\ref{fig-delay-graphs} shows the result of computing these characterizations of
the temporal delay for the distribution parameter
$c \in \{ \sqrt{2}, 2 \}$, and with the estimate of $t_{\text{max}}$
interpolated to subquantization resolution using parabolic
interpolation.
For comparison, the graphs also show the dependencies on the temporal
scale levels for the continuous temporal delay estimates according to
Equations~(\ref{eq-mean-time-caus-limit-kern})
and~(\ref{eq-approx-temp-pos-time-caus-log-distr}).
As can be seen from the graphs, the measures of the temporal delays
are significantly shorter for the distribution parameter $c = 2$
compared to the distribution parameter $c = \sqrt{2}$.
We can also note that the measures of the temporal delays are shorter
for the discrete approximations of the time-causal limit kernel
compared to the continuous time-causal limit kernel.

The reason for the latter difference between the continuous and
the discrete theories is that the time constants $\mu_k$
are determined in a different way from the
differences $\Delta  \tau_k$ in the temporal scale levels for the discrete kernels
than for the continuous kernels; compare (\ref{eq-disc-time-constant})
in the discrete case with the expression
$\mu_k = \sqrt{\Delta  \tau_k}$ in the continuous case.

\begin{figure*}[hbtp]
  \begin{center}
    \begin{tabular}{ccc}
      {\em Input signal\/} \vspace{-8mm} \\
      \includegraphics[width=0.30\textwidth]{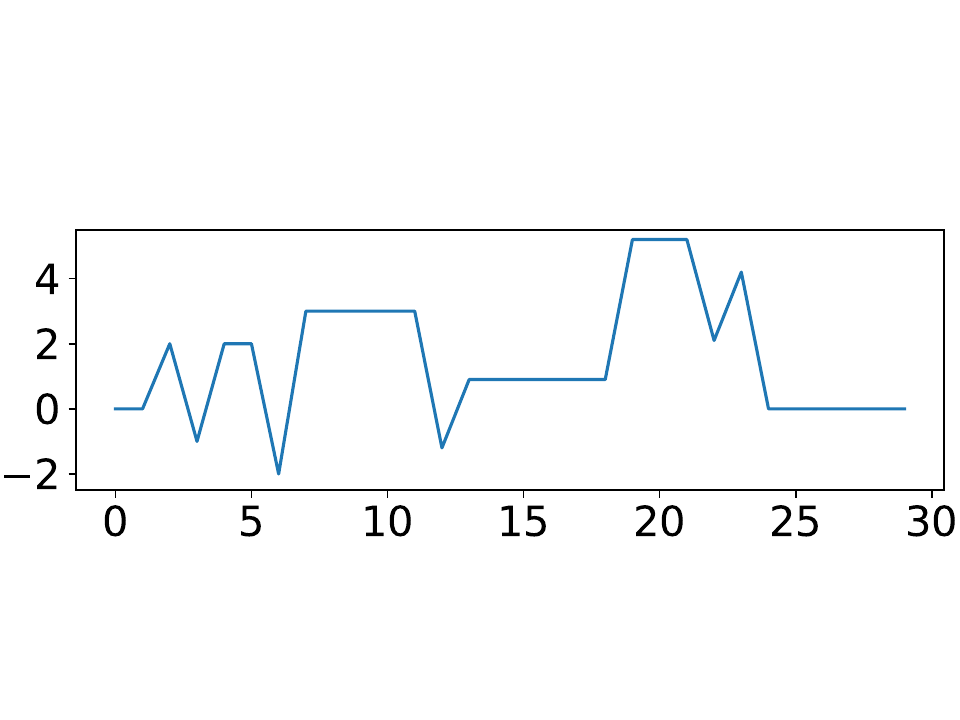}
      \vspace{-6mm} \\ 
      {\em DoT for $\sigma = 1$}
      & {\em DoG for $\sigma = 1$} 
      & {\em DoE for $\sigma = 1$} \vspace{-8mm}  \\
      \includegraphics[width=0.30\textwidth]{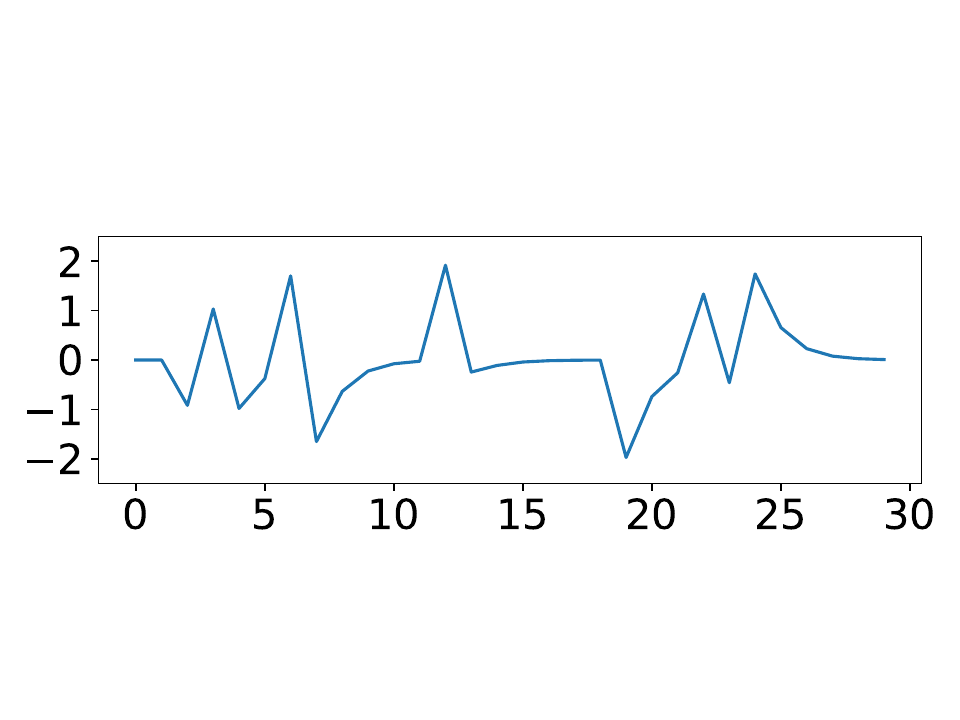}
      & \includegraphics[width=0.30\textwidth]{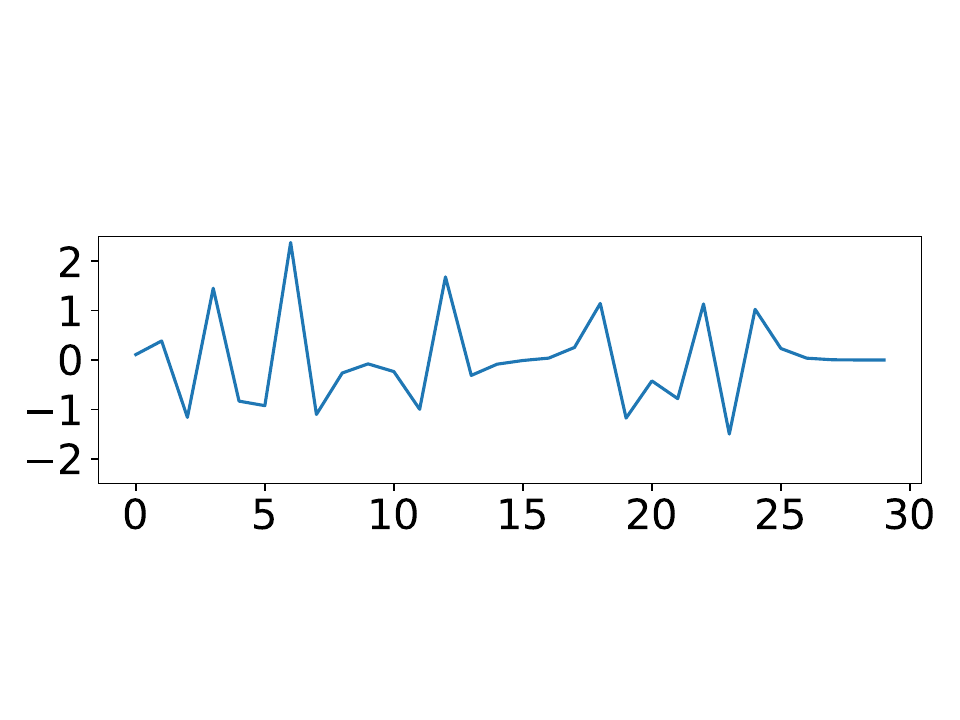}
      & \includegraphics[width=0.30\textwidth]{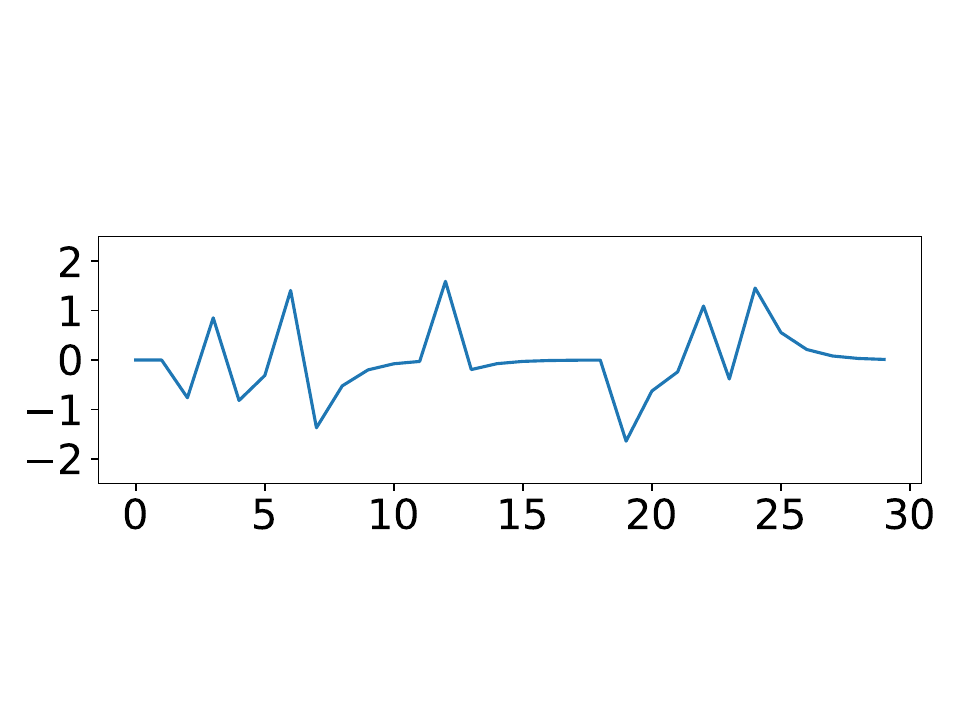}
        \vspace{-6mm} \\      
      {\em DoT for $\sigma = 2$}
      & {\em DoG for $\sigma = 2$} 
      & {\em DoE for $\sigma = 2$} \vspace{-8mm} \\
      \includegraphics[width=0.30\textwidth]{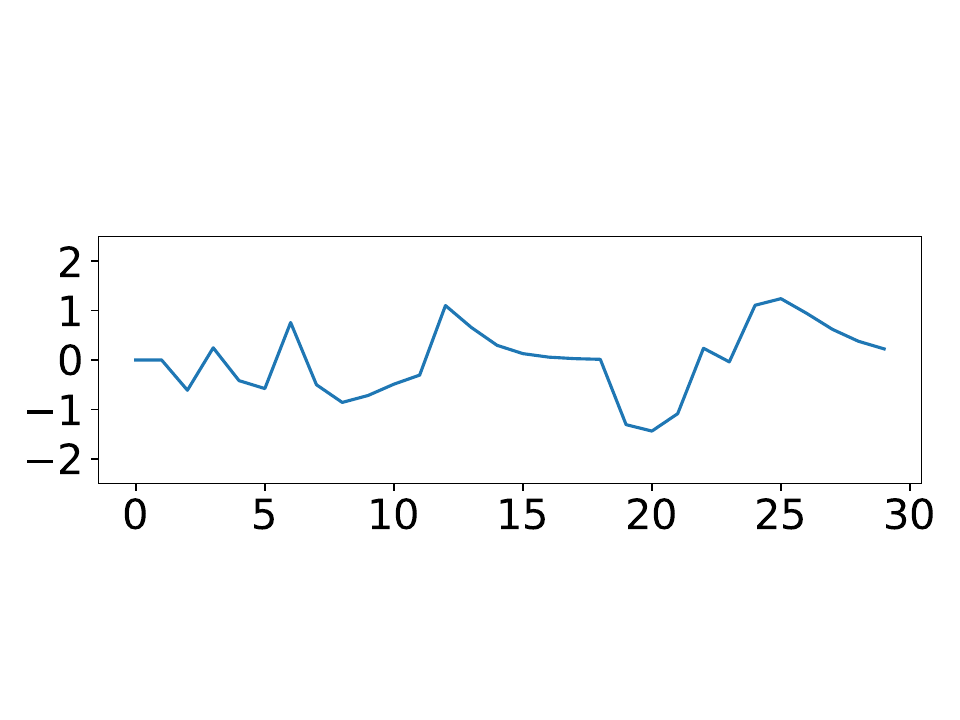}
      & \includegraphics[width=0.30\textwidth]{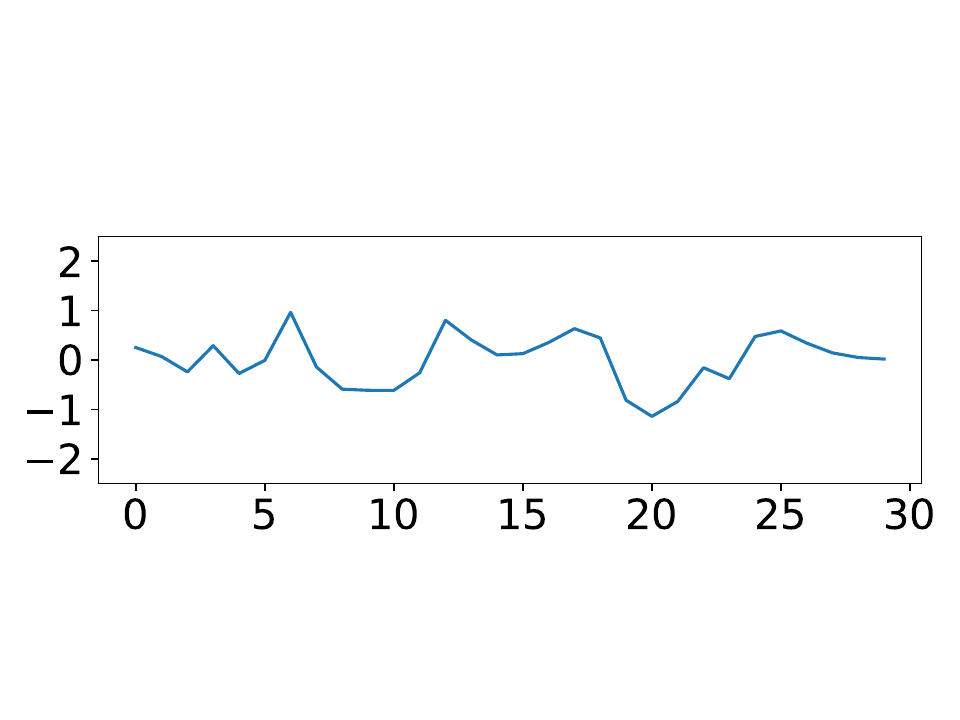}
      & \includegraphics[width=0.30\textwidth]{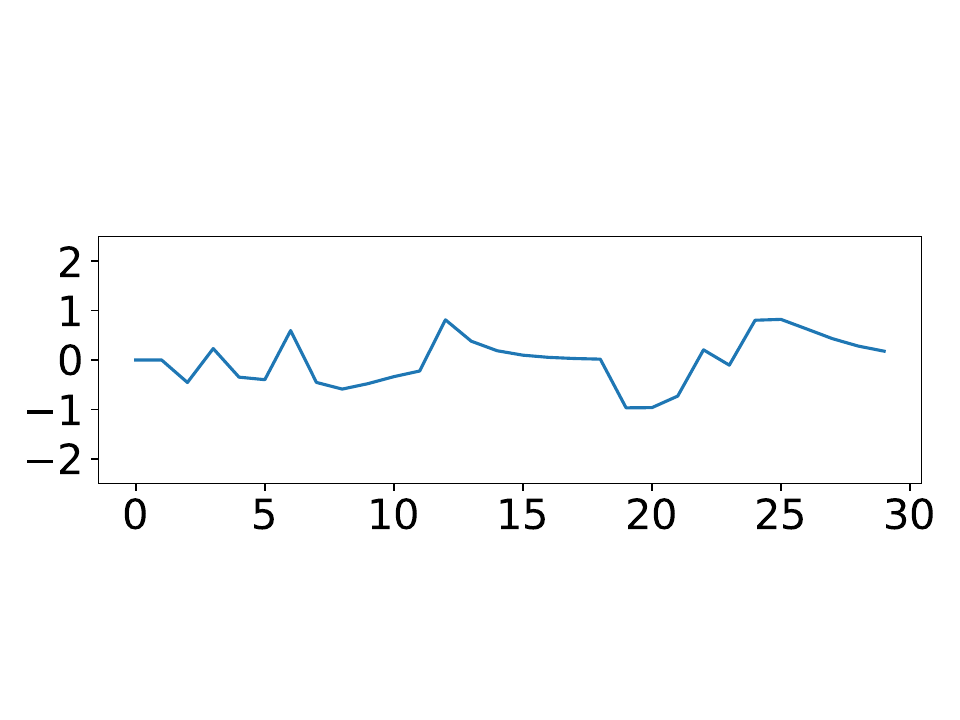}
      \vspace{-6mm} \\      
      {\em DoT for $\sigma = 4$}
      & {\em DoG for $\sigma = 4$} 
      & {\em DoE for $\sigma = 4$} \vspace{-8mm} \\
      \includegraphics[width=0.30\textwidth]{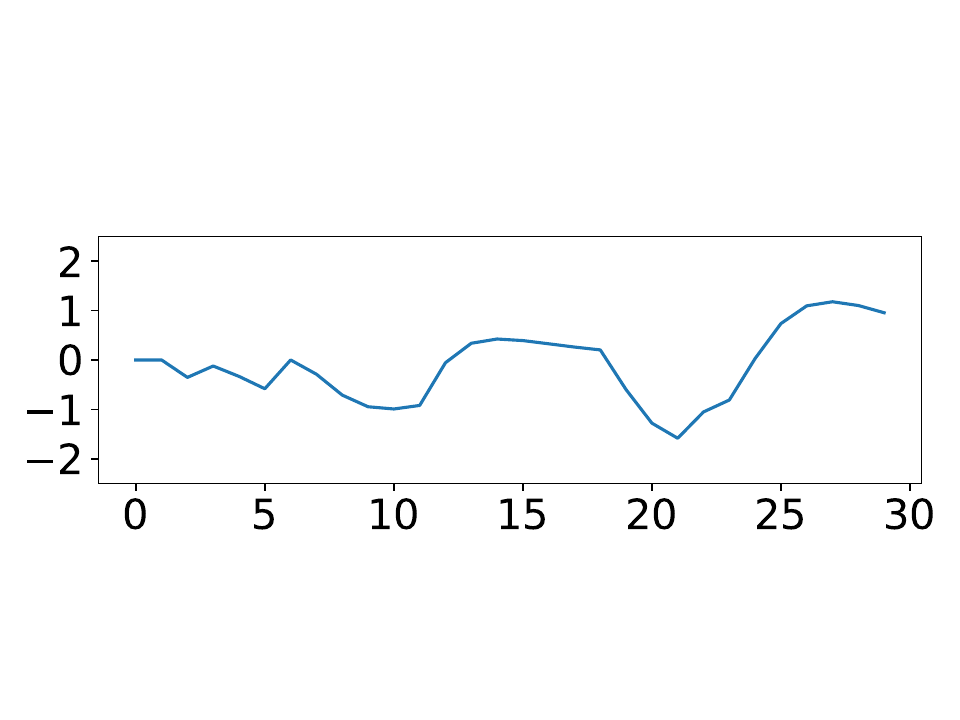}
      & \includegraphics[width=0.30\textwidth]{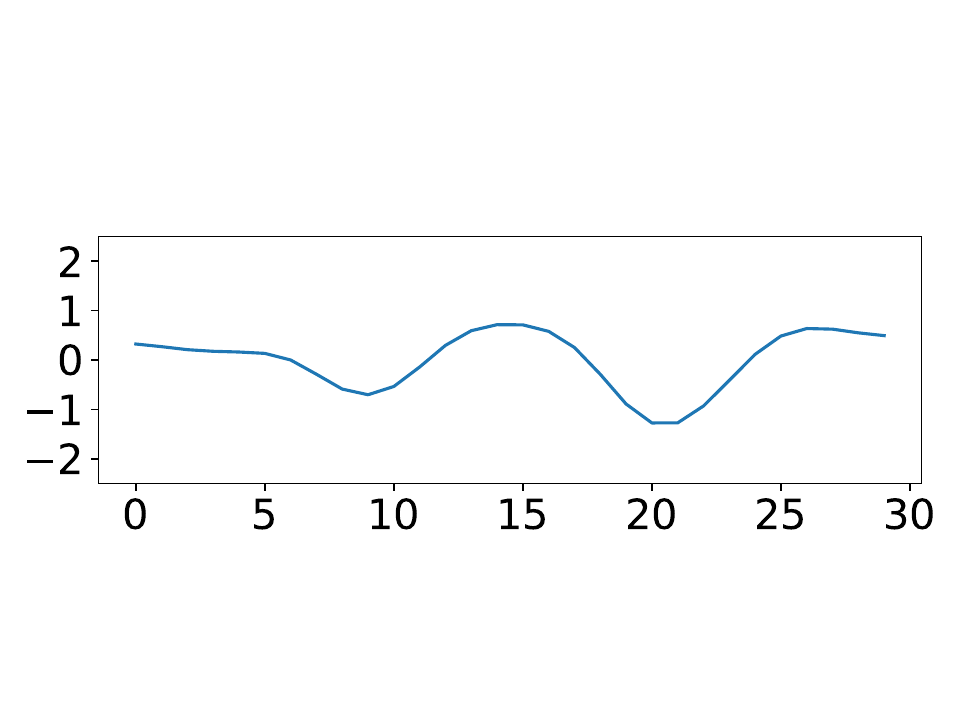}
      & \includegraphics[width=0.30\textwidth]{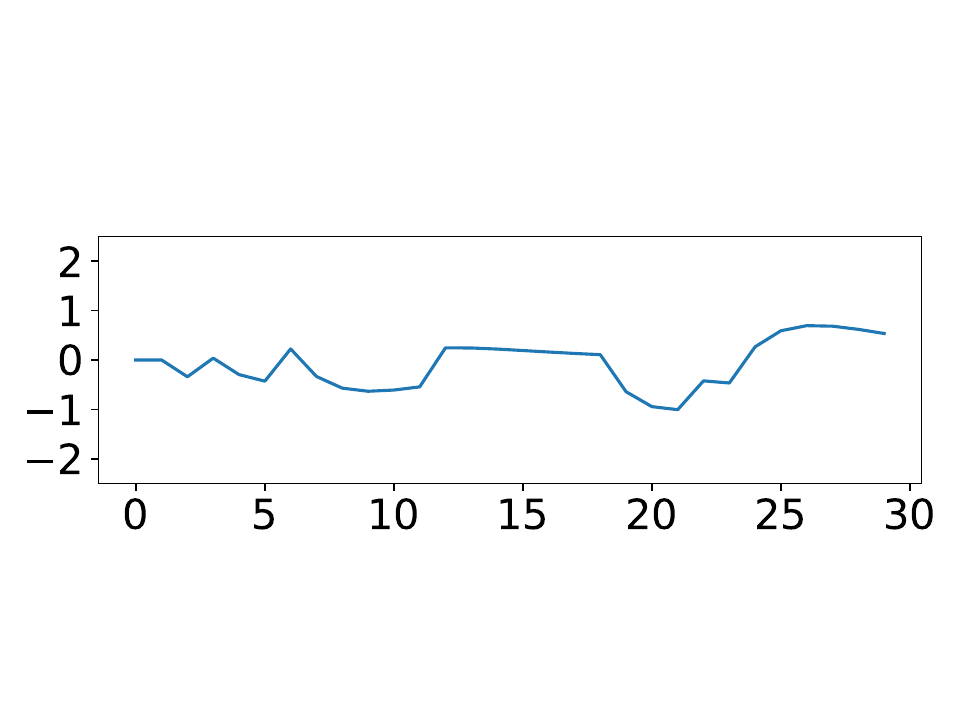}
      \vspace{-6mm} \\      
      {\em DoT for $\sigma = 8$}
      & {\em DoG for $\sigma = 8$} 
      & {\em DoE for $\sigma = 8$} \vspace{-8mm} \\
      \includegraphics[width=0.30\textwidth]{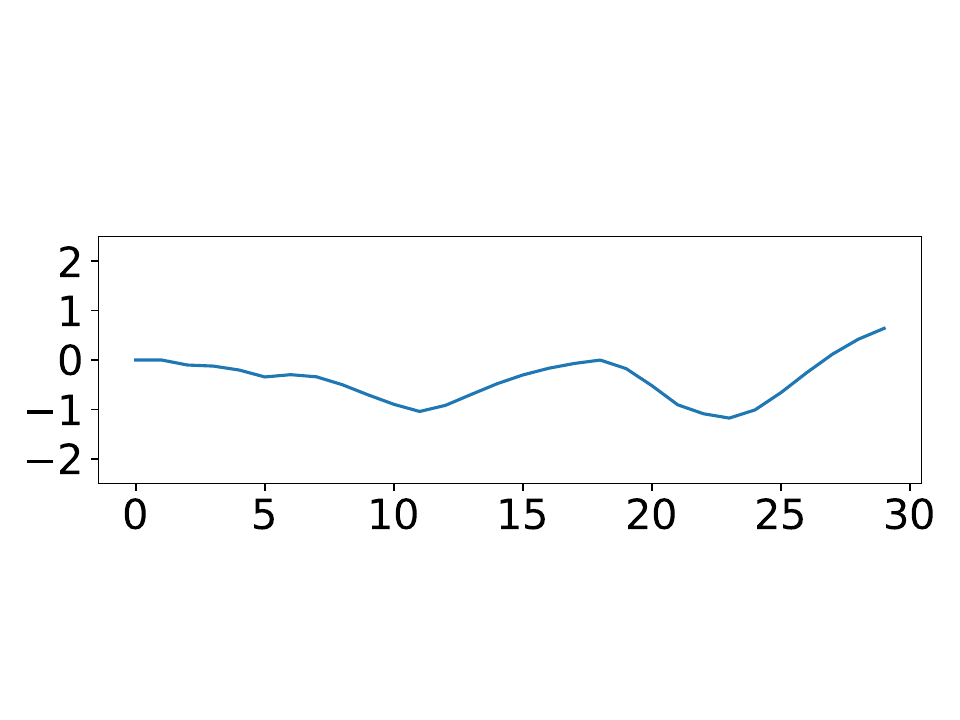}
      & \includegraphics[width=0.30\textwidth]{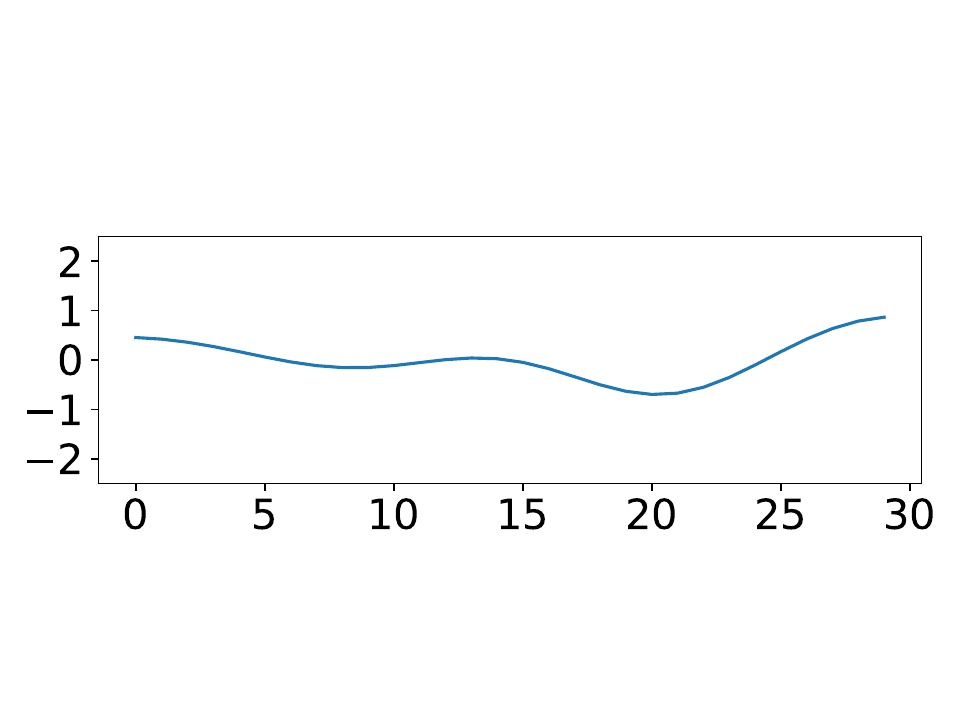}
      & \includegraphics[width=0.30\textwidth]{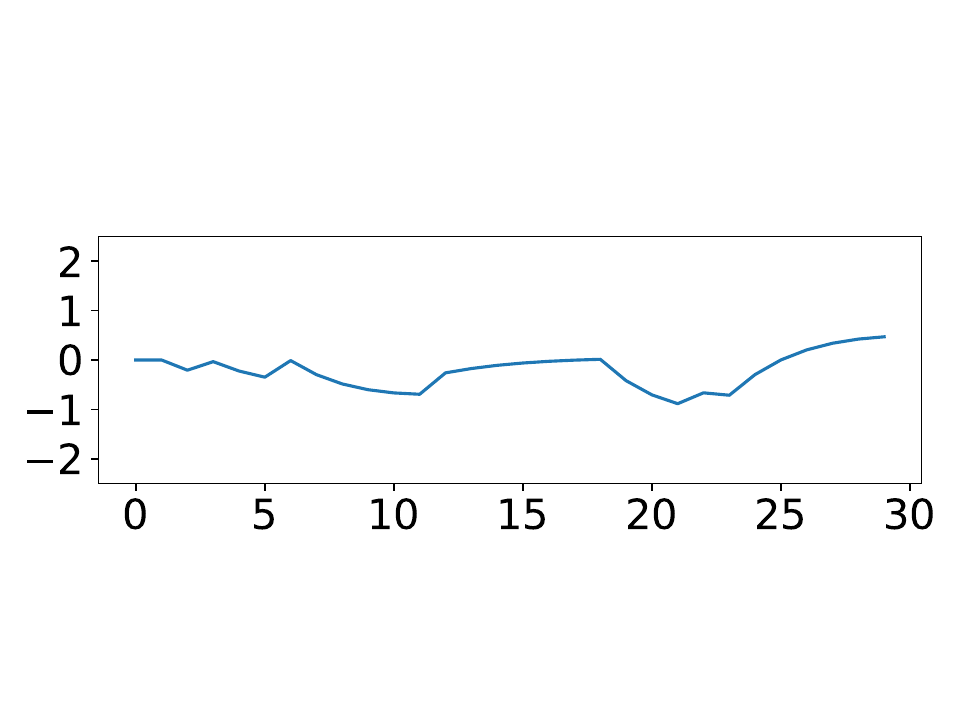}
      \vspace{-6mm} \\      
      {\em $L$ for $\sigma = 8$}
      & {\em $L$ for $\sigma = 8$} 
      & {\em $L$ for $\sigma = 8$} \vspace{-8mm} \\
      \includegraphics[width=0.30\textwidth]{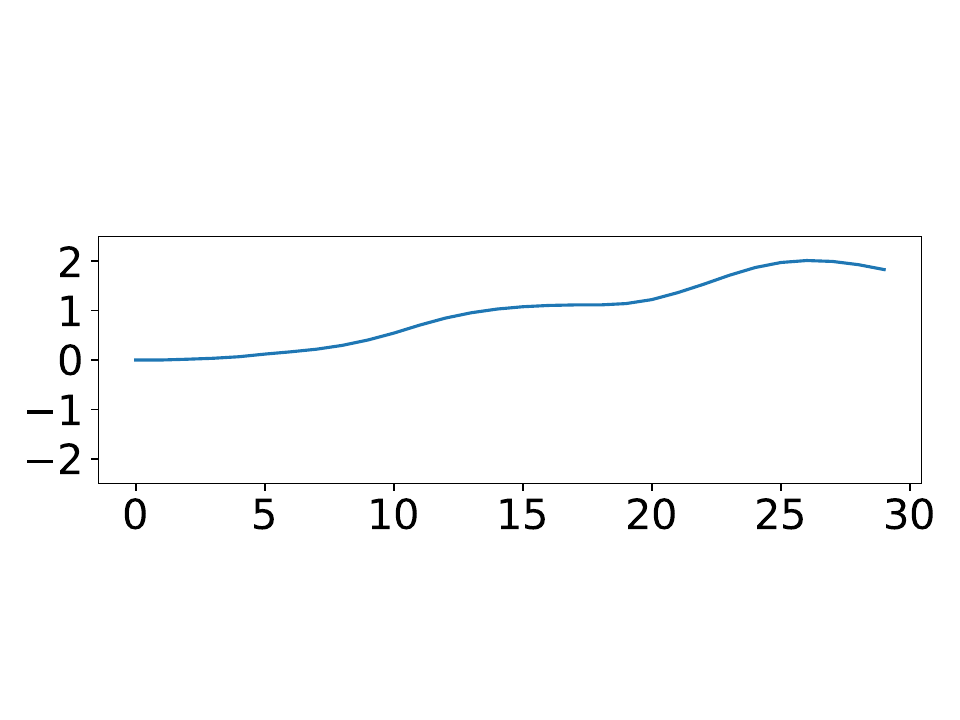}
      & \includegraphics[width=0.30\textwidth]{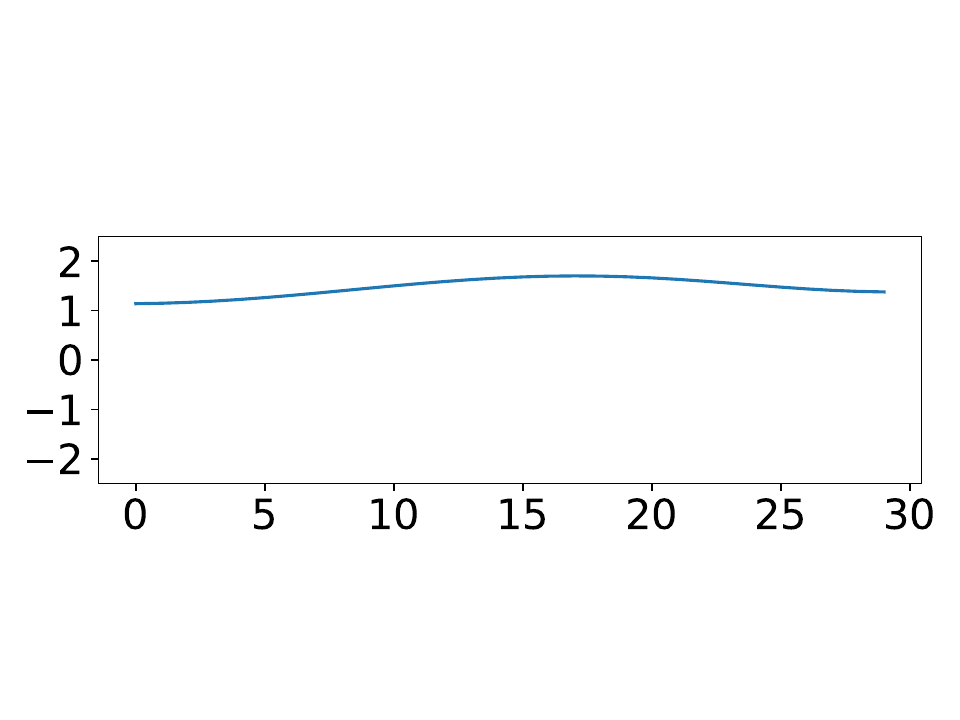}
      & \includegraphics[width=0.30\textwidth]{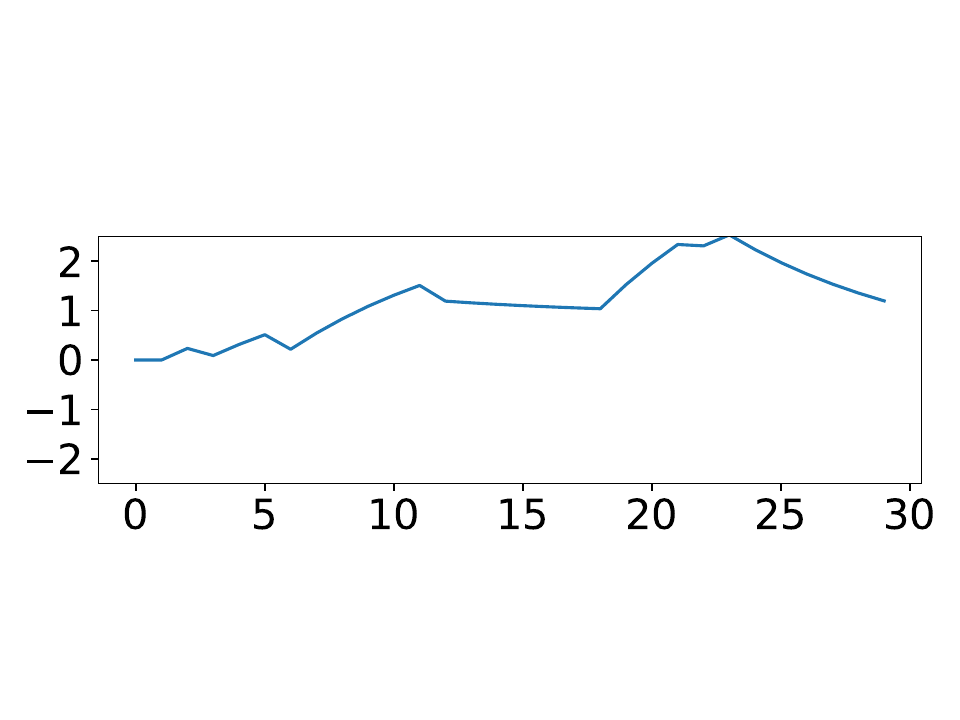}
      \vspace{-6mm} \\      
    {\em reconstructed signal}
      & {\em reconstructed signal} 
      & {\em reconstructed signal} \vspace{-8mm} \\      
      \includegraphics[width=0.30\textwidth]{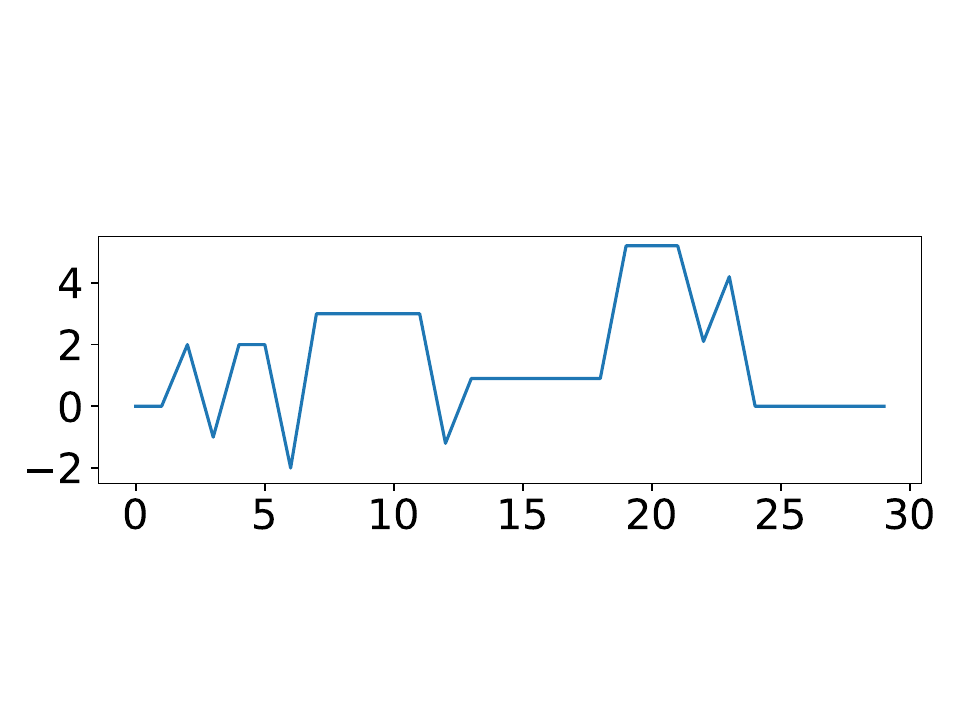}
      & \includegraphics[width=0.30\textwidth]{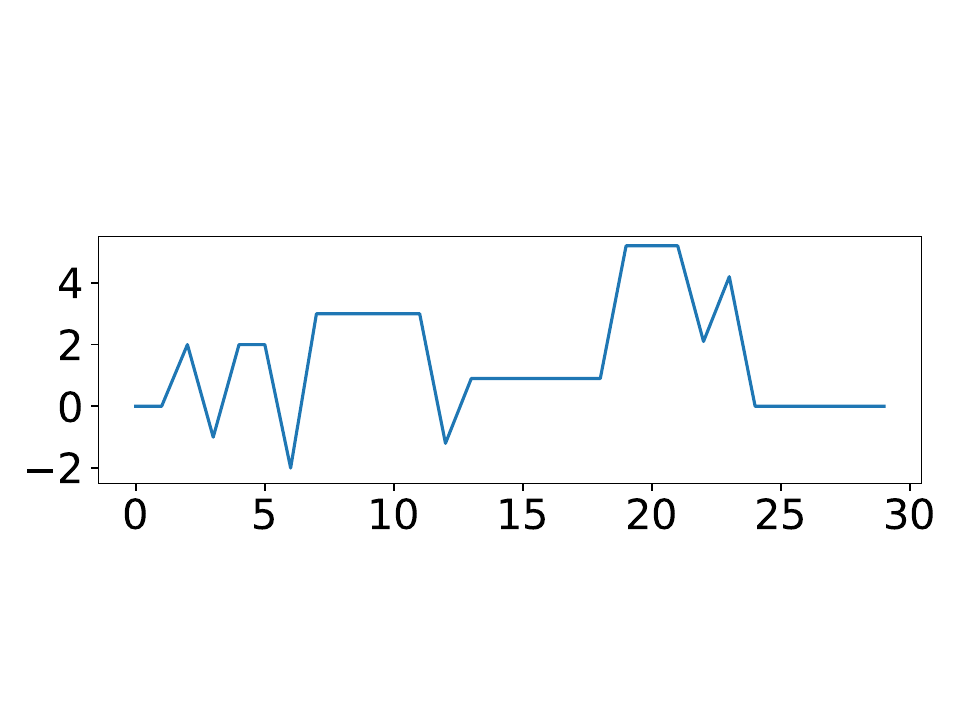}
      & \includegraphics[width=0.30\textwidth]{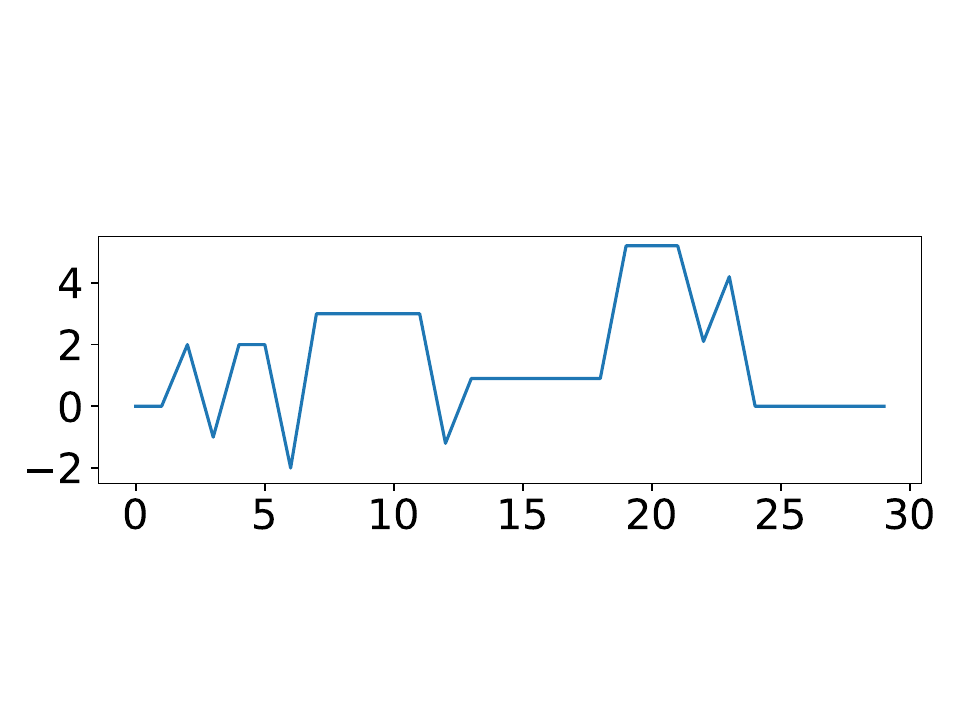}
      \vspace{-6mm} \\      
    \end{tabular}
  \end{center}
  \caption{Bandpass representations computed of the model signal
    ``Blocks'' at different temporal scale levels $\sigma = \sqrt{\tau}$
    using (i)~the time-causal DoT, (ii)~the non-causal DoG and and
    (iii)~the time-causal DoE approaches, including reconstructions of the
    original signal from each type of bandpass representation in combination
    with the scale-space representations $L$ at the coarsest level of scale.}
  \label{fig-bandpass-blocks}
\end{figure*}

\begin{figure*}[hbtp]
  \begin{center}
    \begin{tabular}{ccc}
      {\em Input signal\/} \vspace{-8mm} \\
      \includegraphics[width=0.30\textwidth]{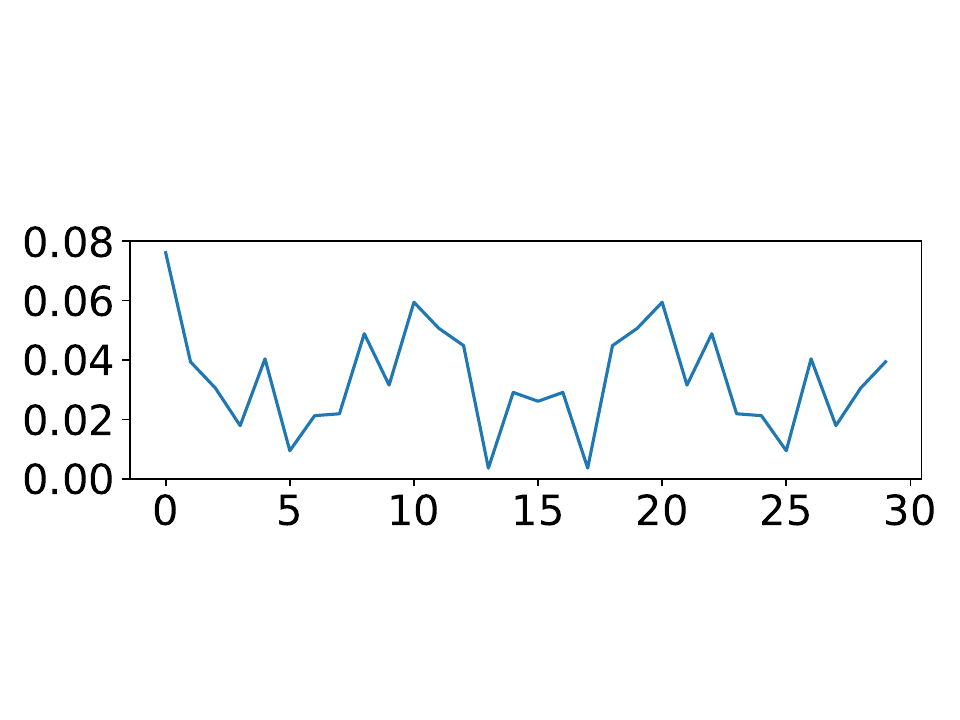}
      \vspace{-6mm} \\ 
      {\em DoT for $\sigma = 1$}
      & {\em DoG for $\sigma = 1$} 
      & {\em DoE for $\sigma = 1$} \vspace{-8mm}  \\
      \includegraphics[width=0.30\textwidth]{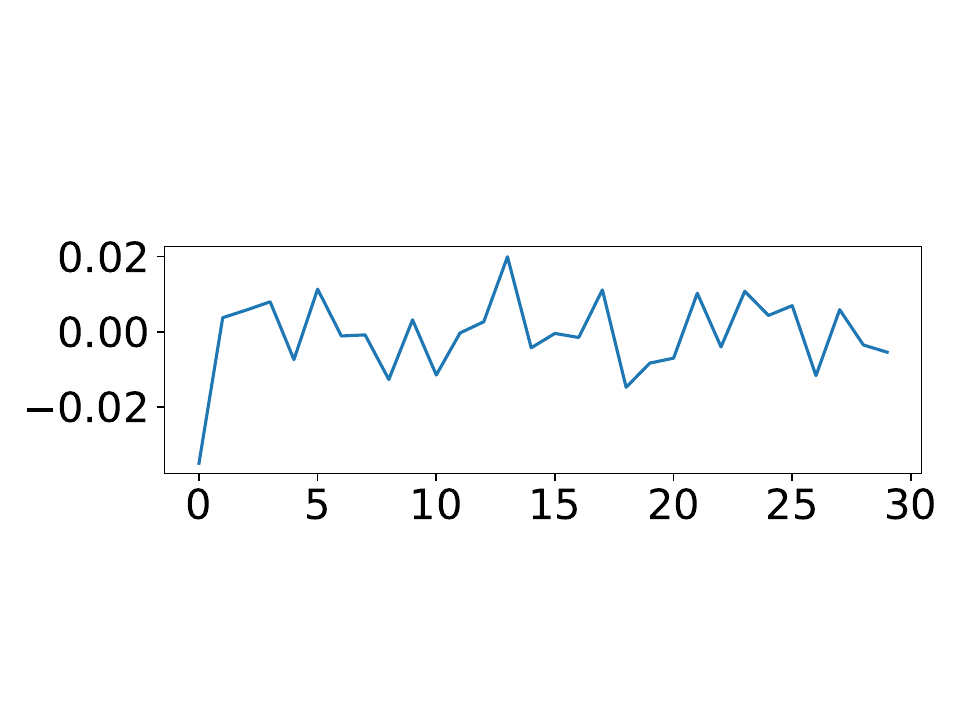}
      & \includegraphics[width=0.30\textwidth]{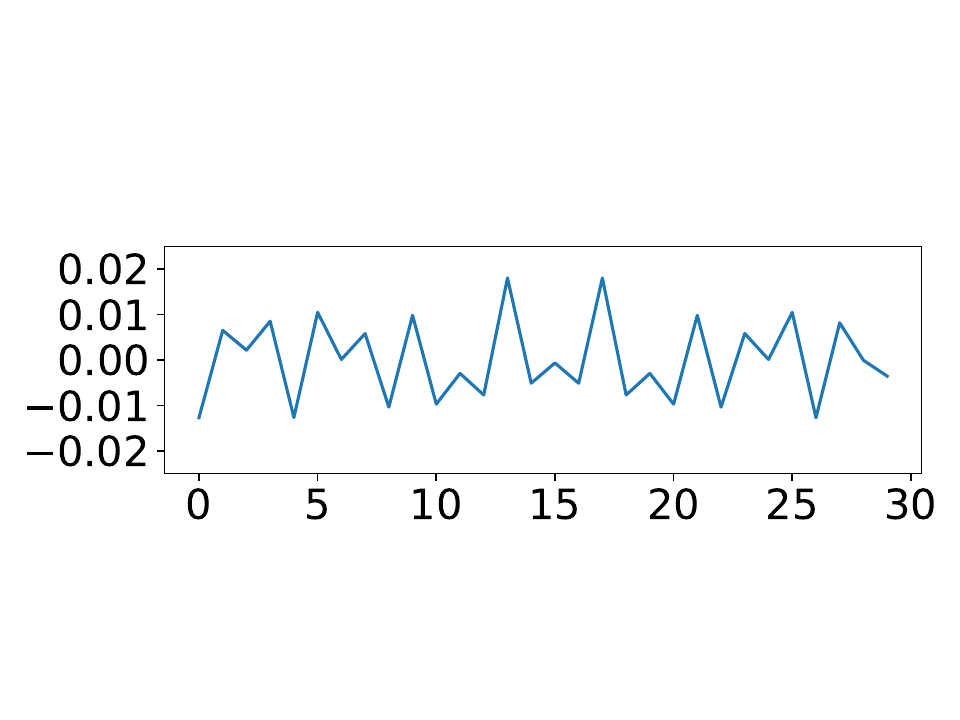}
      & \includegraphics[width=0.30\textwidth]{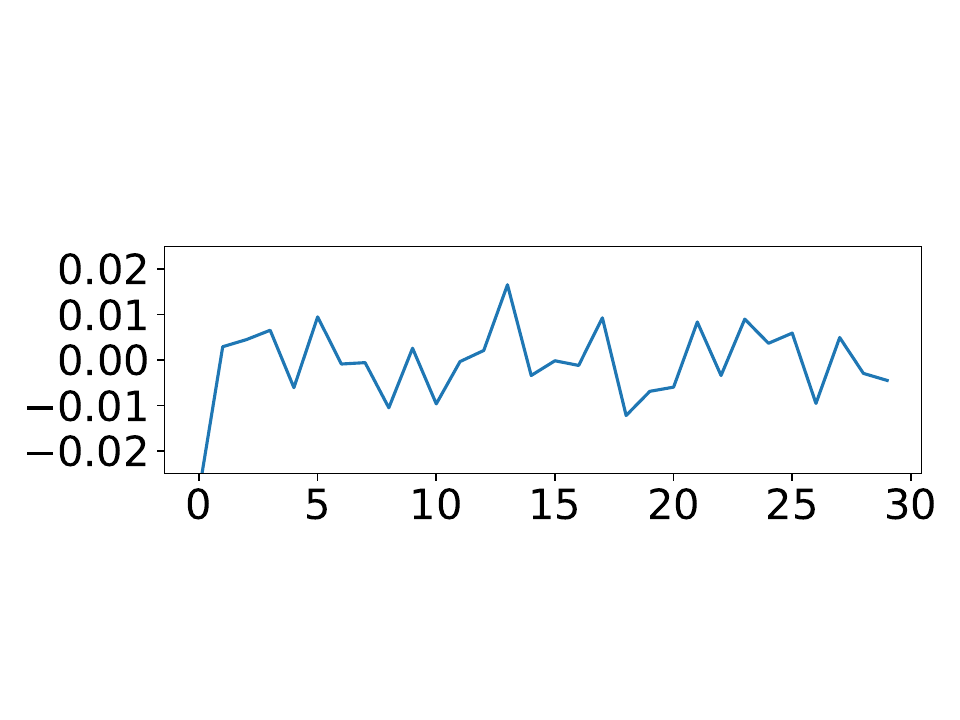}
        \vspace{-6mm} \\      
      {\em DoT for $\sigma = 2$}
      & {\em DoG for $\sigma = 2$} 
      & {\em DoE for $\sigma = 2$} \vspace{-8mm} \\
      \includegraphics[width=0.30\textwidth]{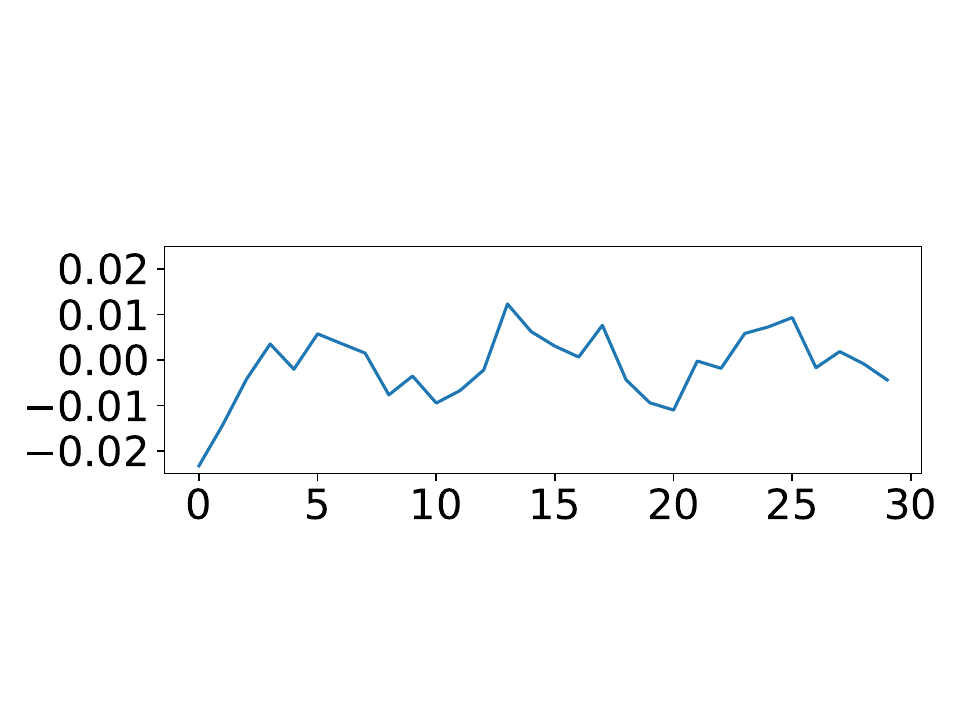}
      & \includegraphics[width=0.30\textwidth]{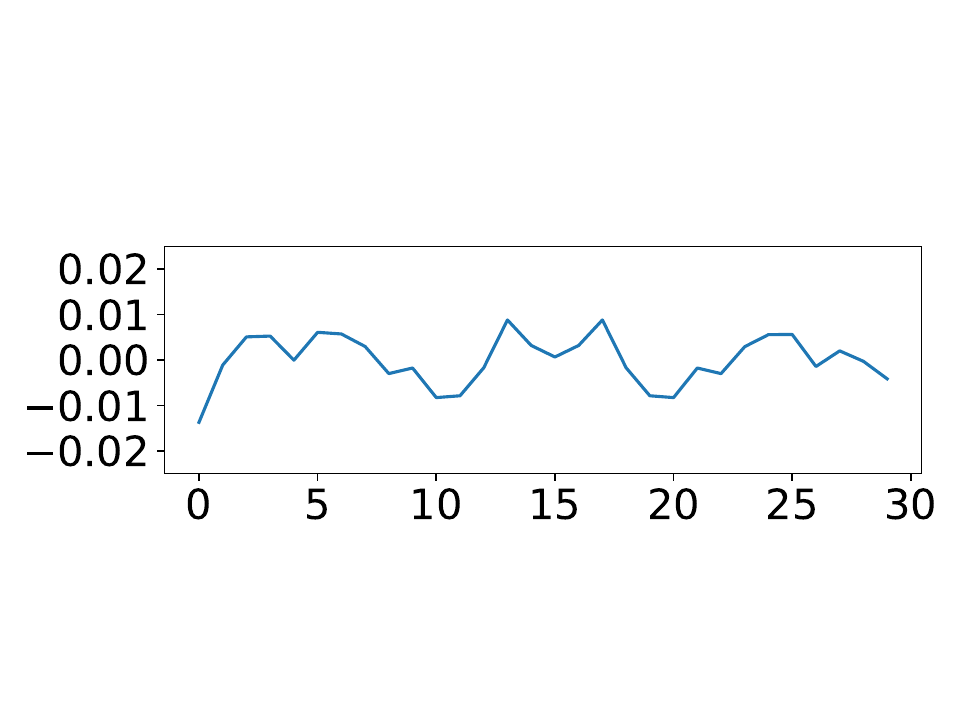}
      & \includegraphics[width=0.30\textwidth]{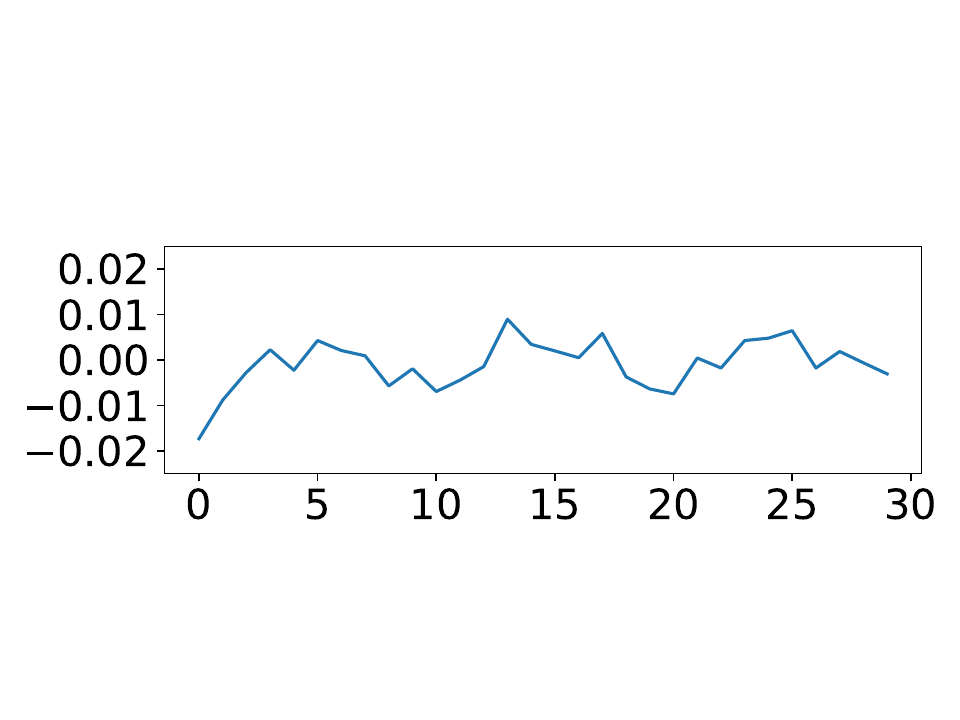}
      \vspace{-6mm} \\      
      {\em DoT for $\sigma = 4$}
      & {\em DoG for $\sigma = 4$} 
      & {\em DoE for $\sigma = 4$} \vspace{-8mm} \\
      \includegraphics[width=0.30\textwidth]{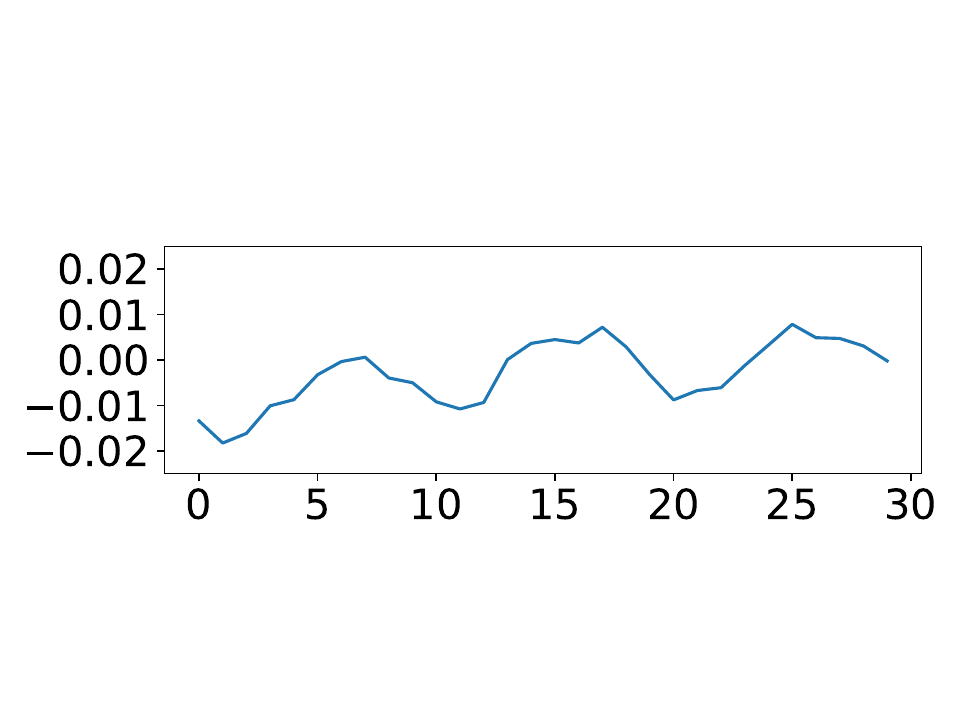}
      & \includegraphics[width=0.30\textwidth]{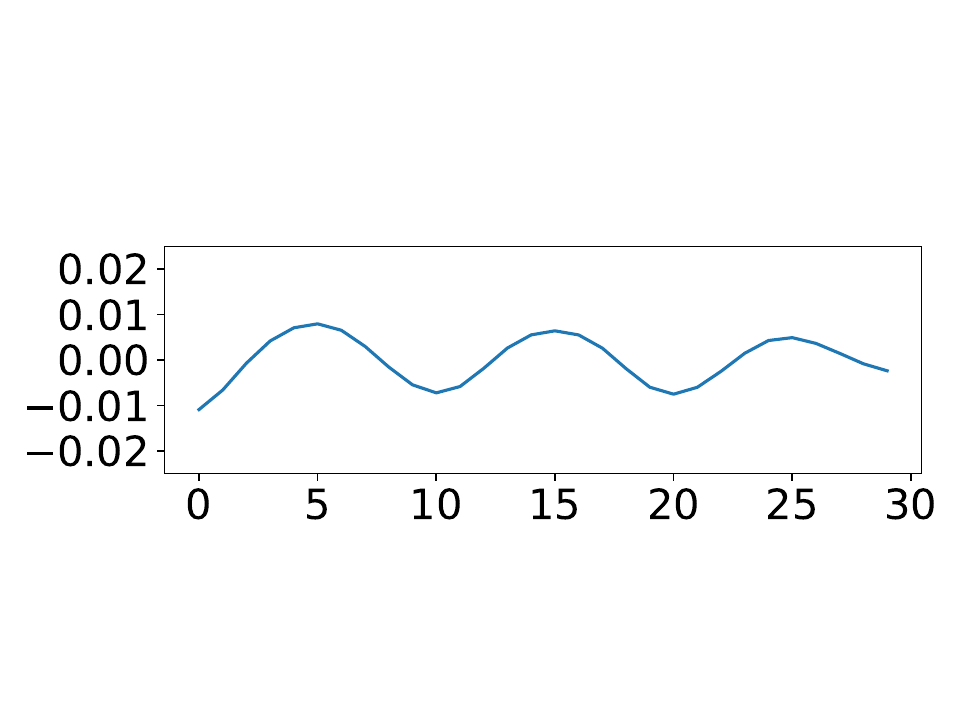}
      & \includegraphics[width=0.30\textwidth]{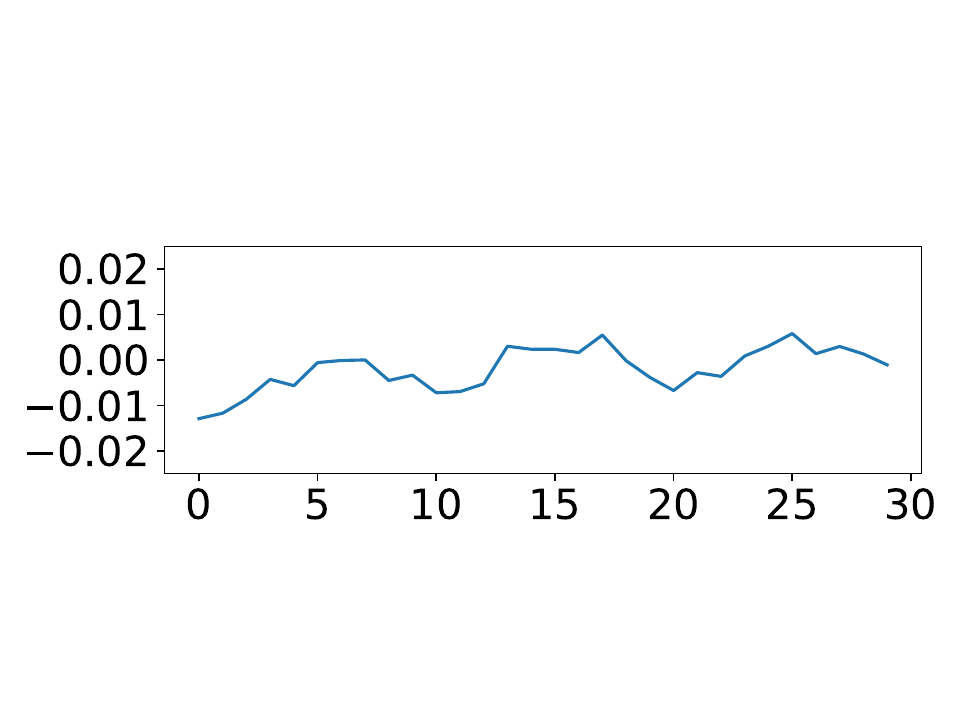}
      \vspace{-6mm} \\      
      {\em DoT for $\sigma = 8$}
      & {\em DoG for $\sigma = 8$} 
      & {\em DoE for $\sigma = 8$} \vspace{-8mm} \\
      \includegraphics[width=0.30\textwidth]{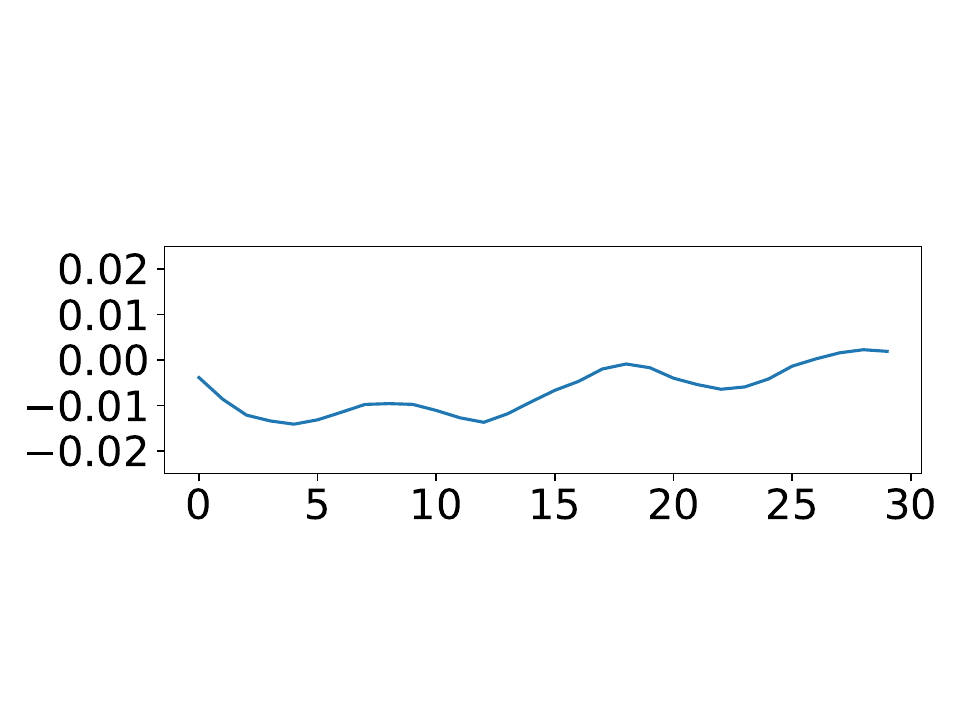}
      & \includegraphics[width=0.30\textwidth]{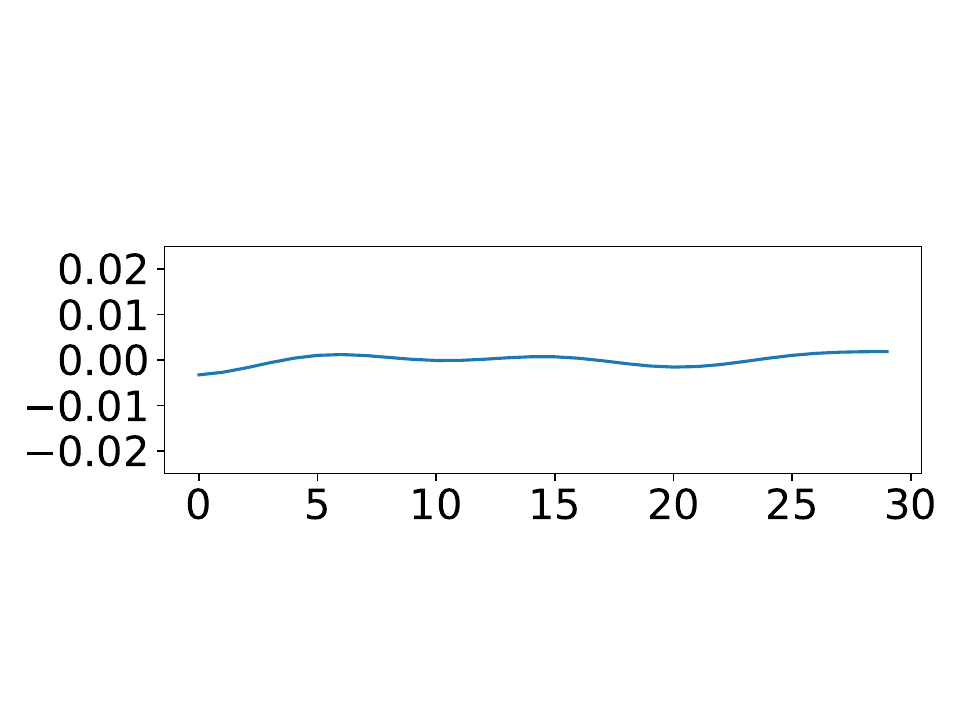}
      & \includegraphics[width=0.30\textwidth]{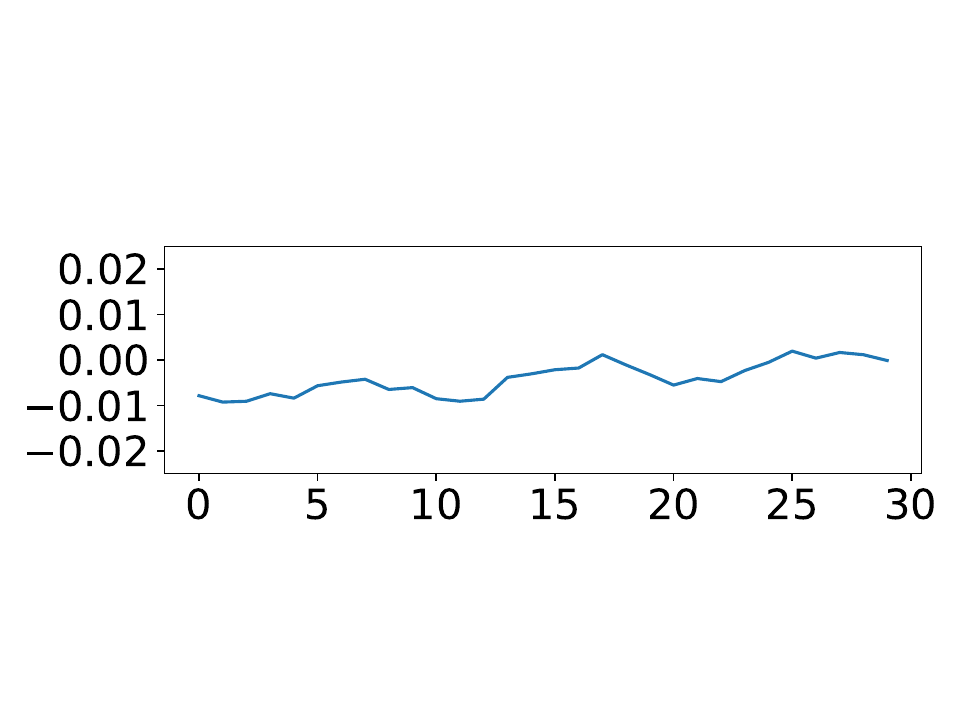}
      \vspace{-6mm} \\      
      {\em $L$ for $\sigma = 8$}
      & {\em $L$ for $\sigma = 8$} 
      & {\em $L$ for $\sigma = 8$} \vspace{-8mm} \\
      \includegraphics[width=0.30\textwidth]{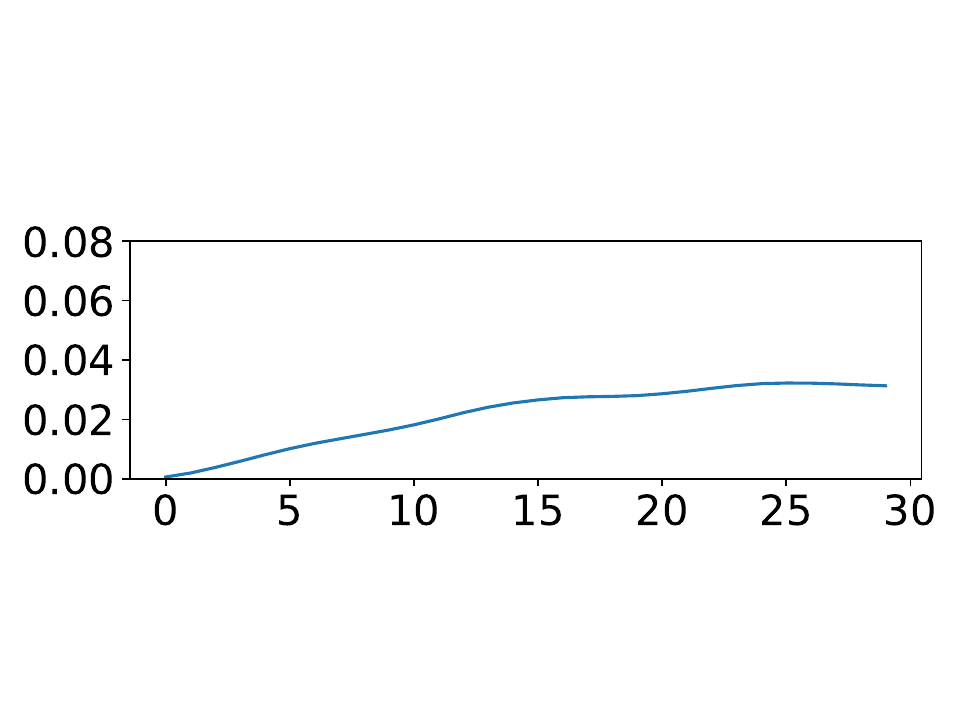}
      & \includegraphics[width=0.30\textwidth]{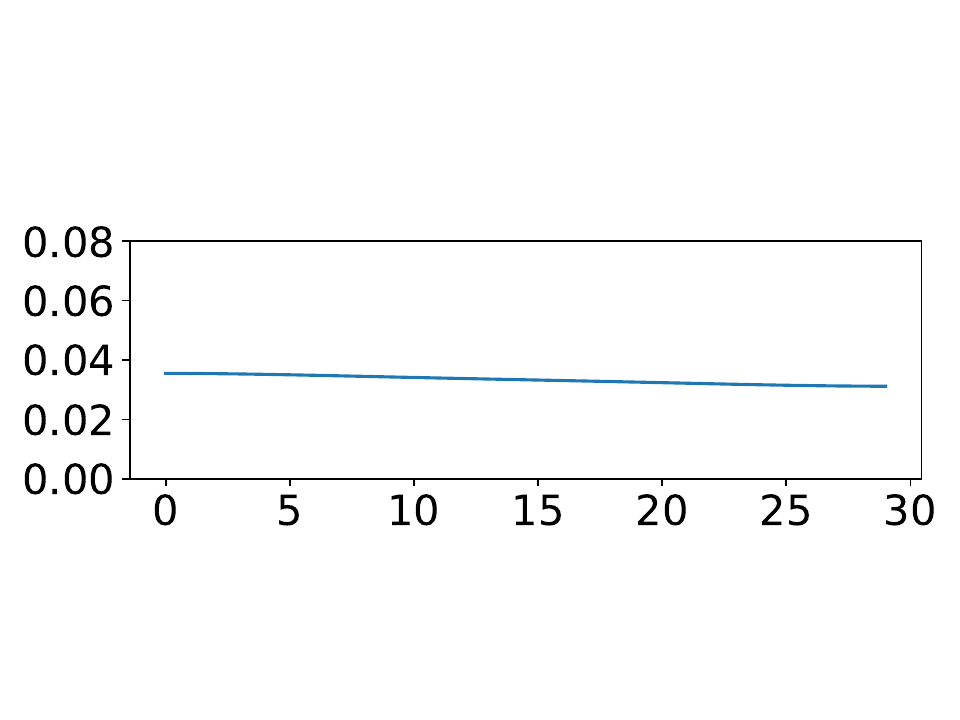}
      & \includegraphics[width=0.30\textwidth]{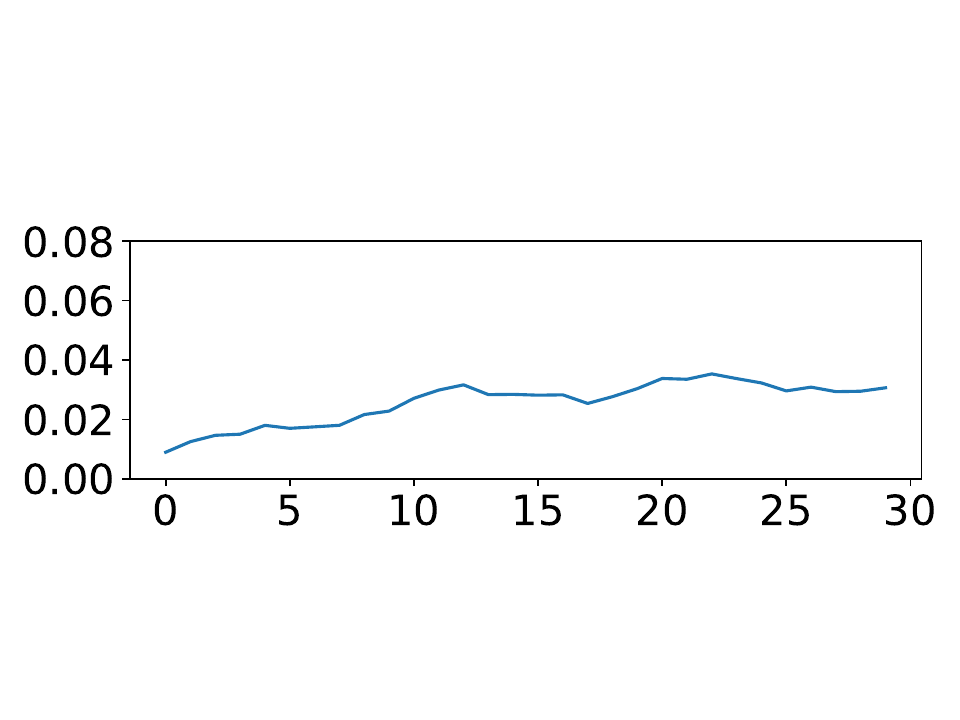}
      \vspace{-6mm} \\      
    {\em reconstructed signal}
      & {\em reconstructed signal} 
      & {\em reconstructed signal} \vspace{-8mm} \\      
      \includegraphics[width=0.30\textwidth]{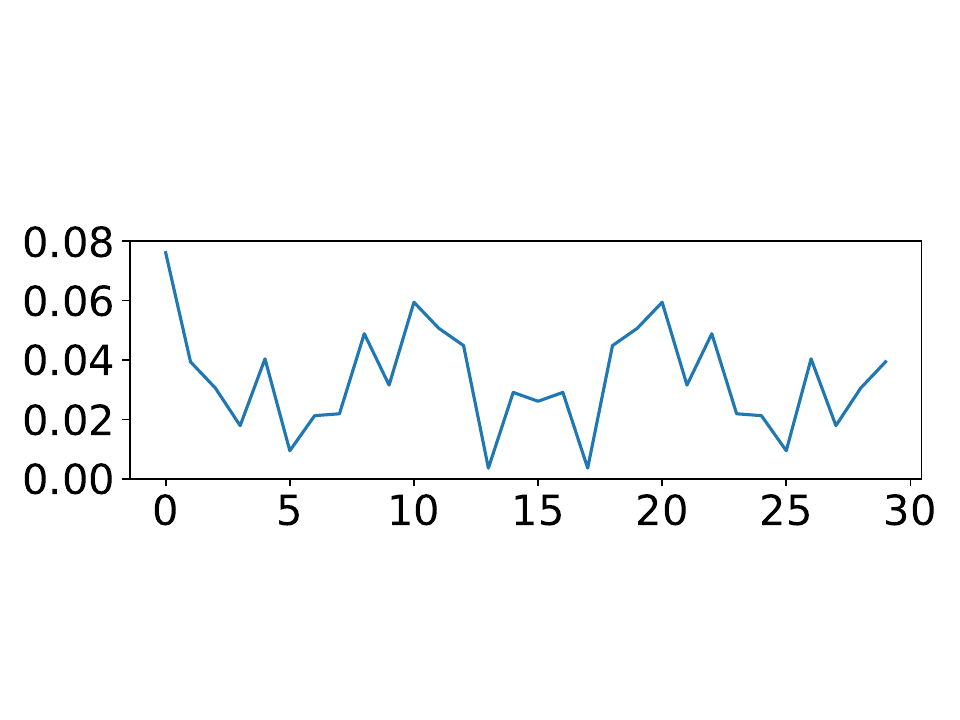}
      & \includegraphics[width=0.30\textwidth]{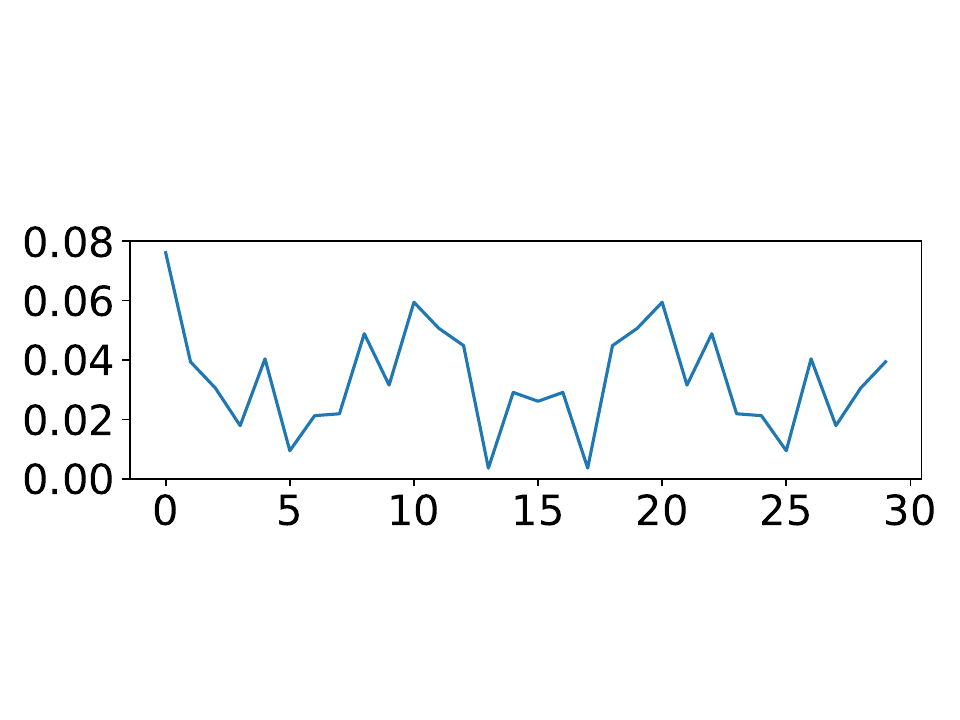}
      & \includegraphics[width=0.30\textwidth]{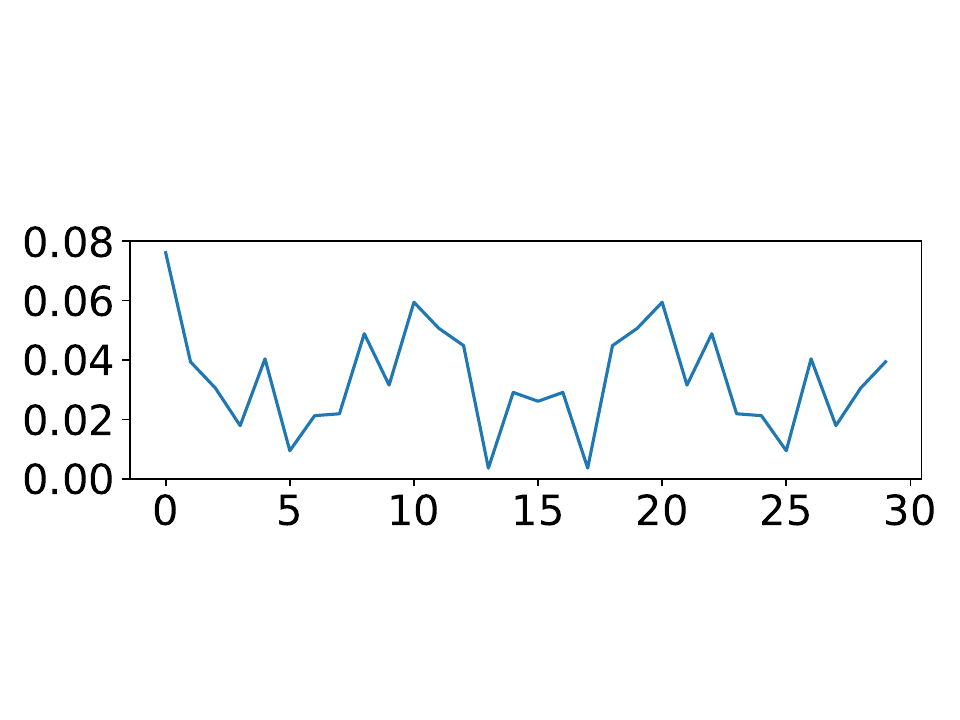}
      \vspace{-6mm} \\      
    \end{tabular}
  \end{center}
  \caption{Bandpass representations computed of the model signal
    ``Riemann'' at different temporal scale levels $\sigma = \sqrt{\tau}$
    using (i)~the time-causal DoT, (ii)~the non-causal DoG and and
    (iii)~the time-causal DoE approaches, including reconstructions of the
    original signal from each type of bandpass representation in combination
    with the scale-space representations $L$ at the coarsest level of scale.}
  \label{fig-bandpass-riemann}
\end{figure*}

\subsection{Experiments}
\label{sec-exps}

\subsubsection{Bandpass decompositions based on time-causal wavelets}

To illustrate the properties of the proposed notion of time-causal
bandpass representation,
Figures~\ref{fig-bandpass-blocks}--\ref{fig-bandpass-riemann}
show the result of computing
the following three types of discrete bandpass representations from two
model signals%
\footnote{The model signals ``Block'' and  ``Riemann'' have been generated with the function
  pywt.data.demo\_signal() in the PyWavelets library for $n = 30$ time
  samples. The reason for choosing such comparably short test signals
  in this case is to be able to represent the main temporal structures over
  temporal scales using as few as 4 bandpass channels, such that the
  visualizations of all the scale channels fit within single column
  for each type of bandpass representation.}
using 4 temporal scale levels according to
$\tau_k = \tau_0 \, c^{2k}$ (\ref{eq-temp-sc-levels}) for
$\tau_0 = 1$ and $c = 2$, with 4 additional
pre-smoothing stages prior to the first scale level used
in the time-causal limit kernel:
\begin{itemize}
\item
  Discrete time-causal bandpass representations $\Delta L_{\text{DoT}}(t;\; \tau_k, c)$
  according to (\ref{eq-disc-time-caus-bandpass-repr}),
  computed from differences between discrete temporal scale-space
  representations obtained by smoothing with discrete analogues of
  the time-causal limit kernel according to
  Section~\ref{sec-disc-anal-time-caus-limit-kern}, and as based on first-order recursive
  filters coupled in cascade as described in 
  Algorithm~\ref{fig-pseudo-code-rec-filt-casc}, with the complementary
  reconstruction performed as described in
  Algorithm~\ref{fig-pseudo-code-decode-bandpass}.
  This approach constitutes the main type of pure
  time-causal bandpass representation proposed in this article.
\item
  Discrete analogues of the non-causal differences-of-Gaussians
  bandpass representation $\Delta L_{\text{DoG}}(t;\; \tau_k)$ according to
  (\ref{eq-diff-temp-sc-from-DoG}), with the underlying continuous convolution
  operations with continuous Gaussian kernels $g(t;\; \tau)$ replaced by discrete
  convolutions with the discrete analogue of the Gaussian kernel
  $T(m;\; \tau)$ according to (\ref{eq-disc-anal-gauss-def}).

  This approach can be seen as an improved method relative to previous
  forms of difference-of-Gaussians bandpass decompositions
  (Burt and Adelson \citeyear{BA83-COM},
  Crowley and Stern \citeyear{Cro84-dolp},
  Birch {\em et al.\/} \citeyear{BirMitBanRehYouCha10-OptComm},
  Li {\em et al.\/} \citeyear{LiYanLiLuXie20-IETImProc},
  Brown {\em et al.\/} \citeyear{BroBlaBerMurJes24-SurfTop}).
\item
  Discrete time-causal differences-of-exponentials
  bandpass representations $\Delta L_{\text{DoE}}(t;\; \tau_k)$
  \begin{equation}
    \Delta L_{\text{DoE}}(t;\; \tau_k) = L(t;\; \tau_k) - L(t; \tau_{k-1})
  \end{equation}
  defined from differences between the temporal scale channels obtained
  by smoothing the input signal with a set of {\em single\/}
  first-order integrators
  \begin{multline}
    \label{eq-norm-update}
    L(t;\; \tau_k) - L(t-1;\; \tau_k) = \\
    = - \frac{1}{1 + \mu_{k}} \,  (L(t-1;\; \tau_k) - f(t))
  \end{multline}
  with the time constants
  $\mu_k$ defined from the variances $\tau_k$ according to
  $\mu_k = (\sqrt{1 + 4 \tau_k}-1)/2$ in analogy with
  (\ref{eq-disc-time-constant}), and corresponding to the natural
  discretizations of the time-causal scale-channel representations
  \begin{equation}
    \label{eq-def-single-sc-chann-repr}
    L(t;\; \tau_k) = h_{\exp}(\cdot;\; \mu_k) * f(\cdot)
  \end{equation}
  obtained from convolution with single truncated exponential kernels 
  $h_{\exp}(t;\; \mu_k)$ according to
  (\ref{eq-trunc-exp-kern}), although for continuous signals the
  relationship between the variances $\tau_k$ and the time constants
  $\mu_k$ is instead given by $\mu_k = \sqrt{\tau_k}$ according to
  (\ref{eq-var-time-caus-limit-kern}). 

  This type of representation can be
  seen as a natural extension to a time-causal bandpass
  representation of the Laplace-transform-related temporal memory
  representations considered by Howard (\citeyear{How24-HandBookHumMem}) and
  Howard {\em et al.\/}\ (\citeyear{HowEsfLeSed25-CompBrainBeh}).
  In contrast to the time-causal temporal scale-space representation
  underlying the differences-of-time-causal-limit-kernels
  bandpass representation $\Delta L_{\text{DoT}}(t;\; \tau_k, c)$,
  the relationships between the
  temporal scale channels underlying this differences-of-exponentials
  bandpass representations $\Delta L_{\text{DoE}}(t;\; \tau_k)$ do,
  however, not obey a similar guaranteed simplifying property from
  finer to coarser levels of temporal scales, as used as the foundation for the theories of
  variation-diminishing continuous and discrete smoothing transformations in
  Sections~\ref{sec-char-1d-smooth-kernels}
  and~\ref{sec-char-1d-disc-smooth-kernels}.

  For this representation, both the bandpass representation and the
  reconstruction from the bandpass representation are computed in a
  structurally similar way as for the discrete bandpass representation
  based on discrete analogues of the time-causal limit kernel.
  The only difference is that the temporal smoothing is performed
  using a single recursive filter as opposed to using a cascade of recursive
  filters. Thereby, the reconstruction from the bandpass
  representation will also be exact for this discrete time-causal
  differences-of-exponentials bandpass representation.
\end{itemize}
As can be seen from the figures, all the three types of bandpass
representations lead to intuitively reasonable decompositions of the
signals in terms of different types of temporal structures at different temporal scales.

Notably, are inherent temporal delays for time-causal bandpass representations.
Such temporal delays constitute an inherent property of any time-causal
signal processing operations, since all processing is restricted to
what has occurred in the past. For the non-causal
differences-of-Gaussians representations, there are, however, not any
temporal delays, because those representations are symmetric relative
to the past and the future.

From the visualizations of the bandpass representation, we can see
that the differences-of-Gaussians
representations are somewhat smoother at coarser temporal scales than
the differences-of-time-causal-limit-kernel representations. An
explanation for this property is that the Gaussian kernel as well as
its discrete analogue are both infinitely divisible smoothing kernels,
which thereby implies very strong smoothing effects given any strictly
positive scale level. The differences-of-time-causal-limit-kernel
representations are, in turn, significantly smoother than the
differences-of-exponentials representations, because the former are
computed from a set of recursive filters coupled in cascade, whereas
the latter are computed from using only a single recursive filter for
each scale channel.

We can also note that the reconstructions $\hat{f}$ of the original
signal $f$ from the bandpass representations in combination with
the scale-space representation at the coarsest scale do fully
reconstruct the original signal, also for the proposed discrete
implementation. Numerically, the reconstruction errors
\begin{equation}
  \label{eq-reconst-error}
  \epsilon = \frac{\| \hat{f} - f \|_2}{\| f \|_2}
\end{equation}
are at the level of machine precision,
as shown in Table~\ref{tab-reconst-error}.

\begin{table}[hbt]
  \begin{tabular}{lccc}
    \hline
    Signal & $\epsilon_{\text{DoT}}$ & $\epsilon_{\text{DoG}}$ & $\epsilon_{\text{DoE}}$ \\
    \hline
    Blocks & $5.9 \cdot 10^{-17}$ & $3.7 \cdot 10^{-17}$ & $4.2 \cdot 10^{-17}$ \\
    Riemann & $8.5 \cdot 10^{-17}$ & $4.0 \cdot 10^{-17}$ & $4.3 \cdot 10^{-17}$ \\
    \hline
  \end{tabular}
  \caption{Reconstruction errors according to (\ref{eq-reconst-error})
    for the three types of pure bandpass representations considered
    here, in terms of (i)~differences-of-time-causal-limit-kernel (DoT)
    representations, (ii)~differences-of-Gaussians (DoG)
    representations and (iii)~differences-of-exponentials (DoE)
    computed for the two model signals shown in
    Figures~\ref{fig-bandpass-blocks}--\ref{fig-bandpass-riemann}.}
  \label{tab-reconst-error}
\end{table}

\begin{figure*}[hbtp]
  \begin{center}
    \begin{tabular}{cc}
      {\em Input signal\/}  \\ 
      \includegraphics[width=0.30\textwidth]{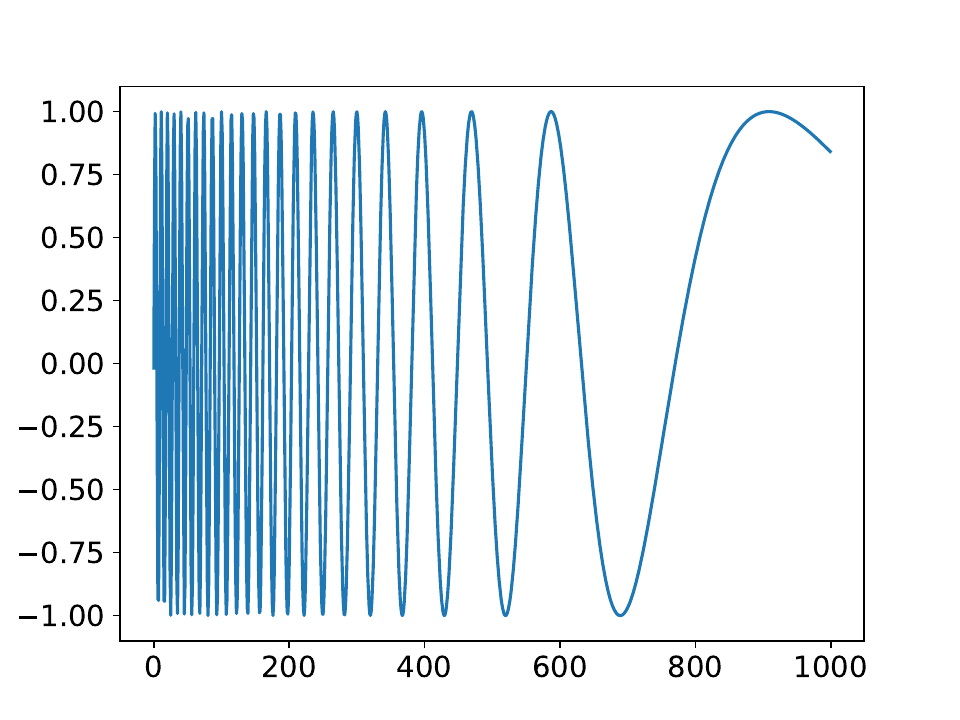}
      \\
      {\em first-order derivative for $c = \sqrt{2}$}
      & {\em first-order derivative for $c = 2$} \\
      \includegraphics[width=0.30\textwidth]{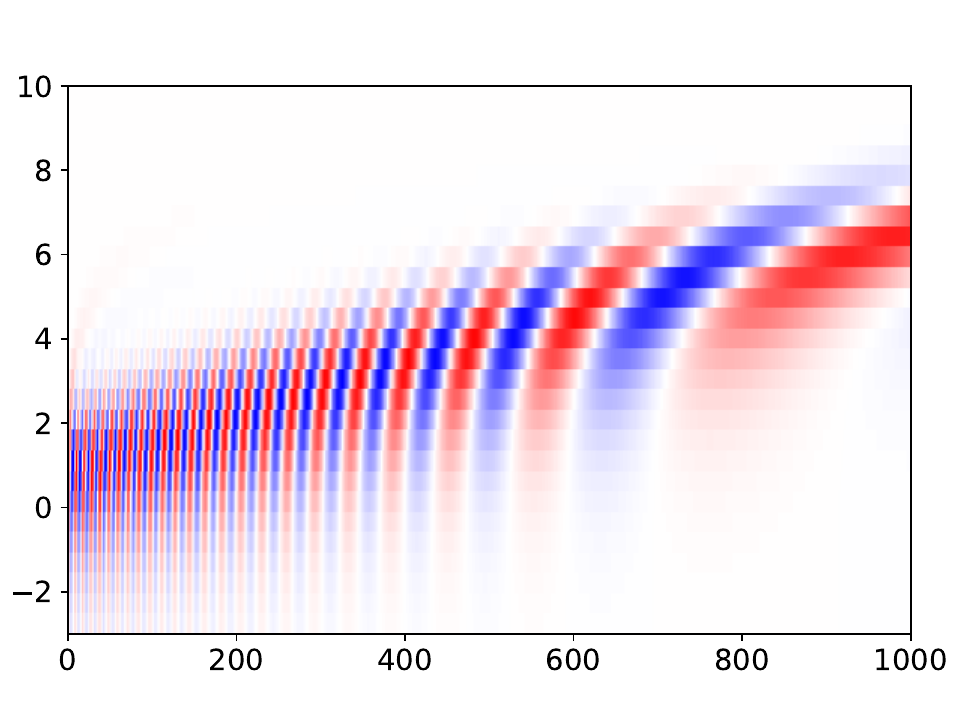}
      &  \includegraphics[width=0.30\textwidth]{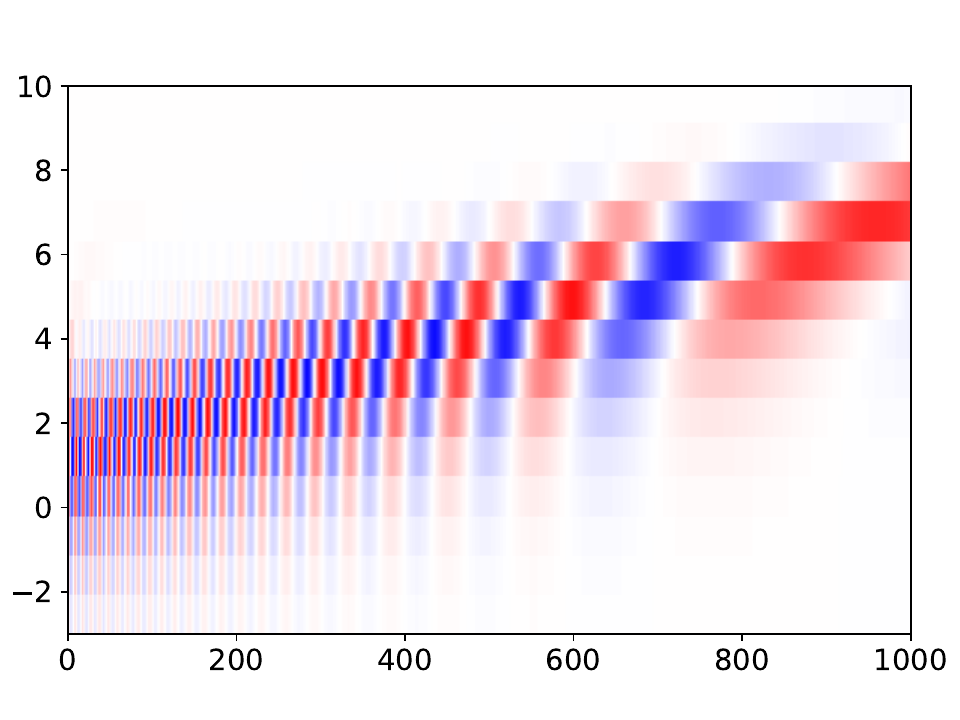} \\
     {\em second-order derivative for $c = \sqrt{2}$}
      & {\em second-order derivative for $c = 2$} \\
      \includegraphics[width=0.30\textwidth]{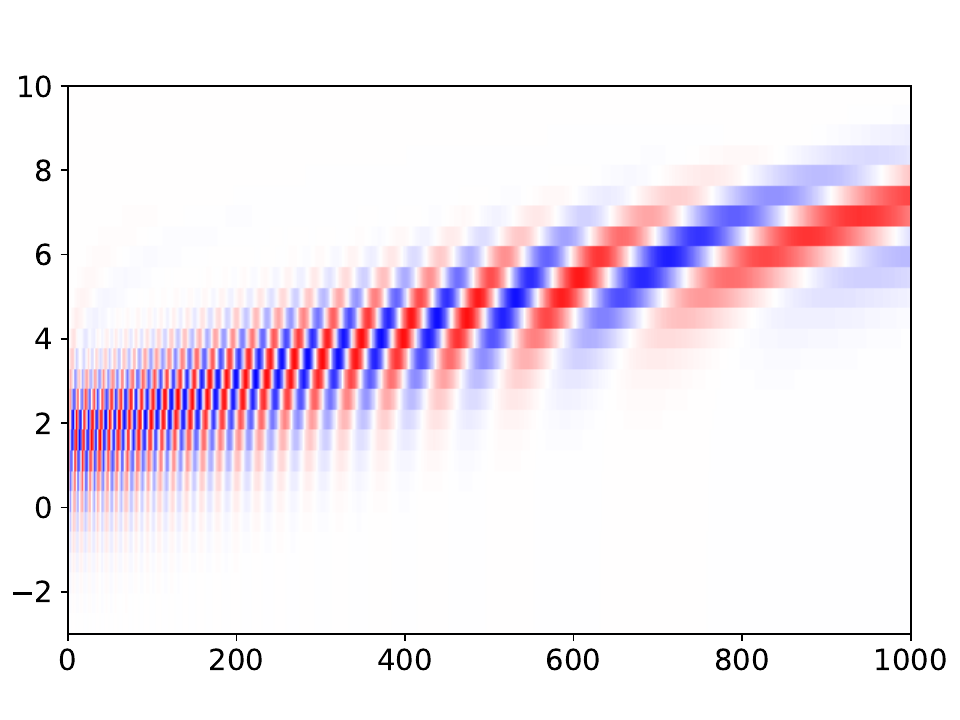}
      &  \includegraphics[width=0.30\textwidth]{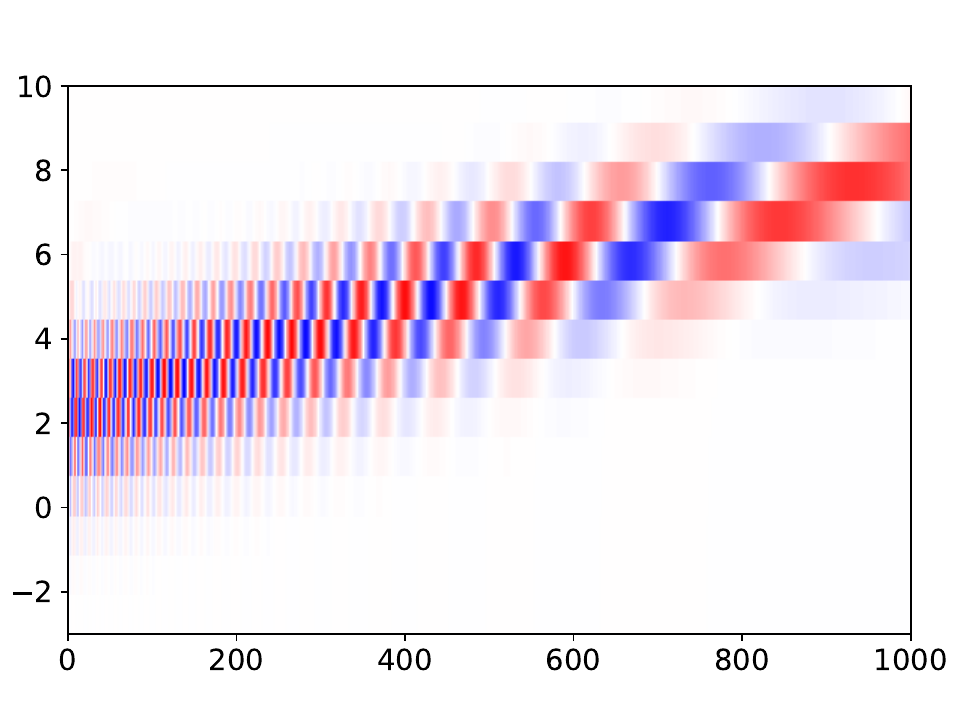} \\
    {\em quasi quadrature for $c = \sqrt{2}$}
      & {\em quasi quadrature for $c = 2$} \\
      \includegraphics[width=0.30\textwidth]{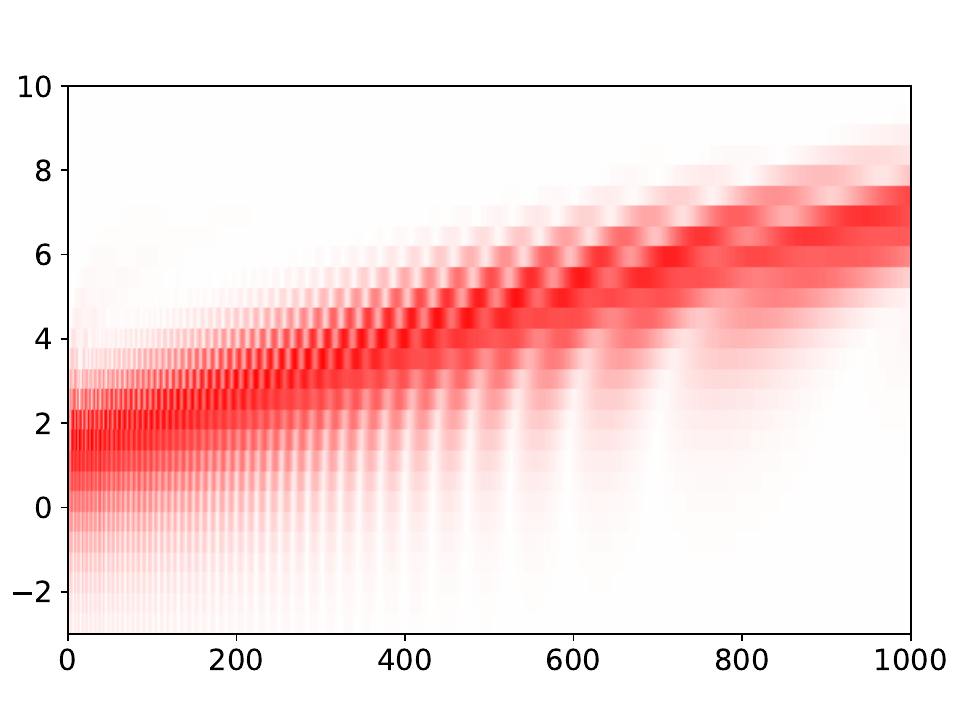}
      &  \includegraphics[width=0.30\textwidth]{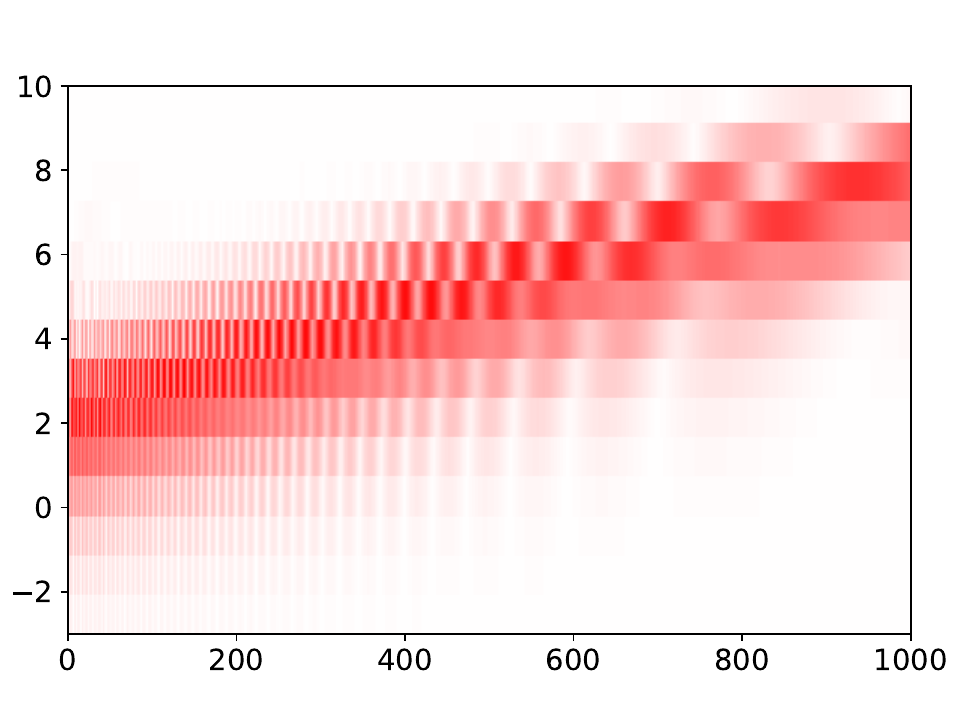} \\
    \end{tabular}
  \end{center}
  \caption{The result of computing discrete approximations of
    convolutions with the temporal derivatives of the time-causal limit
    kernel $L_{\zeta^n}$ for orders $n = 1$ and $n =2$ from
    {\em a sine wave with
    exponentially varying frequency\/}, here over the scale range
    $\sigma = \sqrt{\tau} \in [1/8, 1024]$ and using the scale
    normalization power $\gamma = 1$ corresponding to
    $L_1$-normalization over scales of the underlying temporal
    derivatives of the time-causal limit kernel. In the bottom row, we
    have also shown the result of computing the quasi quadrature measure
    ${\mathcal Q} L = \sqrt{L_{\zeta}^2 + C \, L_{\zeta\zeta}^2}$ for the
    relative weighting parameter $C = 1/\sqrt{2}$.
    As can be seen from the results, the dominant responses in the
    time-causal wavelet representations move from finer to coarser
    scales with increasing local wavelengths in the signal, in
    agreement with the desirable properties of a multi-scale wavelet
    representation, and also in agreement with the scaling
    properties.
    (The reason why the slopes of the red and blue stripes are not
    vertical, but oblique, is because of the different amounts of
    temporal delay at different temporal scales.)
    (Horizontal axes:
    time $t \in [0, 1000]$.
    Vertical axes in the time-causal wavelet representations:
    Effective scale = $\log_2 \sqrt{\tau}$.)
    (In the time-causal wavelet representations, red denotes positive
    values and blue denotes negative values.)}
  \label{fig-sineexpwave-exp}
\end{figure*}

\subsubsection{Scale-selective property of the time-causal wavelet concept}

To demonstrate the ability of the proposed time-causal wavelet
representations to adaptively respond to local structures at different temporal
scales in a signal, let us consider the result of performing a
time-causal wavelet analysis based on explicit temporal derivatives
of the following model signal with
exponentially varying frequency:
\begin{equation}
  f(t) = \sin\left( \exp \left( \frac{b-t}{a} \right) \right)
\end{equation}
for $a = 200$ and $b = 1000$,
see Figure~\ref{fig-sineexpwave-exp} for an illustration.

For this purpose, we computed discrete approximations of the
first-order and second-order temporal derivatives
according to (\ref{eq-def-time-caus-temp-der-disc-wavelet})
over the scale range $\sigma = \sqrt{\tau} \in [1/8, 1024]$
for the scale normalization power $\gamma= 1$ corresponding to
$L_1$-normalization over scales of the underlying temporal
derivatives of the continuous time-causal limit kernel.
The motivation for this choice, is to treat responses at different
scales in a similar manner, so that the magnitudes of the temporal
derivative responses can be readily compared across scales,
as manifested in the scale covariance property
(\ref{eq-transf-prop-sc-norm-temp-ders-limit-kern-perfect-sc-inv}).

To define a feature descriptor that is less sensitive to the local
phase in the signal near the preferred scales, as determined by the
temporal duration of the locally dominant temporal structures,
we also computed the following quasi quadrature measure
(Lindeberg \citeyear{Lin18-SIIMS})
\begin{equation}
  {\mathcal Q} L = \sqrt{L_{\zeta}^2 + C \, L_{\zeta\zeta}^2}
\end{equation}
for the weighting parameter $C = 1/\sqrt{2}$.

As can be seen from the results, the dominant responses in the
time-causal wavelet representations across scales move from finer to coarser
scales with increasing local wavelengths in the signal,
in agreement with the desirable properties of a multi-scale wavelet
representation, and also in agreement with the scaling
properties. Due to the exponential increase of the local wavelengths
in the input signal as function of time, the corresponding
temporal scale estimates ought to also increase exponentially over
time. With the logarithmic parameterization of the temporal scale
parameter used in the figures, we experimentally obtain scale dependencies
that are well described as linear on the logarithmic scale,
thus with a very good match between
the continuous theory and the discrete implementation.

The quantization effects are notably stronger when using a sparser
sampling of the temporal scale levels, as induced by setting
the distribution parameter to $c = 2$, while the quantization effect
are much lower when using a denser sampling of the temporal
scale levels, as induced by setting the distribution parameter
to $c = \sqrt{2}$.

In these respects, the results from this experiment are in agreement
with the theoretical results presented in the previous sections,
and demonstrate how the proposed family
of time-causal wavelet representations can locally capture and reflect
the temporal scales of locally dominant temporal structures
in a time-causal analysis of temporal signals.

\begin{figure*}[hbtp]
  \begin{center}
    \begin{tabular}{cc}
      {\em Input signal\/}  \\ 
      \includegraphics[width=0.30\textwidth]{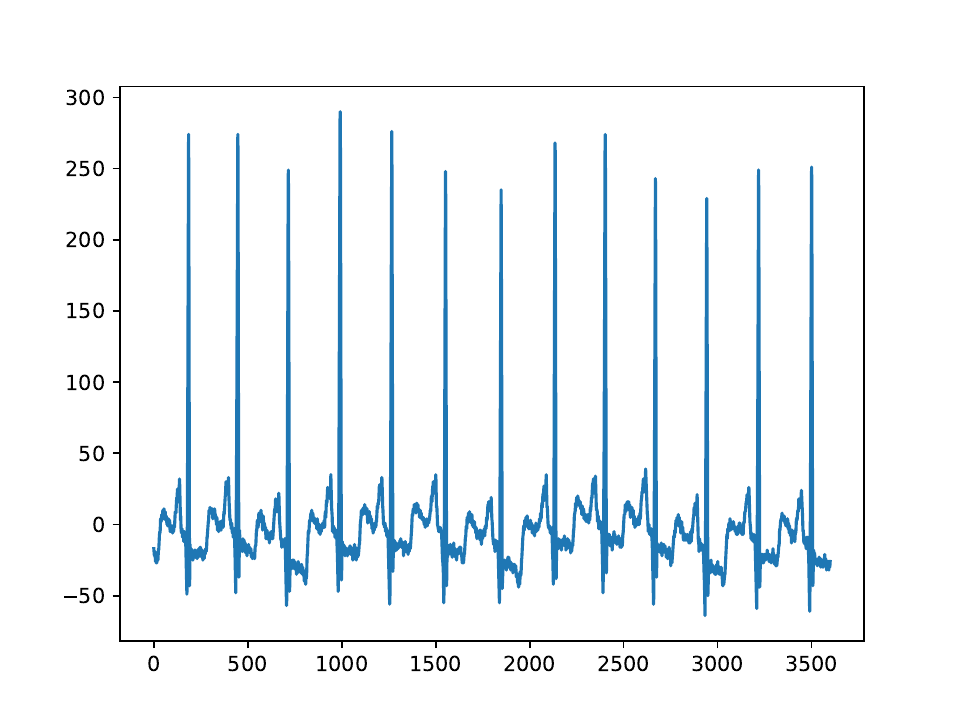}
      \\
      {\em first-order derivative for $c = \sqrt{2}$}
      & {\em first-order derivative for $c = 2$} \\
      \includegraphics[width=0.30\textwidth]{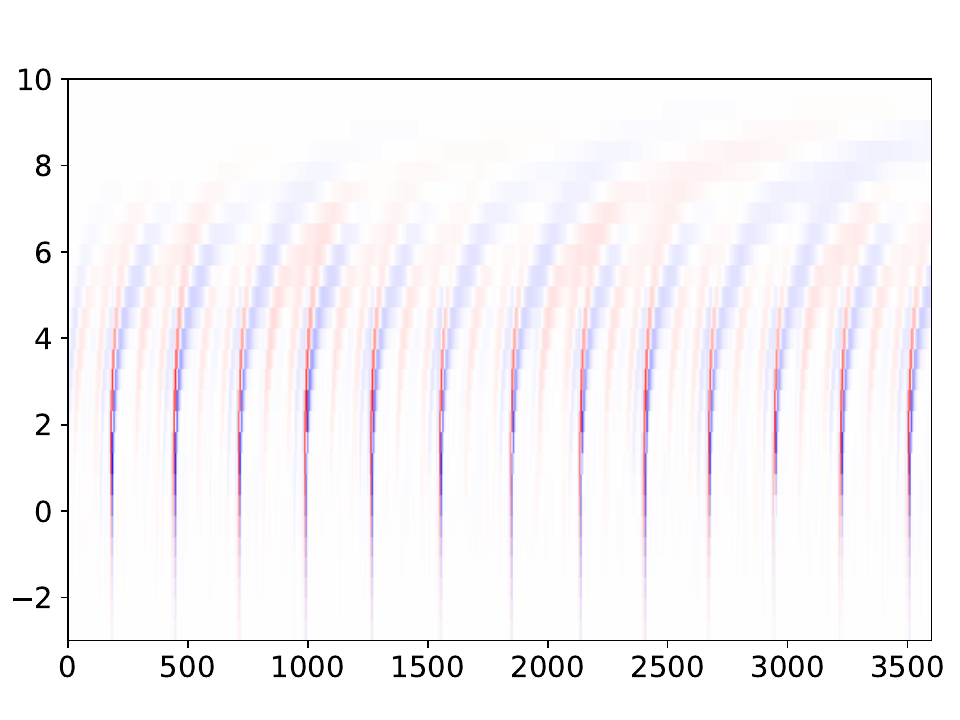}
      &  \includegraphics[width=0.30\textwidth]{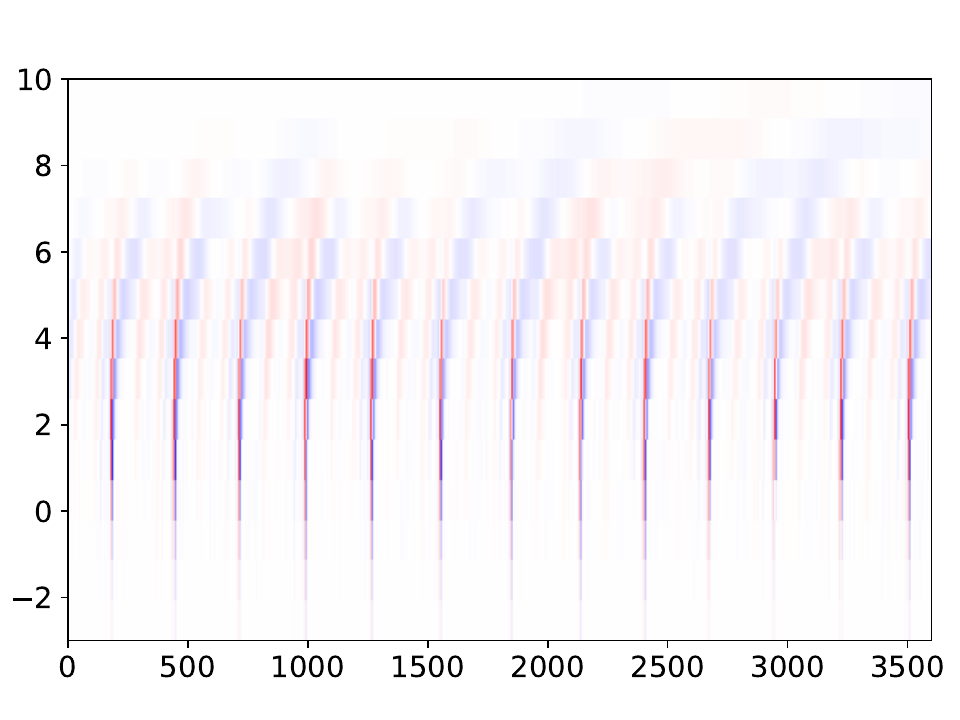} \\
     {\em second-order derivative for $c = \sqrt{2}$}
      & {\em second-order derivative for $c = 2$} \\
      \includegraphics[width=0.30\textwidth]{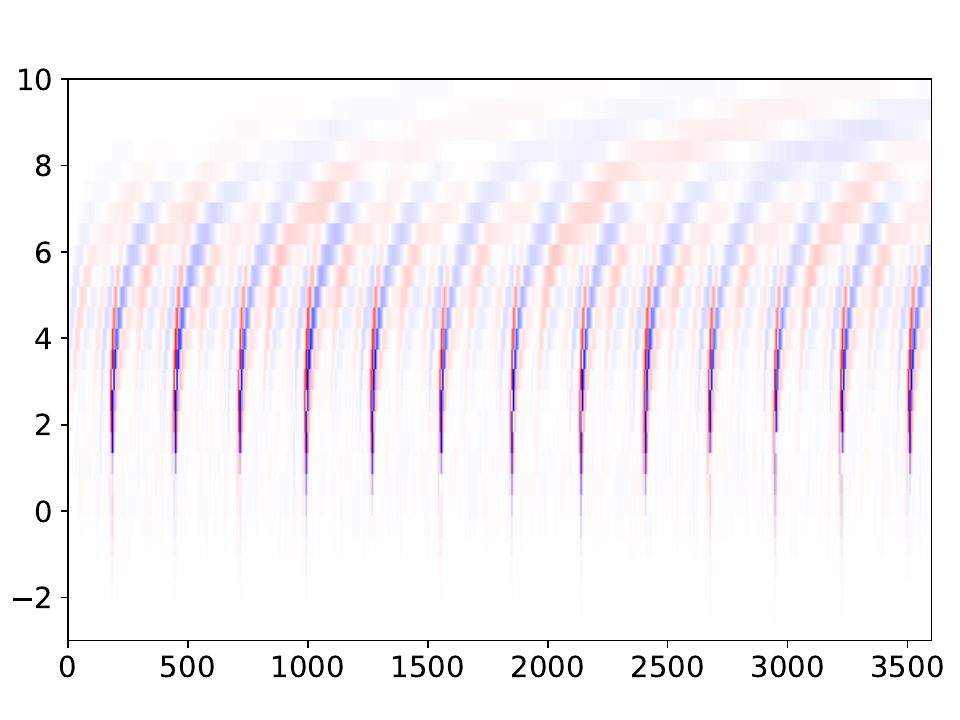}
      &  \includegraphics[width=0.30\textwidth]{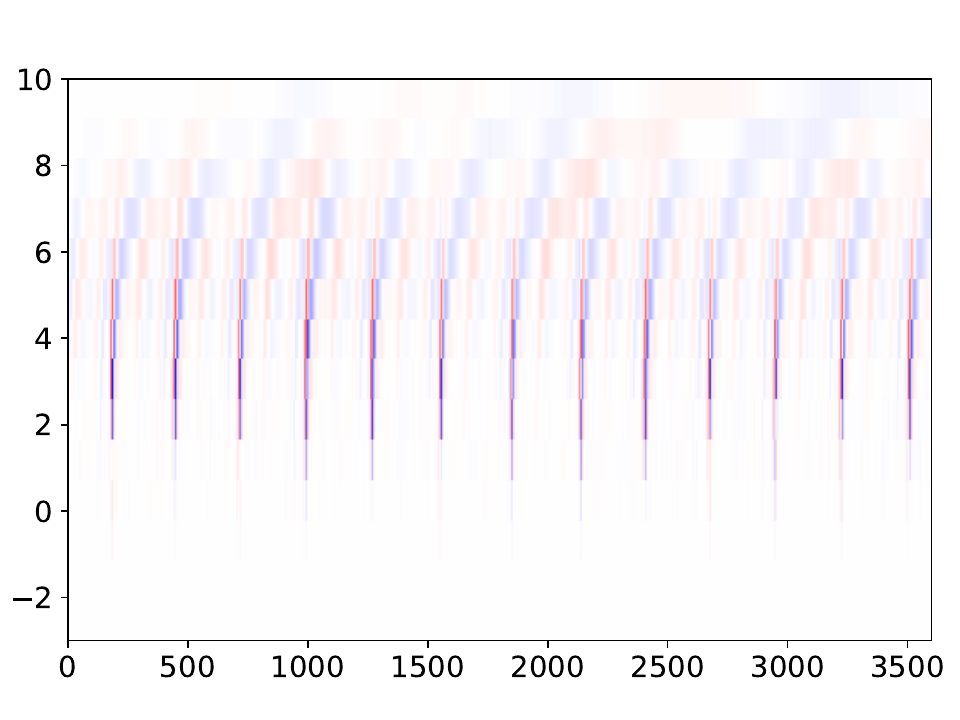} \\
    {\em quasi quadrature for $c = \sqrt{2}$}
      & {\em quasi quadrature for $c = 2$} \\
      \includegraphics[width=0.30\textwidth]{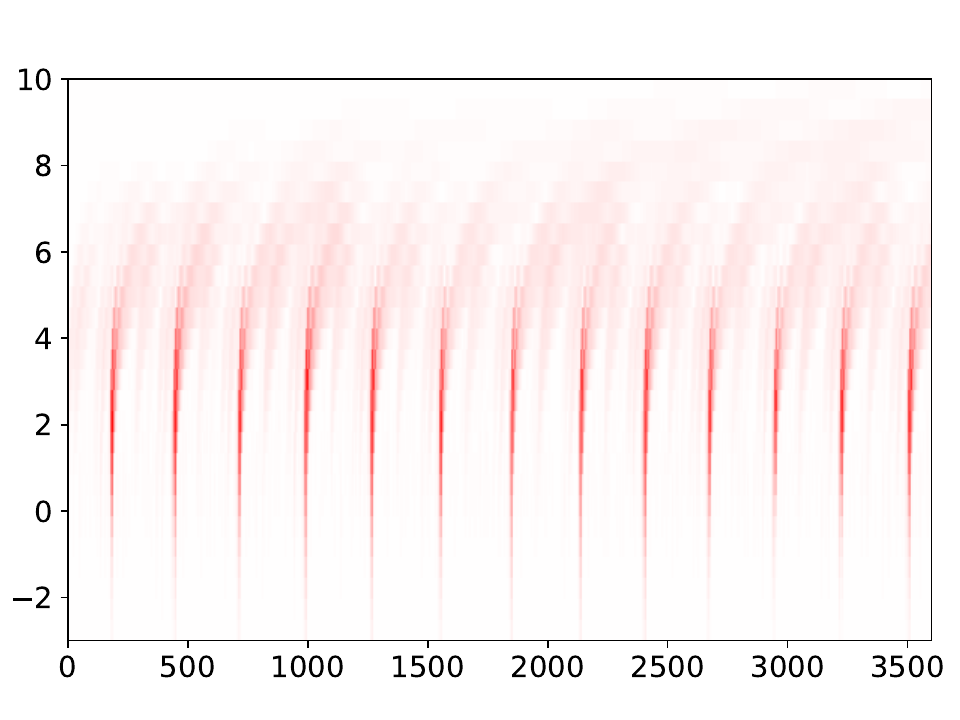}
      &  \includegraphics[width=0.30\textwidth]{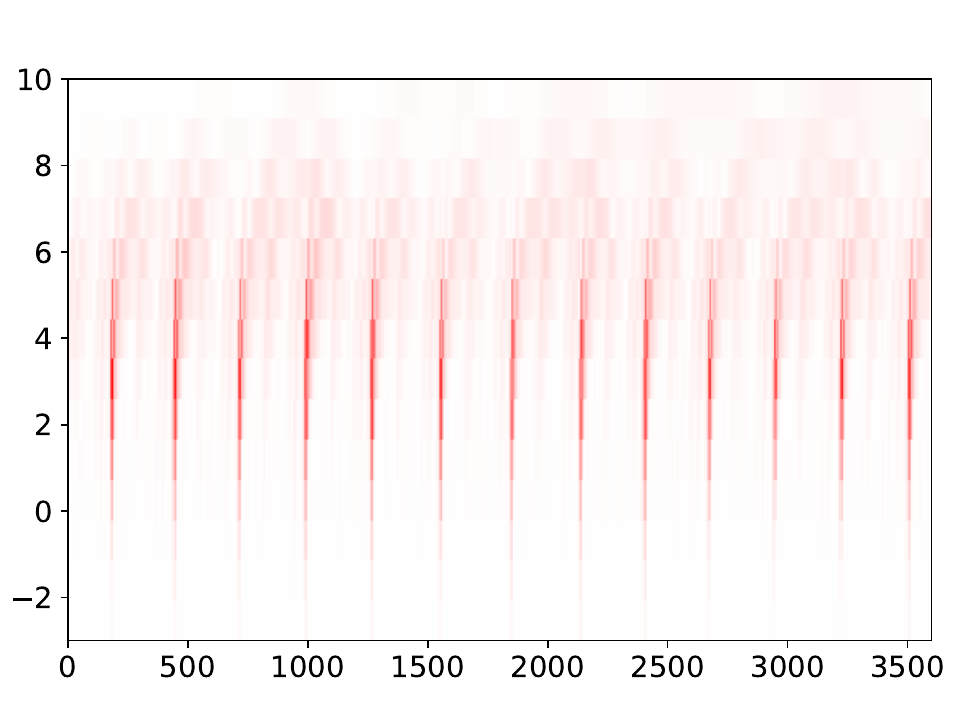} \\
    \end{tabular}
  \end{center}
  \caption{The result of computing discrete approximations of
    convolutions with the temporal derivatives of the time-causal limit
    kernel $L_{\zeta^n}$ for orders $n = 1$ and $n =2$ from
    {\em an electrocardiography signal\/}, here over the scale range
    $\sigma = \sqrt{\tau} \in [1/8, 1024]$ and using the scale
    normalization power $\gamma = 1$ corresponding to
    $L_1$-normalization over scales of the underlying temporal
    derivatives of the time-causal limit kernel. In the bottom row, we
    have also shown the result of computing the quasi quadrature measure
    ${\mathcal Q} L = \sqrt{L_{\zeta}^2 + C \, L_{\zeta\zeta}^2}$ for the
    relative weighting parameter $C = 1/\sqrt{2}$.
    (The reason why the slopes of the red and blue stripes are not
    vertical, but oblique, is because of the different amounts of
    temporal delay at different temporal scales.)
    (Horizontal axes:
    time $\in [0, 3600]$.
    Vertical axes in the time-causal wavelet representations:
    Effective scale = $\log_2 \sqrt{\tau}$.)
    (In the time-causal wavelet representations, red denotes positive
    values and blue denotes negative values.)
  (The values on the vertical axis of the signal have been shifted to make the DC
  level approximately zero.)}
  \label{fig-ecg-exp}
\end{figure*}

\begin{figure*}[hbtp]
  \begin{center}
    \begin{tabular}{cc}
      {\em Input signal\/}  \\ 
      \includegraphics[width=0.30\textwidth]{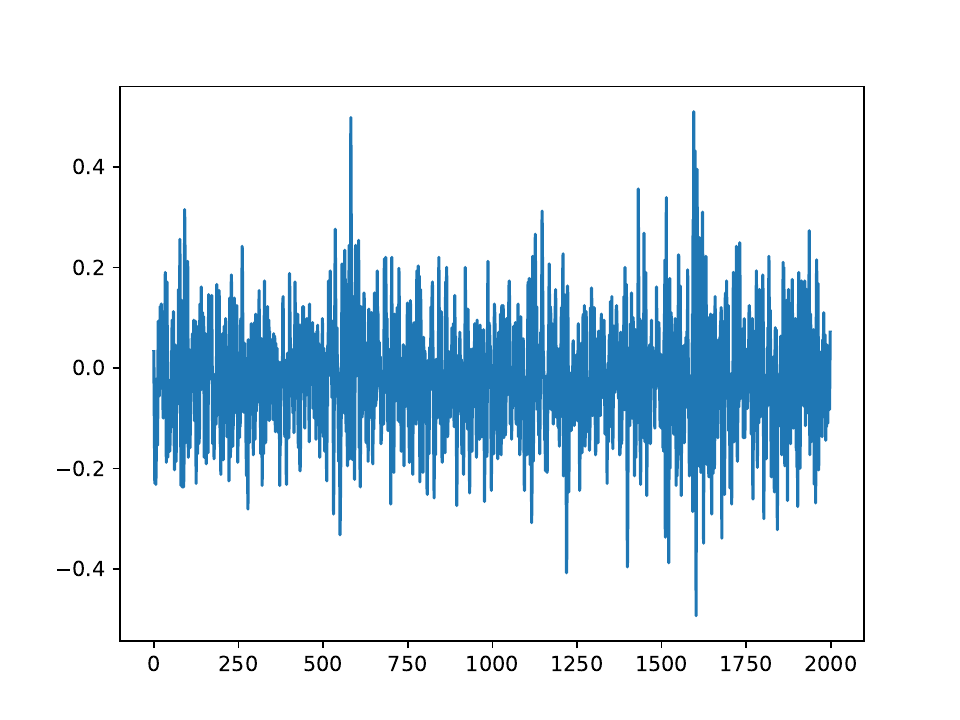}
      \\
      {\em first-order derivative for $c = \sqrt{2}$}
      & {\em first-order derivative for $c = 2$} \\
      \includegraphics[width=0.30\textwidth]{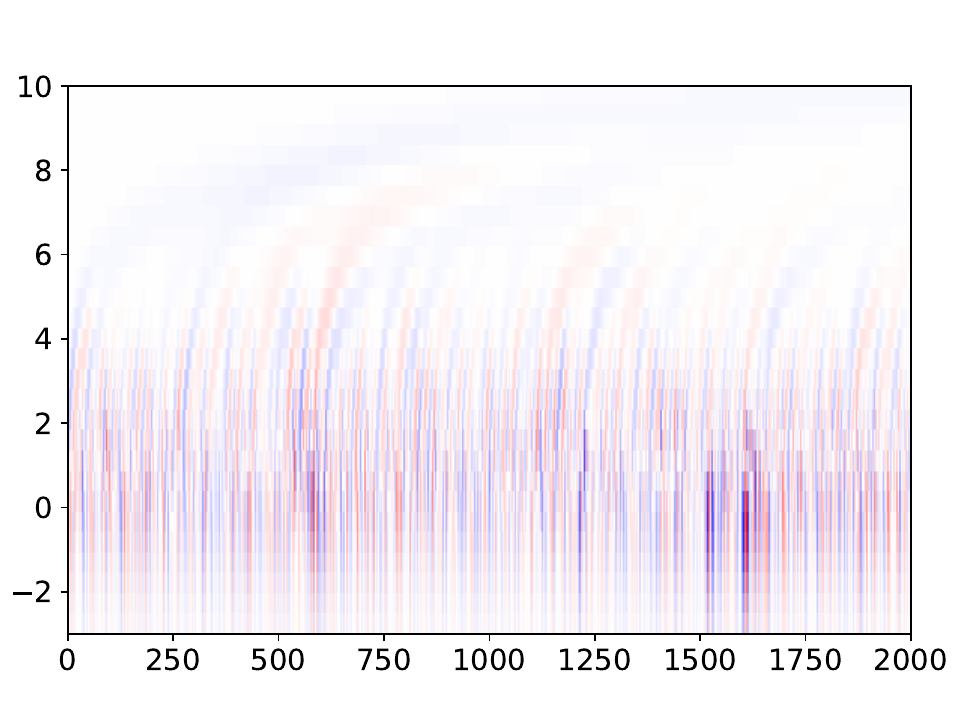}
      &  \includegraphics[width=0.30\textwidth]{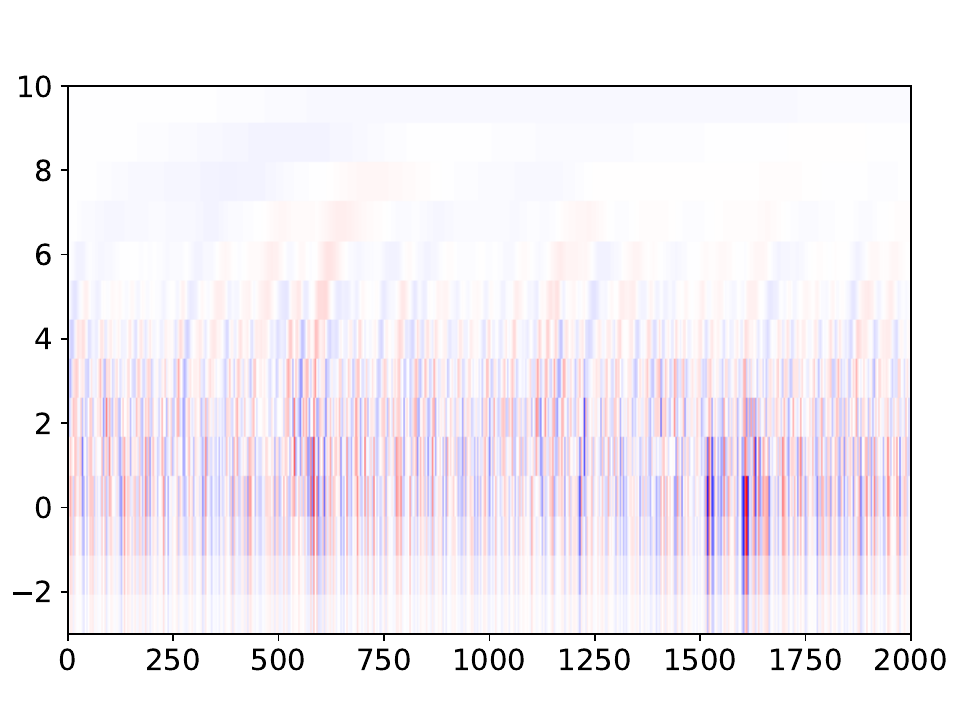} \\
     {\em second-order derivative for $c = \sqrt{2}$}
      & {\em second-order derivative for $c = 2$} \\
      \includegraphics[width=0.30\textwidth]{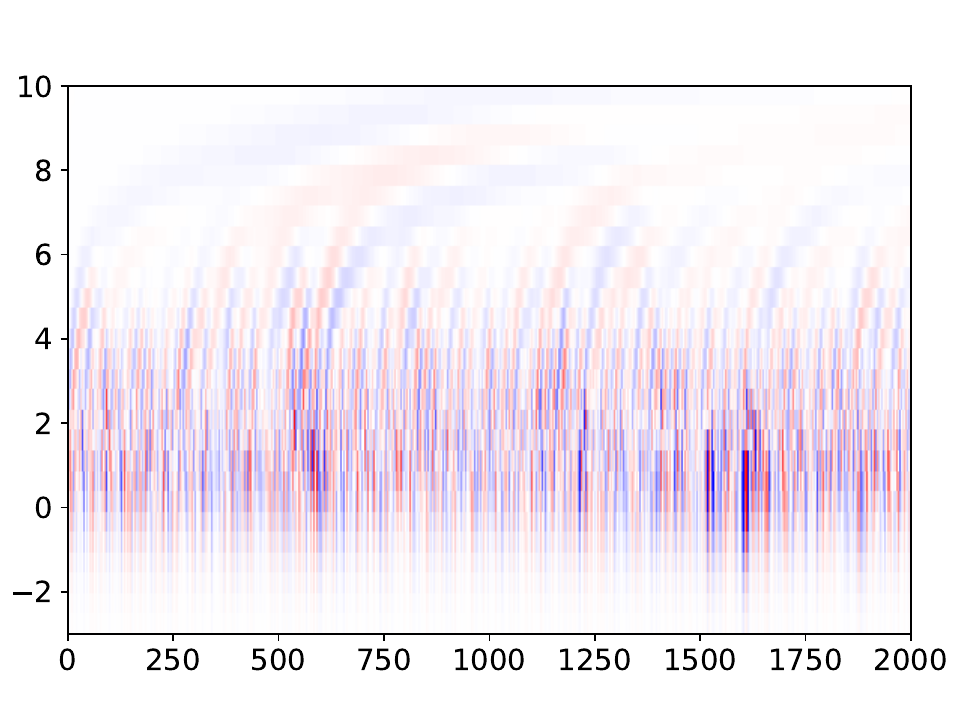}
      &  \includegraphics[width=0.30\textwidth]{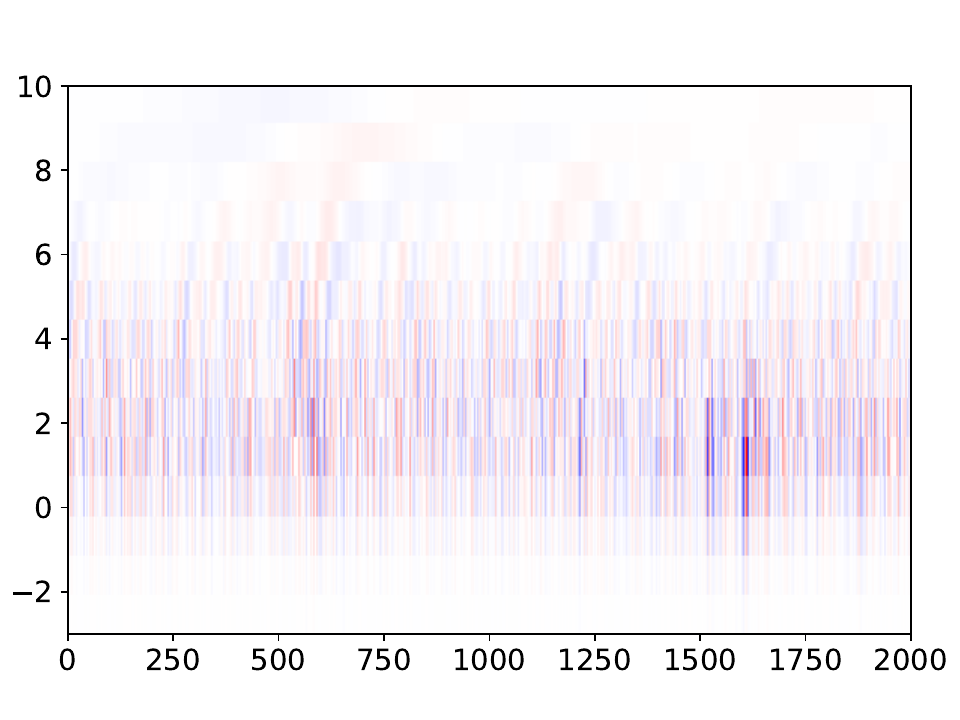} \\
    {\em quasi quadrature for $c = \sqrt{2}$}
      & {\em quasi quadrature for $c = 2$} \\
      \includegraphics[width=0.30\textwidth]{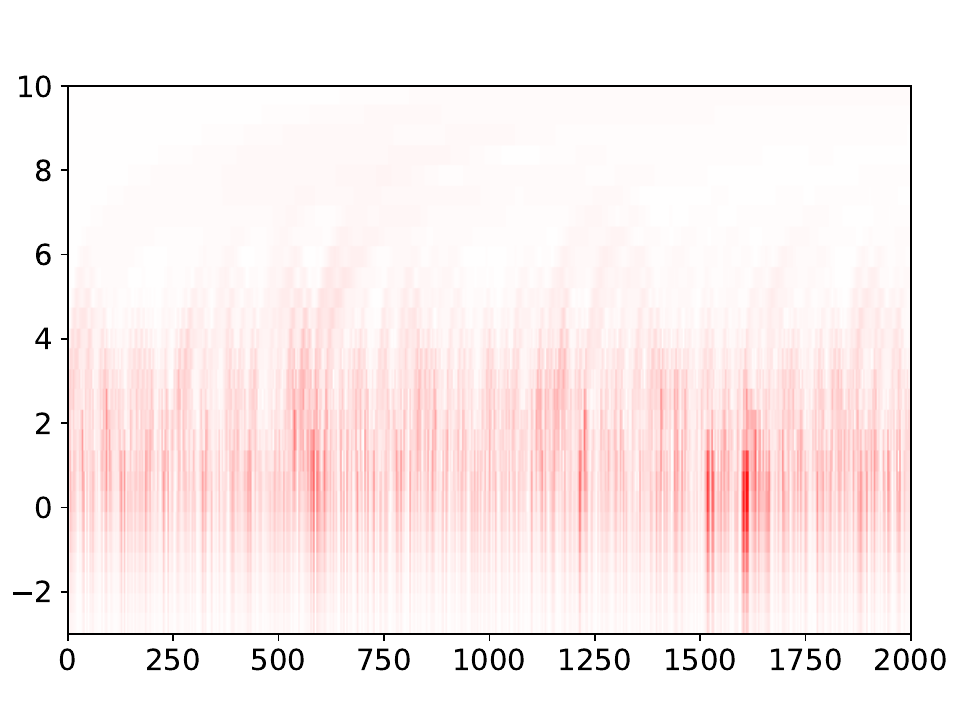}
      &  \includegraphics[width=0.30\textwidth]{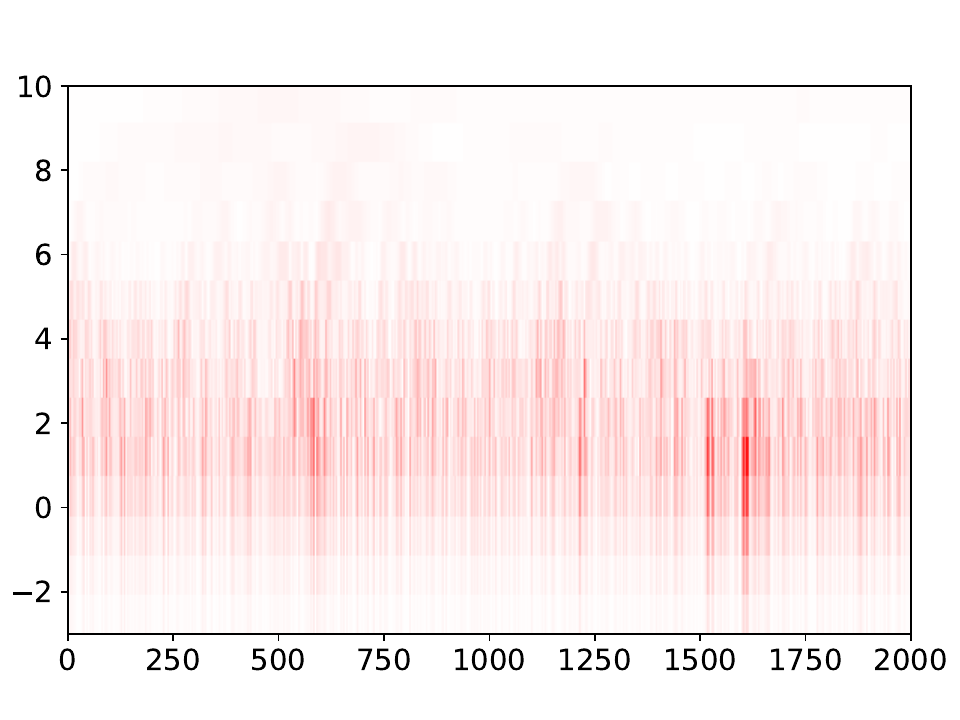} \\
    \end{tabular}
  \end{center}
  \caption{The result of computing discrete approximations of
    convolutions with the temporal derivatives of the time-causal limit
    kernel $L_{\zeta^n}$ for orders $n = 1$ and $n =2$ from 
    {\em a vibration signal\/}, here over the scale range
    $\sigma = \sqrt{\tau} \in [1/8, 1024]$ and using the scale
    normalization power $\gamma = 1$ corresponding to
    $L_1$-normalization over scales of the underlying temporal
    derivatives of the time-causal limit kernel. In the bottom row, we
    have also shown the result of computing the quasi quadrature measure
    ${\mathcal Q} L = \sqrt{L_{\zeta}^2 + C \, L_{\zeta\zeta}^2}$ for the
    relative weighting parameter $C = 1/\sqrt{2}$.
    (Horizontal axes:
    time $\in [0, 2000]$.
    Vertical axes in the time-causal wavelet representations:
    Effective scale = $\log_2 \sqrt{\tau}$.)
    (In the time-causal wavelet representations, red denotes positive
    values and blue denotes negative values.)
  (The values on the vertical axis of the signal have been shifted to make the DC
  level approximately zero.)}
  \label{fig-vib-exp}
\end{figure*}

\begin{figure*}[hbtp]
  \begin{center}
    \begin{tabular}{cc}
      {\em Input signal\/}  \\ 
      \includegraphics[width=0.30\textwidth]{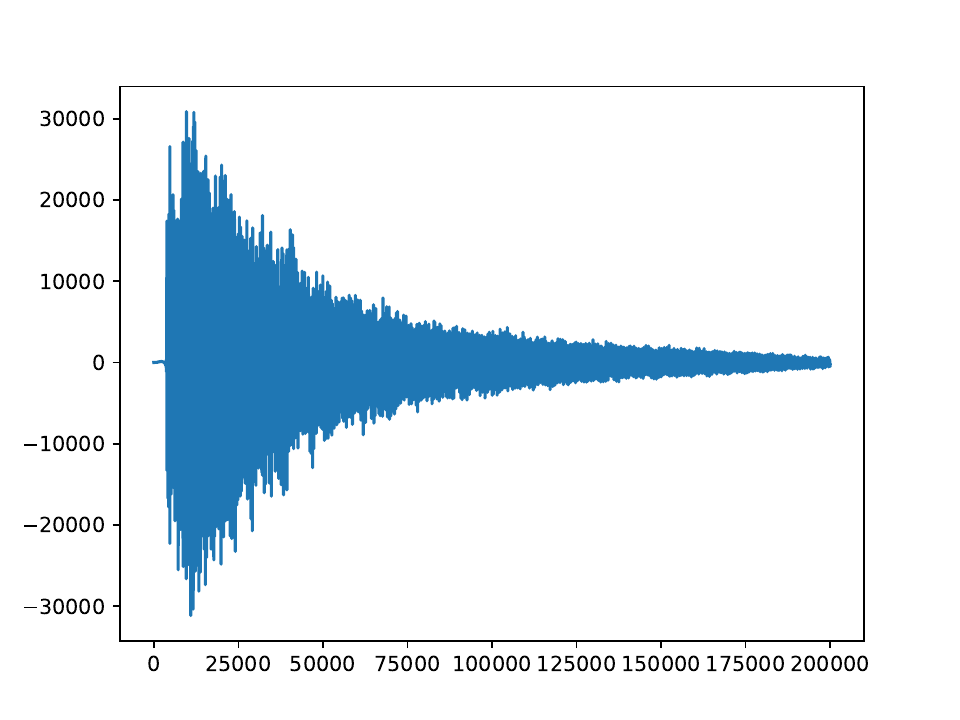}
      \\
      {\em first-order derivative for $c = \sqrt{2}$}
      & {\em first-order derivative for $c = 2$} \\
      \includegraphics[width=0.30\textwidth]{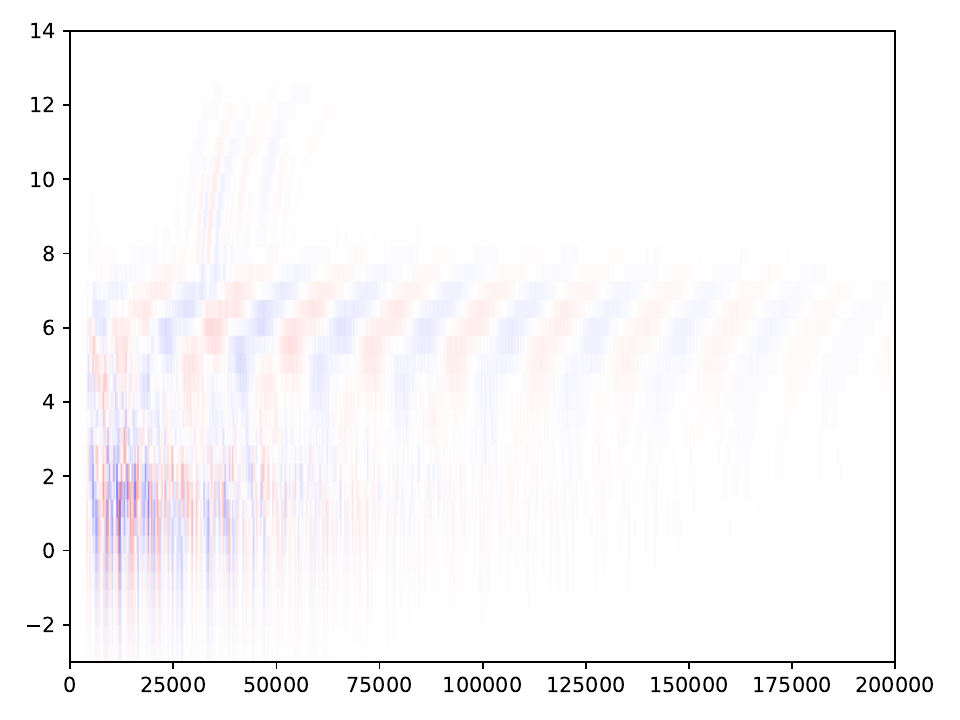}
      &  \includegraphics[width=0.30\textwidth]{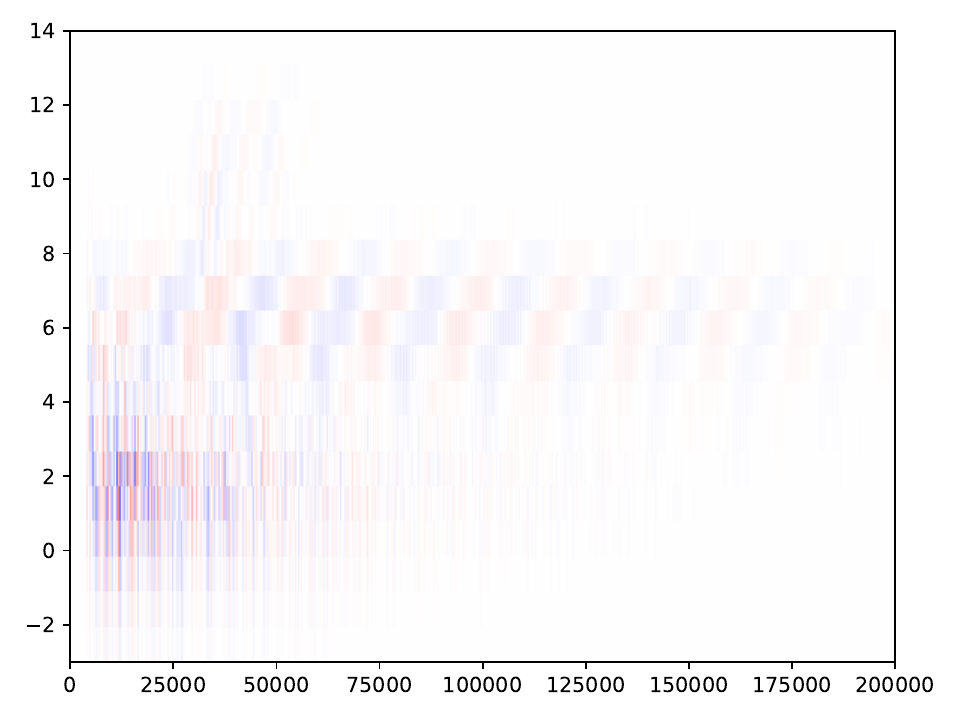} \\
     {\em second-order derivative for $c = \sqrt{2}$}
      & {\em second-order derivative for $c = 2$} \\
      \includegraphics[width=0.30\textwidth]{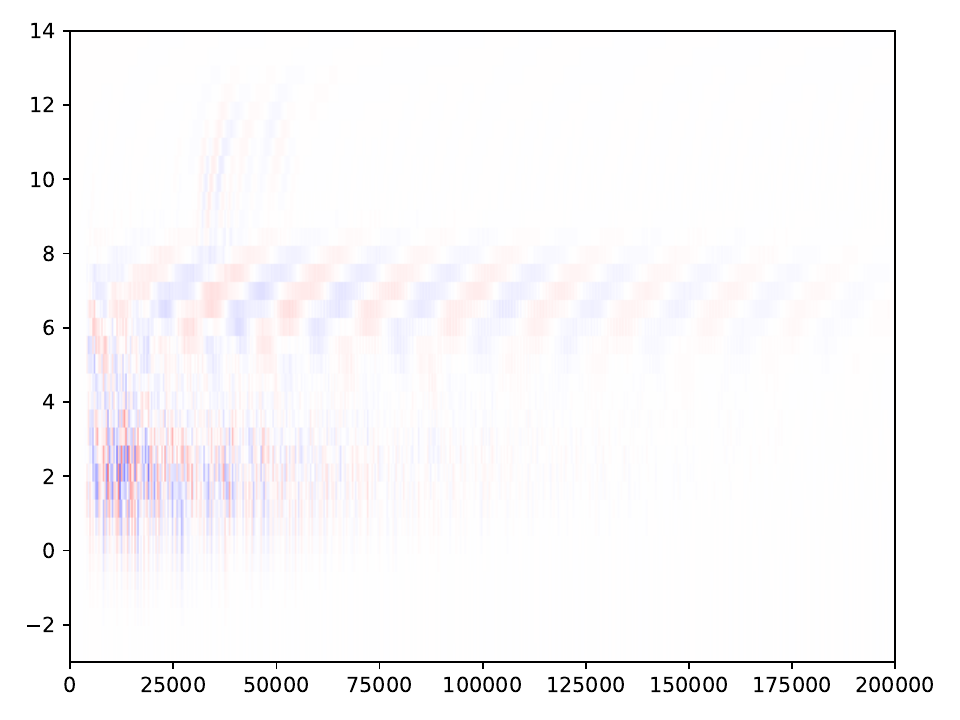}
      &  \includegraphics[width=0.30\textwidth]{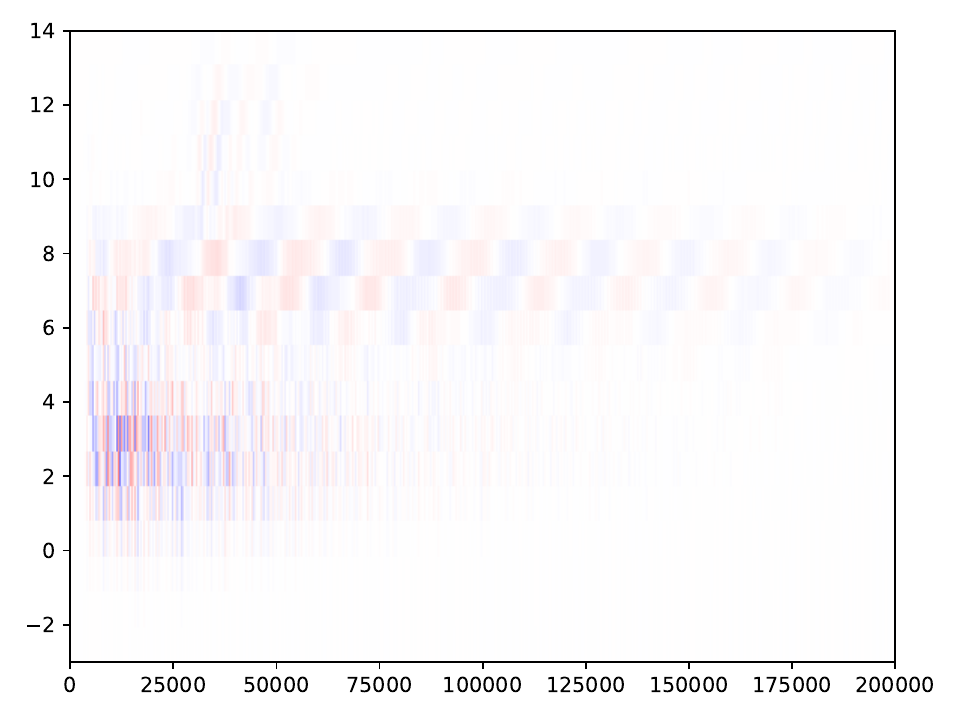} \\
    {\em quasi quadrature for $c = \sqrt{2}$}
      & {\em quasi quadrature for $c = 2$} \\
      \includegraphics[width=0.30\textwidth]{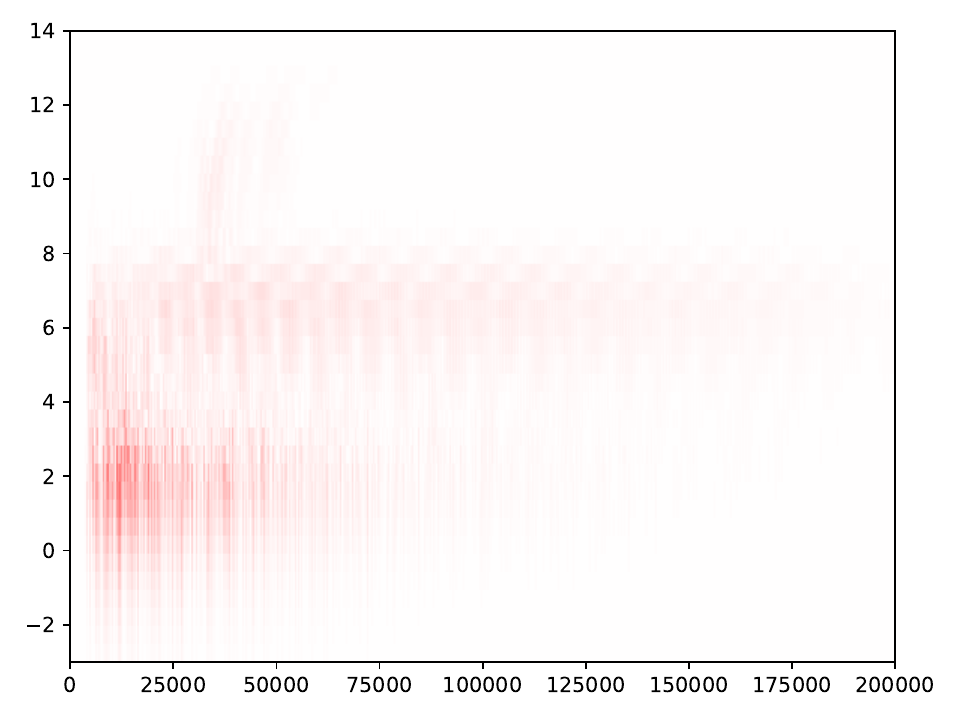}
      &  \includegraphics[width=0.30\textwidth]{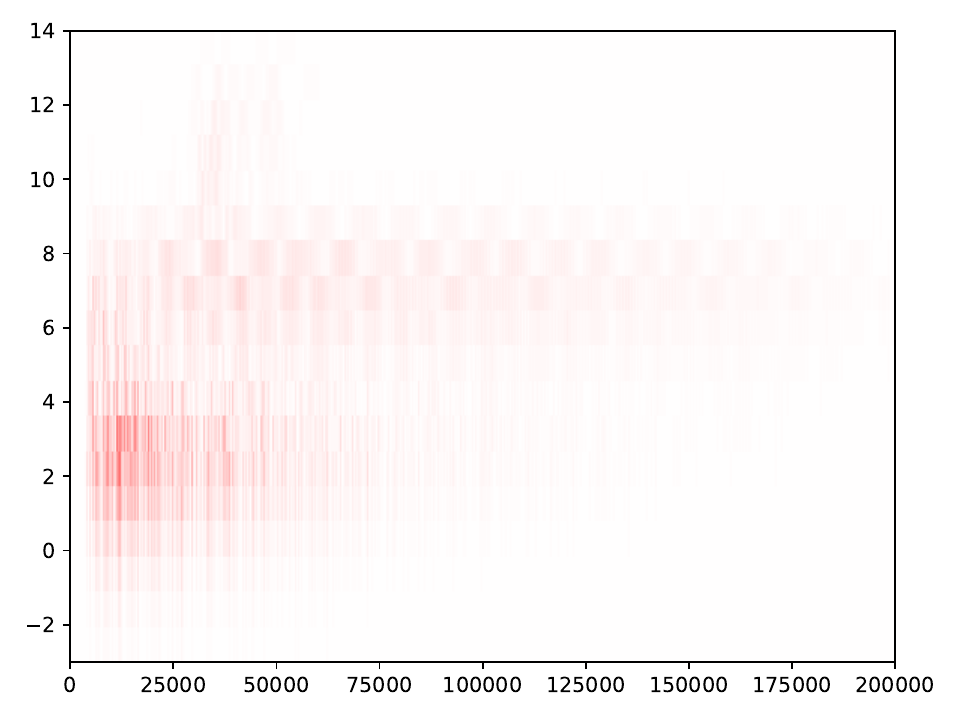} \\
    \end{tabular}
  \end{center}
  \caption{The result of computing discrete approximations of
    convolutions with the temporal derivatives of the time-causal limit
    kernel $L_{\zeta^n}$ for orders $n = 1$ and $n =2$ from
    {\em an auditory signal with a percussion sound from a cymbal\/},
    here over the scale range
    $\sigma = \sqrt{\tau} \in [1/8, 16384]$ and using the scale
    normalization power $\gamma = 1$ corresponding to
    $L_1$-normalization over scales of the underlying temporal
    derivatives of the time-causal limit kernel. In the bottom row, we
    have also shown the result of computing the quasi quadrature measure
    ${\mathcal Q} L = \sqrt{L_{\zeta}^2 + C \, L_{\zeta\zeta}^2}$ for the
    relative weighting parameter $C = 1/\sqrt{2}$.
    (Horizontal axes:
    time $\in [0, 150000]$ relative to a sampling frequency of 44.1~kHz.
    Vertical axes in the time-causal wavelet representations:
    Effective scale = $\log_2 \sqrt{\tau}$.)
    (In the time-causal wavelet representations, red denotes positive
    values and blue denotes negative values.)
  (The values on the vertical axis of the signal have been shifted to make the DC
  level approximately zero.)}
  \label{fig-cymbal-exp}
\end{figure*}

\subsubsection{Time-causal wavelet representations of real-world
  signals}

Let us also illustrate the effects of applying the time-causal wavelet
analysis based on explicit temporal derivatives
to three real-world signals:

Figure~\ref{fig-ecg-exp} shows such results for an electrocardiography signal
from the PhysioNet service (http://www.physionet.org) from
the MIT-BIH Arrhythmia database, item ``NSR/100m (0).mat'' from
(Plawiak \citeyear{Pla17-ECGdata}).
As can be seen from the results, there are strong values in the
wavelet transforms near the transients in the signal. At coarser
temporal scales, the temporal delays become longer. The temporal
delays are, however, significantly shorter when the distribution
parameter is $c = 2$ compared to using $c = \sqrt{2}$.

Figure~\ref{fig-vib-exp} shows such results from a signal with
measurements of vibrations from the NASA Bearing Dataset
(https://www.kaggle.com/datasets/vinayak123tyagi/bearing-dataset)
(1st test recorded 2003.11.25.23.39.56).
As can be seen from the results, there are peaks in the energy of the
wavelet representations near the temporal moments where there are
also major peaks in signal.

Figure~\ref{fig-cymbal-exp} shows corresponding results for an auditory percussion
sound of a cymbal
(from https://pixabay.com/sound-effects/search/cymbal/ item ``Cymbal Crash'').
As we can see, the time-causal wavelet transform
largely responds over three ranges of temporal scales, with different
durations of the responses over these ranges of temporal scales,
and as corresponding to vibrations in the cymbal over different ranges
of frequencies.
In this way, the proposed time-causal wavelet transform leads to an
intuitive separation of the temporal variations in the signal over
different temporal scales.

\section{Summary and discussion}
\label{sec-summ-concl}

We have presented a theory for how the notions of temporal causality
and temporal recursivity can be introduced in wavelet representations.

For this purpose, we have built the time-causal wavelet analysis on a
special temporal smoothing kernel, denoted the time-causal limit
kernel, and which constitutes the result of coupling an infinite set
of truncated exponential kernels in cascade, with specially chosen
time constants to obtain temporal scale covariance and self-similarity
across scales. Specifically, we have in Sections~\ref{sec-methods}
and~\ref{sec-time-caus-wavelets}
proposed to use temporal derivatives of this kernel as mother
wavelets, with complementary normalization in a suitable $L_p$-norm.
We have analyzed the continuous scaling properties of this
representation, and also shown that the temporal derivatives of the
time-causal limit kernel obey the admissibility condition for a mother
wavelet. We have also argued that the temporal derivatives of the
time-causal limit kernel can by theoretical arguments be regarded as
canonical time-causal analogues of the derivatives of the regular
Gaussian kernel, which for example, constitute the foundation for the
Ricker / Mexican hat wavelet.

Since the time-causal limit kernel or its temporal derivatives do
not have any compact closed-form explicit expressions, we have for
this purpose specifically focused on what can be stated theoretically about these
kernels, regarding their recurrence relations across scales, their scaling
properties across scales, and regarding bandpass
wavelet representations defined from differences between time-causal wavelet
representation at adjacent discrete temporal scales.

Computationally, the proposed bandpass wavelet representations
specifically allow from much more straightforward reconstruction of
the original signal from the wavelet transforms.
Whereas a standard (usually non-causal) wavelet representation
requires reconstruction from the general inversion formula
reproduced in Equation~(\ref{eq-def-inv-wavelet-transf}),
when using the bandpass wavelet representation according
to Equation~(\ref{eq-bandpass-wavelet-repr}), the original
signal can instead be recovered from a mere summation of the
bandpass wavelet channels according to
Equation~(\ref{eq-reconstr-temp-scsp-time-caus}).
Notably, this opens up to real-time manipulation of local
contents in the signal, by restricting the reconstruction to subsets
of temporal subbands, that is over subsets of the temporal scale
channels.

As we have shown in
Equation~(\ref{eq-rel-bandpass-kernel-1st-temp-der}), there is a
direct relationship between the underlying computationally equivalent
difference-of-time-causal-limit-kernels (DoT) kernel and the
{\em first-order\/} temporal derivative of the time-causal limit kernel.
This relationship is specifically qualitatively different from the corresponding
relationship for non-causal wavelets based on derivatives of the
Gaussian kernel, where the
corresponding bandpass kernel does according to
Equation~(\ref{eq-reconstr-ricker-wavelet}) instead correspond to the
{\em second-order\/} derivative of the non-causal Gaussian kernel.

Then, we have in Section~\ref{sec-disc-approx}
turned to the problem of numerically
approximating the time-causal limit kernel and its temporal
derivatives for discrete signals.
In this regard, first-order recursive filters constitute the natural
analogue to the first-order integrators, that equivalently describe the
temporal smoothing with truncated exponential kernels, and which constitute
the main computational primitives in the continuous theory.
Thereby, we have describe a canonical way to compute discrete
time-causal wavelet representations in
Section~\ref{sec-disc-wavelet-repr},
and also shown how this representation relates to bandpass
representations, as can also be defined from continuous wavelets.
Notably, the proposed implementation method is fully time-recursive,
and does therefore lend itself directly to real-time processing
on regular signal processing architectures.
Specifically, as described in
Algorithm~\ref{fig-pseudo-code-rec-filt-casc}, the temporal smoothing
can be implemented in terms of a low number (often 4-8) first-order recursive
filters coupled in cascade.
The wavelet transforms are then obtained by applying compact temporal
difference operators according to Equation~(\ref{eq-temp-der-approx-molecules})
to the temporally smoothed data, complemented by complementary scale
normalization. Thereby, real-time computations can
be achieved on very low-capacity computational architectures. 

These theoretical properties do then constitute the foundation for
treating temporal structures in the input signal at different temporal
scales in a self-similar manner for different types of applications.
In Section~\ref{sec-char-scaling-props}
and Section~\ref{sec-exps}, we have specifically analyzed
how well the ideal scaling properties of the continuous theory are
transferred to a discrete implementation, with special attention to
the ability of the proposed time-causal wavelet representations to
reveal locally dominant structures at different temporal scales in
the input signals.

In summary, these characterizations clearly demonstrate how the proposed family
of time-causal wavelet representations can locally capture and reflect
the temporal scales of locally dominant temporal structures based on a
fully time-causal analysis of the input signals.

Based on these results, and the above mentioned principled theoretical
foundations, we propose this notion of time-causal wavelet
analysis as a valuable multi-purpose tool for signal processing tasks,
where streams of signals are to be processed and analyzed in real
time. Beyond real-time monitoring and
time-series analysis and prediction, such applications
may concern purposes of real-time feedback for
{\em e.g.\/}\ control-loop systems, and as time-causal and
scale-covariant temporal basis
functions in learning-based approaches. Specifically, we argue that the
proposed time-causal wavelet representations could be suitable
for signals that may contain local variations over
a rich span of temporal scales, and more generally for analysing
physical or biophysical temporal phenomena, where a fully
time-causal analysis is called for to be physically realistic,
by not in any way accessing data from the future,
as otherwise often done for pre-recorded data.

Concerning further applications of the theory, we have intentionally
in this presentation focused mainly on theoretical
and generic properties of the proposed families of time-causal wavelet
representations, and not here aimed at developing more application-oriented
experiments beyond the proof-of-concept studies in Sections~\ref{sec-exps}
and~\ref{sec-char-scsel-props-blob-edge}.
The motivation for this is that we want the results to be as
timeless as possible, and not depending on specific properties of the
experimental data, which may be different for different application domains,
and may furthermore warrant the inclusion of additional more specialized
computational mechanisms, as well as integration with other signal
processing approaches to build full-fledge applications.

Let us, however, emphasize that temporal derivatives of the
time-causal limit kernel, as used as the mother wavelets in the here
proposed time-causal wavelet representations, have been used
for handling the temporal domain in models for spatio-temporal
receptive fields for purposes of video analysis in classical computer vision
(Jansson and Lindeberg \citeyear{JanLin18-JMIV},
  Lindeberg \citeyear{Lin18-JMIV,Lin18-SIIMS}).
Those results clearly demonstrate that the numerical properties of the
temporal derivatives of the time-causal limit kernel
can be used for computing highly useful results on real-world temporal
data.

Temporal derivatives of the time-causal limit kernel have also
been used for modelling the temporal component in biological
visual receptive fields (Lindeberg \citeyear{Lin21-Heliyon}),
and leading to provable covariance properties under geometric
image transformations (Lindeberg \citeyear{Lin25-JMIV}).
A precursor to the time-causal limit kernel, in terms of finite sets of
truncated exponential kernels coupled in cascade, with logarithmically
distributed temporal scale levels, has furthermore been
used for modelling the temporal component in auditory receptive fields
(Lindeberg and Friberg \citeyear{LinFri15-PONE}). In this work, we
instead make use of the temporal derivatives of the time-casual limit
kernel for developing a time-causal wavelet theory, and do also
describe structural relations to non-causal wavelet representations
based on derivatives of the Gaussian kernel.

Closely related work of using single truncated exponential kernels for
modelling neural computations in biological perception, memory
processes and temporal learning has in turn been performed by
Jacques (\citeyear{JacTigSarHowSed22-ICML}),
Holt {\em et al.\/}\ (\citeyear{HolQiaSch22-ICML}),
Howard {\em et al.\/}\ (\citeyear{HowEsfLeSed23-arXiv}),
Howard (\citeyear{How24-HandBookHumMem}) and
Howard {\em et al.\/}\ (\citeyear{HowEsfLeSed25-CompBrainBeh}),
with close relations to earlier work on temporal scale-space
representations by Lindeberg and Fagerstr{\"o}m
(\citeyear{LF96-ECCV}).
Compared to the zero-order truncated exponential functions used as
temporal basis functions for learning by Howard {\em et al.\/}
and Holt {\em et al.\/},
the time-causal wavelets in this paper obey simplifying cascade
properties over temporal scales and are also DC-balanced.
For biological support  of the logarithmic distribution of the temporal
scale levels used both here, according to
Equation~(\ref{eq-temp-sc-levels}), and in those works,
see Cao {\em et al.\/}\ (\citeyear{CaoBlaChaHasHow22-eLife}).

The raw zero-order (non-differentiated) time-causal limit kernel has
additionally been successfully used as a replacement of the non-causal
Gaussian window function in time-causal time-frequency analysis.
For the corresponding notion of a time-causal and time-recursive
analogue of Gabor filtering, corresponding to a time-causal analogue
of the Morlet wavelet
(Morlet {\em et al.\/} \citeyear{MorAreFouGla82I-GeomPhys,MorAreFouGla82II-GeomPhys}),
see Lindeberg (\citeyear{Lin25-TIT}).
Compared to such time-causal time-frequency analysis,
the time-causal wavelets developed in this paper require significantly
less computations and lead to shorter temporal delays, 
and with a better ability to handle rapid transients.
For more general notions of a non-causal wavelets based on Hermite
transforms, and which constitute extensions of the wavelets based on
first- and second-order derivatives of the Gaussian kernel
to higher orders of differentiation, see
Markett (\citeyear{Mar93-ApproxTh}),
Glaeske (\citeyear{Gla00-BanCentr})
and Pandey and Phukan (\citeyear{PanPhu20-JAnalAppl}).

In this treatment, we have, however, solely focused on temporal
derivatives of orders 1 and 2, and
leave the topic to future work of characterizing the properties
of time-causal wavelets based on temporal derivatives of the
time-causal limit kernel for higher orders of temporal differentiation.

\section*{Acknowledgements}

Python code, that implements the temporal smoothing operation
underlying the formulation of the discrete temporal wavelets, is available
in the pytempscsp package, available at GitHub:
\begin{quote}
  \url{https://github.com/tonylindeberg/pytempscsp}
\end{quote}
as well as through PyPi:
\begin{quote}
  \tt pip install pytempscsp
\end{quote}

\bibliographystyle{abbrvnat}

{\footnotesize
\bibliography{bib/defs,bib/tlmac}
}

\end{document}